\documentclass[a4paper,11pt]{article}

\usepackage{iftex}
\ifPDFTeX\pdfoutput=1\fi
\usepackage{jheppub}
\usepackage{xcolor}
\usepackage{amsmath,amssymb,graphicx,booktabs,array,microtype}
\usepackage{longtable}
\usepackage{mathrsfs}
\usepackage{placeins}

\graphicspath{{images/}}

\makeatletter
\def\@fpheader{\relax}
\makeatother

\newcommand{\Dh}{\hat\Delta}
\newcommand{\Dc}{\mathcal D}
\newcommand{\CP}{\mathbb{CP}}

\newcommand{\Hilb}{\mathcal H}
\newcommand{\Tr}{\operatorname{Tr}}
\newcommand{\Ct}{C_{\rm tilt}}
\newcommand{\Nm}{N_m}
\title{Monodromy Defects and Bulk Renormalization Group Flows on the Fuzzy Sphere}

\author[a]{Kimyeong Lee}
\author[a]{Joydeep Naskar}

\affiliation[a]{Beijing Institute of Mathematical Sciences and Applications, Huaibei Town, Huairou District, Beijing 101408, China}

\emailAdd{klee@bimsa.cn}
\emailAdd{joydeepnaskar@bimsa.cn}

\abstract{
A monodromy around a defect specifies a symmetry transformation, but the low energy properties of the defect depend on  its interaction with   bulk degrees of freedom and also itself. We study this dependence on the fuzzy sphere. Along an anisotropy flow, an $O(3)$ half-twist approaches the $O(2)$ half-twist when its twisted components stay light and disappears when they become massive. At large $N$, untwisted spectators give a positive correction to the defect entropy, fixed by the conformal weight of the free defect. The surviving $O(2)$ monodromy admits a relevant charged perturbation that pins the line. Pinning before or after the easy-plane bulk flow gives matching infrared junction data. A pin of the untwisted component survives only the opposite, easy-axis flow. Walls between opposite pinnings test the stability of the candidate ordered line. They are relevant on the pinned $O(2)$ monodromy lines studied and favoured to be relevant, with a smaller margin, on the ordinary $O(2)$ line, while the ordinary $O(3)$ and $O(4)$ walls lie close to marginality. We separate the wall dimension into a tilt stiffness and an angular shape, comparing them with a Ward identity, a two-loop expansion and a large-$N$ saddle. In the fully twisted large-$N$ model, winding weakens the condensate and lowers the wall dimension and the entropy released by pinning. The resulting entropy comparison permits disordering into the unpinned line wherever the leading wall estimate is relevant. Finally, we construct the regular half-twist on the fuzzy four-sphere as a starting point for analogous questions in five dimensions.}

\keywords{Renormalization Group, Scale and Conformal Symmetries, Boundary Quantum Field Theory, 1/N Expansion}
\hypersetup{pdftitle={Monodromy Defects and Bulk Renormalization Group Flows on the Fuzzy Sphere},pdfauthor={Kimyeong Lee, Joydeep Naskar}}
\begin{document}
\maketitle
\raggedbottom
\clearpage

\section{Introduction}\label{sec:intro}
A symmetry defect is defined by what happens to bulk  fields that are carried around it. For a monodromy defect the rule is simple: the fields will be  transformed by a fixed element of the global symmetry~\cite{Billo:2013jda,Gaiotto:2013nva,Soderberg:2017oaa,Giombi:2021uae,Bianchi:2021snj,Gimenez-Grau:2021wiv}. However, symmetry specifies only this transformation, or the twist. The defect line considered here carries some internal degrees of freedom and so remains dynamical.  Its spectrum, entropy and response to perturbations depend on the interaction with  bulk fields and  also itself. In this paper we ask how these ingredients select the long-distance state of a monodromy defect. A line can change in two ways, since the bulk can flow underneath it, and also  the line can flow on its own. We study these two kinds of flow in three stages, using their different entropy constraints to organize the comparison. A flow on the line must lower the defect entropy, while a bulk flow is not bound by this rule, so the low-lying line spectrum and the defect entropy carry different information along the two.

\begin{figure}[!htbp]\centering
\includegraphics{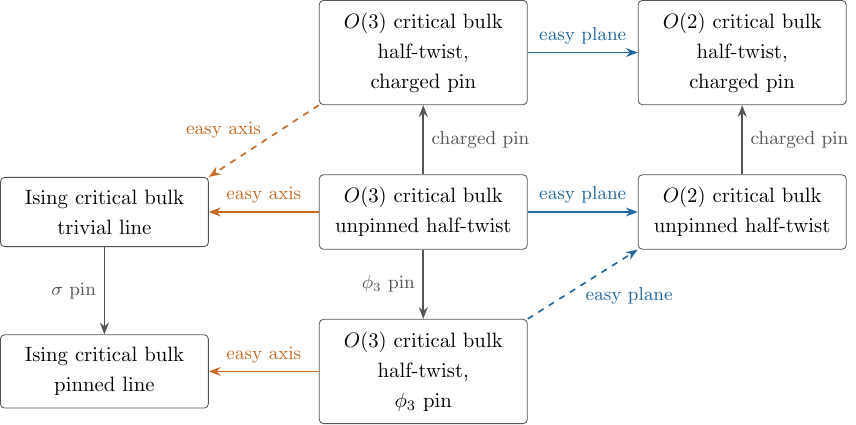}
\caption{Bulk flows and pins. Horizontal arrows follow the anisotropy, while vertical arrows pin the line at fixed bulk criticality. The upper square compares pinning before and after the easy-plane flow. A pin of the untwisted component instead survives the easy-axis flow. Dashed arrows indicate proposed endpoints reached after the bulk flow makes the pinned field massive and a residual, symmetry-preserving line coupling flows away. Figures~\ref{fig:branches} and~\ref{fig:pins} test these connections; finite-size remnants of strong pins are discussed in Appendix~\ref{app:pinbranches}.}
\label{fig:story}
\end{figure}

The first stage is a bulk renormalization group (RG) flow that removes some of the fields on which the symmetry acts. If the twist still acts on the light fields, we expect an infrared monodromy, while if it acts only on massive fields, the light theory should no longer detect it. To make this concrete we take the three-dimensional $O(3)$ Wilson--Fisher theory, whose order parameter has components $(\phi_1,\phi_2,\phi_3)$, with the twist
\begin{equation}
 g_\pi=\operatorname{diag}(-1,-1,+1),
 \qquad(\phi_1,\phi_2,\phi_3)\mapsto(-\phi_1,-\phi_2,\phi_3),
 \label{eq:axialtwist}
\end{equation}
and perturb it by an anisotropy that commutes with the twist. The easy-plane branch of the flow keeps $(\phi_1,\phi_2)$ light and leads toward the $O(2)$ theory, whereas the easy-axis branch leaves only the untwisted Ising field.\footnote{The competition between the $O(2)$ and Ising fixed points under such anisotropies is a classic multicritical problem~\cite{Pelissetto:2000ek,Calabrese:2002bm,Hasenbusch:2011zwv}.} The fuzzy sphere places the symmetry flux and the bulk deformation in the same finite Hamiltonian~\cite{Zhu:2022gjc,Hu:2023xak,Hu:2023ghk,He:2026ong}.\footnote{The fuzzy sphere has since been applied to many three-dimensional theories and their flows, including $O(N)$ models~\cite{Guo:2025odn}, fermionic theories~\cite{Zhou:2025kng} and the flow from $SO(5)$ to $O(4)$~\cite{Yang:2025stu}, as well as to conformal defects and boundaries~\cite{Zhou:2023fqu,Zhou:2024dbt,Dedushenko:2024nwi}.} We can therefore follow the defect along both branches and compare the endpoints with independently constructed infrared defects (Fig.~\ref{fig:story}). The easy-plane endpoint is not yet the end of the story. Its lowest charged mode has dimension below one, so a line coupling that breaks the $U(1)$ symmetry is relevant there, and the third stage below continues from this endpoint.

The second stage concerns the fields that the twist does not touch. An untwisted spectator carries none of the monodromy, but it interacts with the twisted fields, and through these interactions it can change the defect entropy. The entropy of a line decreases along RG flows on the line when the bulk is held critical~\cite{Affleck:1991tk,Friedan:2003yc,Casini:2016fgb,Kobayashi:2018lil,Cuomo:2021rkm}, whereas a bulk flow that removes a spectator lies outside that theorem~\cite{Green:2007wr,Fredenhagen:2009tn,Gaiotto:2012np}. The defect entropy is therefore a natural place to look for physics beyond the surviving symmetry. At large $N$ the spectators compensate the deficit of $\langle\phi_a\phi_a\rangle$ left by the twisted pair. This constraint also governs the response to pinning, where a condensate contributes to the balance of fluctuations.

The third stage is a coupling on the line itself, to the relevant charged mode of the $O(2)$ monodromy, which pins the line. Kravchuk and Radcliffe identify the endpoint of this flow as a monodromy pinning defect~\cite{Kravchuk:2025evf}, which becomes the ordinary pinned line when the twist is removed. Pinning raises the question whether a line defect can order. Opposite pinnings define two line conditions joined by a domain wall, and the candidate ordered line built from them is stable only if such walls are irrelevant~\cite{Lanzetta:2025xfw}.\footnote{Line and surface defects of the $O(N)$ model and the phases they support are discussed from several angles in Refs.~\cite{Allais:2014fqa,Cuomo:2022xgw,Aharony:2022ntz,Nahum:2022fqw,Rodriguez-Gomez:2022gbz,Rodriguez-Gomez:2022gif,Krishnan:2023cff,Trepanier:2023tvb,Giombi:2023dqs,Raviv-Moshe:2023yvq,Cuomo:2023qvp}, and the closely related ordering at a boundary of the $O(N)$ model in Refs.~\cite{Metlitski:2020cqy,Padayasi:2021sik}.} We connect the two kinds of flow by asking whether pinning before or after a bulk flow reaches the same line. We compare pins of the twisted pair and of the untwisted component along both branches, testing whether each pin survives when its field becomes massive. The wall dimensions test stability against wall proliferation, while the defect entropy constrains the possible endpoint.

\subsection{Summary of results}\label{sec:results}
Throughout, $N$ counts vector components and $\Nm$ counts untwisted orbitals on $S^2$. The results follow these three stages, with a fourth item opening the higher-dimensional problem.
\begin{enumerate}
\item \textbf{The monodromy follows the light fields.} On the easy-plane branch the $O(3)$ half-twist flows to the $O(2)$ half-twist. Its low-lying defect spectrum and the normalized coupling of the order parameter to the lowest charged defect mode agree with those of an independently tuned $O(2)$ Hamiltonian at every size we compared. Near four dimensions this coupling is the same for every model at first order. We compute its second-order correction, which distinguishes the two endpoints [Eq.~\eqref{eq:betaeps2}]. On the easy-axis branch, once the twisted flavours are frozen out and the local potential at the flux is removed, the line cannot be distinguished from no defect in the tested channels.

\item \textbf{Untwisted spectators raise the defect entropy.} For an $O(N)$ model with one twisted complex component, the large-$N$ gap equation forces the $N-2$ untwisted spectators to enhance their fluctuations near the defect, through the singular branch of their lowest line mode. The resulting correction to the defect entropy is positive and fixed by the conformal weight of the free twisted scalar. For a twist under which a charged field acquires the phase $e^{2\pi\mathrm i\alpha}$ around the defect,
\begin{equation*}
 \log\mathfrak g_N(\alpha)=\log\mathfrak g_{\rm free}(\alpha)
 +\frac{64}{3\pi N}\,\mathfrak h_{\rm free}(\alpha)+O(N^{-2}),
\end{equation*}
where $\mathfrak g$ is the expectation value of a circular defect, so that $\log\mathfrak g$ is the defect entropy, and $\mathfrak g_{\rm free}$ and $\mathfrak h_{\rm free}$ are the defect expectation value and conformal weight of the free twisted complex scalar.\footnote{The cancellation of the fluctuation trace in every circle-mode sector is proved in Appendix~\ref{app:sector}. The coefficient then follows if the large-momentum expansion holds uniformly, which we check on the tabulated integrand, and if the isotropic cutoff gives the covariant circular-defect entropy, which dimensional regularization supports near four dimensions. The direct traces agree with the formula (Appendix~\ref{app:largeM}).} A bulk flow that removes one spectator therefore ends on the defect with the larger entropy, so a field that carries none of the monodromy still changes the defect.\footnote{This comparison assumes that the flow connects the corresponding saddles. For the three-dimensional $O(3)\to O(2)$ flow, the large-$N$ expansion does not determine the sign or magnitude of the entropy change. Near four dimensions the real part of the increase is positive through second order, but its continuation to three dimensions is not controlled (Sec.~\ref{sec:entropy}).}

\item \textbf{Pinning, walls and the fate of ordered lines.} Pinning before or after the easy-plane flow gives matching infrared junction data. A pin of the untwisted component instead survives only the easy-axis flow. The family obtained by charged pinning connects to the ordinary pinned line as the twist is removed. We measure the dimension $\Delta_{+-}$ of the wall between opposite pinnings and separate it as $\Delta_{+-}=2\Ct\mathfrak r$ into a tilt stiffness $\Ct$ and an angular shape $\mathfrak r$. A Ward identity measures the stiffness, while our two-loop calculation and large-$N$ junction saddle test the shape. The twist lowers the measured wall dimension. In $O(2)$ the walls destabilize the candidate ordered monodromy line, and their relevance on the ordinary pinned line is favoured with a smaller margin. The ordinary $O(3)$ and $O(4)$ walls lie close to marginality, while walls become irrelevant at sufficiently large $N$. In the fully twisted large-$N$ model, the constraint relates the condensate to the entropy released by pinning. Wherever the leading wall estimate is relevant, the entropy permits the candidate ordered line to disorder into the unpinned line. Including the next-order entropy correction preserves this compatibility with the leading wall estimate at every twist evaluated.

\item \textbf{A starting point in five dimensions.} We construct the regular half-twist around a great $S^2$ of the fuzzy four-sphere as a finite Hermitian projected Hamiltonian, providing a starting point for bulk flow, spectator dressing and local order on a three-dimensional defect.
\end{enumerate}

Several of these questions have been approached before. Monodromy defects on the fuzzy sphere were studied by Qian~\cite{Qian:2026lrp} and by Tang and Zhu~\cite{Tang:2026kzf} in parallel with our three-dimensional construction and fixed-point calculations, and we use their results as independent benchmarks in Appendix~\ref{app:three} and to identify the endpoints in the main text. Kravchuk and Radcliffe~\cite{Kravchuk:2025evf} identified the monodromy pinning defect, and we add its realization on the lattice, its junction observables and their dependence on the twist. Yang, Hu, Chen and Zhu~\cite{Yang:2026qtk} measured the wall of the ordinary pinned line on the fuzzy sphere and found it relevant in $O(2)$, with a threshold near $N=3$. Our $O(2)$ result supports theirs, while the separation into a Ward stiffness and an angular shape and the more cautious reading of $O(3)$ and $O(4)$ are new. Cuomo, He and Komargodski~\cite{Cuomo:2024psk} computed the internal-angle junction of pinned lines at one loop, the starting point of our two-loop and large-$N$ shape.

\FloatBarrier

\subsection{Organization of the paper}
Section~\ref{sec:defects} introduces the monodromy, the bulk deformation and the observables, and Sec.~\ref{sec:flows} follows the two branches of the bulk flow. Section~\ref{sec:entropy} computes the spectator dressing and its consequence for the endpoint entropies. Section~\ref{sec:pinmono} follows the pinning flow and its relation to the bulk flow, develops the junction method on ordinary pinned lines, and then studies how the twist changes the walls and their entropy constraint. Section~\ref{sec:five} gives the four-sphere construction that starts the five-dimensional problem, and Sec.~\ref{sec:discussion} summarizes the results and discusses open questions. The appendices supply the three-dimensional Hamiltonians and benchmarks (Appendix~\ref{app:three}), conformal perturbation theory and trajectory diagnostics (Appendices~\ref{app:cpt} and~\ref{app:open}), the entropy and junction derivations (Appendices~\ref{app:largeM} and~\ref{app:pinning}), and the four-sphere details (Appendix~\ref{app:s4}). Numerical coverage and notation are collected in Appendices~\ref{app:numerics} and~\ref{app:notation}.\footnote{The codes associated with this paper can be found in this \href{https://github.com/naskarjoydeep/fuzzy-sphere-defects}{\texttt{GitHub repository}.}}

\section{Monodromy and the bulk deformation}\label{sec:defects}
\subsection{What the symmetry fixes}
The bulk theories in this paper are critical points of the $O(N)$ model in $d$ spacetime dimensions\footnote{The flow and pinning calculations on the fuzzy two-sphere study $d=3$, with $\epsilon$-expansion comparisons at $\epsilon=1$. Section~\ref{sec:five} and Appendix~\ref{app:s4} give the separate four-sphere construction and tests for $d=5$.},
\begin{equation}
 S=\int \mathrm d^dx\left[\tfrac12\partial_\mu\phi_a\partial^\mu\phi_a+\tfrac{\lambda_4}4(\phi_a\phi_a)^2\right],\qquad a=1,\ldots,N,
 \label{eq:bulkaction}
\end{equation}
with the quadratic coupling tuned to criticality. In $d=4-\epsilon$ the Wilson--Fisher fixed point lies at $\lambda_4=8\pi^2\epsilon/(N+8)+O(\epsilon^2)$, and $N=1$ is the Ising model. A monodromy defect $\Dc$ adds no term to this action. It is a condition on the fields, and we first recall what the symmetry alone determines. In local coordinates $(y,r,\theta)$, with $y\in\mathbb R^{d-2}$ along the defect and $(r,\theta)$ transverse to it, a field $\phi_q$ of charge $q$ under the rotations of a twisted plane obeys
\begin{equation}
 \phi_q(y,r,\theta+2\pi)=e^{2\pi\mathrm iq\alpha}\phi_q(y,r,\theta),
 \qquad s\in\mathbb Z+q\alpha.
 \label{eq:monodromy}
\end{equation}
Here $\alpha$ is the twist, and $s$ is the ``transverse spin",  the charge under rotations about the defect. The monodromy fixes the allowed transverse spins, but defects with the same twist can still differ in how the fields behave as $r\to0$~\cite{Giombi:2021uae,Bianchi:2021snj}. We write $\hat\psi_s$ for a defect primary of transverse spin $s$ and $\Dh_s$ for its dimension. For a free scalar, retaining $r^{|s|}$ gives the regular defect, with
\begin{equation}
 \Dh_s=\frac{d-2}{2}+|s|,
 \label{eq:freespec}
\end{equation}
while the alternative behaviour $r^{-|s|}$, where it is allowed, defines a different boundary condition. In an interacting theory the corresponding exponents belong to the physical fields and have to be determined dynamically.

For the $O(3)$ flow we deform the fixed point by the traceless tensor that distinguishes the $(1,2)$ plane from the third direction, together with the singlet that retunes the thermal direction,
\begin{equation}
 S\to S+\mu_t\int \mathrm d^3x\,t_W+\mu_{\mathcal E}\int \mathrm d^3x\,\mathcal E.
 \label{eq:anisoaction}
\end{equation}
Here $t_W$ and $\mathcal E$ are the tensor and singlet primaries, whose free-field forms are $\phi_1^2+\phi_2^2-2\phi_3^2$ and $\phi_a\phi_a$. The tensor dimension $\Delta_t$~\cite{Chester:2020iyt} gives the scaling exponent $y_t=3-\Delta_t$ of the anisotropy. Both signs of the anisotropy preserve Eq.~\eqref{eq:axialtwist}. On the lattice the anisotropy is the coupling $g$ of Eq.~\eqref{eq:amp_ops} below, and in our convention $g<0$ keeps the twisted pair light and $g>0$ keeps the untwisted component light, while the thermal coupling is retuned along the way to remain near criticality.

\subsection{Realizing and measuring the defect}\label{sec:method}
Radial quantization sends the line to the two poles of $S^2$ on each slice of the cylinder. The fuzzy sphere replaces the one-particle space by a monopole lowest Landau level with $\Nm$ orbitals~\cite{Wu:1976ge,Haldane:1983xm}, and projected interacting fermions then realize the critical bulk theory~\cite{Zhu:2022gjc,Hu:2023xak,Han:2023yyb,Zhou:2023qfi,He:2026ong}. A fractional symmetry flux shifts the charged orbital grids by $q\alpha$, which imposes the holonomy before the interaction is diagonalized.\footnote{This is the symmetry analogue of threading a magnetic flux through the system, as in the argument for the quantized Hall conductance~\cite{Laughlin:1981jd}.} We select the regular one-particle domain at the poles.

In the vector realization, $\mathsf f_0$ is a singlet fermion and $\mathsf f_a$, $a=1,\ldots,N$, form a vector, so that the order parameter is a bilinear between them. For $N=3$ the deformation reads
\begin{equation}
 H(h,g)=H_{\rm WF}(h)+g\widehat W,\qquad
 \widehat W=\sum_m(\mathsf n_{m1}+\mathsf n_{m2}-2\mathsf n_{m3}),\qquad
 \partial_hH=\sum_m\mathsf n_{m0}.
 \label{eq:amp_ops}
\end{equation}
Here $H_{\rm WF}(h)$ is the projected vector Hamiltonian of Eq.~\eqref{eq:vectorH}, $m$ labels the orbitals, $\mathsf n_{ma}=\mathsf f_{ma}^\dagger\mathsf f_{ma}$ is the occupation of orbital $m$ in flavour $a$, and $h$ is the thermal coupling, which multiplies the singlet density and is tuned to criticality without the defect (Appendix~\ref{app:three}).\footnote{The contact interactions of the vector models are not normal ordered, which leaves a one-body term localized at the flux. Appendix~\ref{app:three} defines it together with the subtracted Hamiltonian, which we use for the free-defect control, the easy-axis branch, the spectator comparisons and every pinned monodromy line, while the unpinned easy-plane data use the Hamiltonian as written.} We then use the same couplings with the flux. The energy unit comes from the light order-parameter component $\phi_{\rm L}$. With $E_n$ the energy levels on the sphere, $E_0$ the lowest, and $E_{\phi_{\rm L}}$ and $E_{\partial\phi_{\rm L}}$ the levels of $\phi_{\rm L}$ and of its first descendant,
\begin{equation}
 \gamma_B=E_{\partial\phi_{\rm L}}-E_{\phi_{\rm L}},\qquad
 \bar E_n=\frac{E_n-E_0}{\gamma_B},\qquad
 \bar E_{\phi_{\rm L}}=0.518942.
 \label{eq:flowtune}
\end{equation}
The last value serves both branches, since the Ising, $O(2)$ and $O(3)$ order-parameter dimensions are nearly equal~\cite{Kos:2016ysd,Chester:2019ifh,Chester:2020iyt}.\footnote{The flow trajectories are defined by this finite-size prescription. Independent bulk levels and first-order conformal perturbation theory test it in Appendices~\ref{app:cpt} and~\ref{app:trajectory}.}

Defect gaps then estimate scaling dimensions, and normalized density matrix elements estimate bulk-to-defect coefficients independently of the energy calibration. We also use symmetry generators to identify the tilt and displacement families (Appendices~\ref{app:ward} and~\ref{app:tiltnorm}). Free-defect and interacting benchmarks are collected in Appendix~\ref{app:three}, and the sizes reached by exact diagonalization and by the density-matrix renormalization group (DMRG)~\cite{White:1992zz} are listed in Appendix~\ref{app:numerics}.

\section{Following the light fields}\label{sec:flows}
\subsection{The easy-plane endpoint}
\begin{figure}[t]\centering
\includegraphics[width=\textwidth]{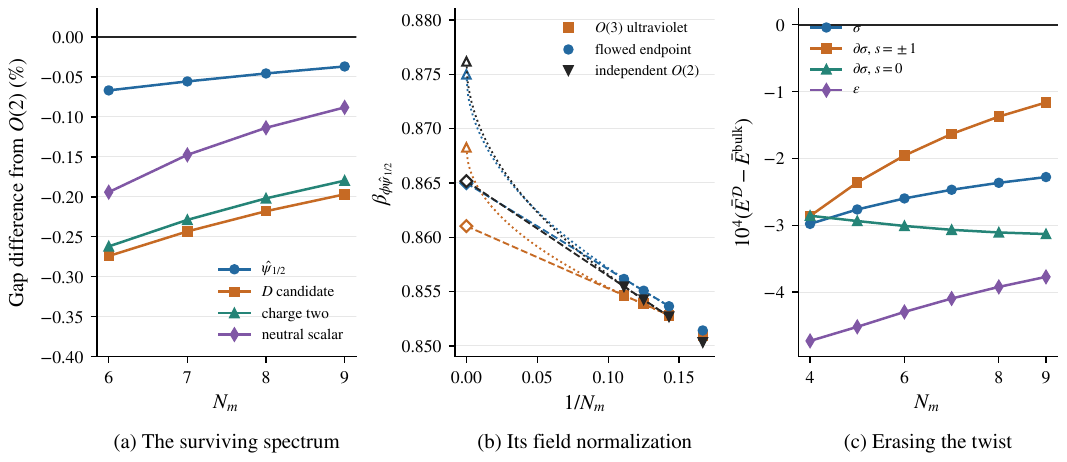}
\caption{Identifying the infrared defect. (a) Relative differences of four flowed gaps at $g=-0.09$ from the independently tuned $O(2)$ half-twist, using each model's bulk energy unit. (b) The normalized bulk-to-defect coefficient of the lowest charged mode, measured in the ultraviolet, along the flow at $g=-0.09$, and in the independent $O(2)$ model at the same sizes, plotted against $1/\Nm$. Fits use $\Nm=7$--$9$ with $\Nm^{-1}$ (dashed) or $\Nm^{-1/2}$ (dotted) corrections. Open diamonds and triangles at zero show the respective intercepts; filled markers are measurements. (c) Signed differences between the easy-axis line levels and their defect-free values at the same couplings, in the frozen limit $g=3$ with the local potential subtracted. The vertical scale magnifies these residuals by $10^4$. In (a,c), lines join measured sizes. Figure~\ref{fig:flowcontrols} gives the crossover and local-potential controls.}
\label{fig:branches}
\end{figure}

On the easy-plane branch the natural candidate for the endpoint is the $O(2)$ half-twist. We test it against an independently tuned $O(2)$ vector Hamiltonian at the same $\Nm$, with each model's bulk gap fixing its own energy unit. The charge-one, charge-two, displacement-candidate and neutral gaps of the flowed defect agree with those of the independent model far more closely than either agrees with the one-loop expansion of the $O(2)$ half-twist [Table~\ref{tab:endpoint}(a)], so the comparison is sharper than perturbation theory could make it. The agreement holds at every matched size [Fig.~\ref{fig:branches}(a)], including the neutral channel, where a spectator composite rises out of the low-energy spectrum and a different excitation becomes the lowest scalar.

\begin{table}[t]\centering\small
\begin{tabular}{llll}\toprule
\multicolumn{4}{l}{(a) Gaps at $\Nm=8$, in each model's bulk unit}\\\midrule
channel & flowed endpoint & independent $O(2)$ & one loop, $\epsilon=1$\\\midrule
charge one & 0.9402 & 0.9406 & 0.90\\
charge two & 2.0135 & 2.0176 & 1.90\\
displacement candidate & 2.0478 & 2.0522 & 2\\
neutral & 1.8963 & 1.8984 & 1.80\\\midrule
\multicolumn{4}{l}{(b) Charge-one gap of the $O(2)$ half-twist, displacement units}\\\midrule
lattice & \multicolumn{3}{l}{$0.9065$ at $\Nm=14$; extrapolated near $0.88$}\\
comparisons & \multicolumn{3}{l}{one loop $0.900$; large-$N$ full twist $0.786$; Ref.~\cite{Tang:2026kzf} $0.89(1)$}\\
shift along the flow & \multicolumn{3}{l}{$-0.009$ to $-0.012$ over $\Nm=4$--$9$; one loop $-0.009$}\\
\bottomrule\end{tabular}
\caption{The easy-plane endpoint. (a) Gaps along the flow at $g=-0.09$, in the independently tuned $O(2)$ model, and from the one-loop expansion with contact shifts [Eqs.~\eqref{eq:oneloop} and~\eqref{eq:contactpred}]. Figure~\ref{fig:branches}(a) compares the other sizes. (b) The lowest charged gap with the displacement fixed at two, its continuum comparisons and its shift along the flow. Coupling comparisons are collected in Table~\ref{tab:coupling}; Appendix~\ref{app:trajectory} gives the crossover and extrapolation details.}
\label{tab:endpoint}
\end{table}

A matching spectrum, however, leaves the normalization of the fields untested. We therefore also measure the bulk-to-defect coefficient $\beta_{\phi\hat\psi_{1/2}}$, the coefficient of the lowest charged defect mode $\hat\psi_{1/2}$ in the defect expansion of the order parameter $\phi$. It follows from the equatorial density overlap divided by the corresponding bulk overlap (Appendix~\ref{app:estimators}). Along the flow it agrees with the independent $O(2)$ value at every size [Fig.~\ref{fig:branches}(b)]. The two series extrapolate to the same limit, while the ultraviolet $O(3)$ series extrapolates slightly lower [Table~\ref{tab:coupling}]. The spectrum and the coupling thus identify the same endpoint, beyond what the symmetry alone requires.

The near-equality of the couplings has a simple origin near four dimensions. At one loop the interaction acts on a defect mode of fixed nonzero transverse spin only through the tadpole, a potential proportional to $1/r^2$ that changes the radial exponent of the mode. In four dimensions this exponent does not enter the coefficient, so at $O(\epsilon)$ the coupling is that of a free field in $4-\epsilon$ dimensions (Appendix~\ref{app:estimators}),
\begin{equation}
 \beta_{\phi\hat\psi_s}=1-\frac\epsilon4\left[\operatorname{dg}(|s|+1)+\gamma_{\rm E}\right]+O(\epsilon^2),
 \label{eq:betaeps}
\end{equation}
for every model and every twist, where $\operatorname{dg}=(\log\Gamma)'$ is the digamma function and $\gamma_{\rm E}$ is Euler's constant. At $\epsilon=1$ and $|s|=\tfrac12$ this is the one-loop value quoted for the Ising twist in Ref.~\cite{Tang:2026kzf}, which lies above the free field in three dimensions [Table~\ref{tab:coupling}]. The differences between models first appear at $O(\epsilon^2)$, which we compute in Appendix~\ref{app:estimators}. At $\epsilon=1$ this order places the ultraviolet $O(3)$ half-twist slightly below the $O(2)$ one, with the sign of the lattice offset and a somewhat smaller size.

The approach to the endpoint is consistent with the expected scaling variable $\mathcal Y=|g|\Nm^{y_t/2}$. In units with the displacement candidate fixed at two, the charge-one gap shifts along the flow by about the one-loop endpoint difference [Table~\ref{tab:endpoint}(b)], and the normalized crossover curves approximately collapse [Fig.~\ref{fig:flowcontrols}(a) and (b), Appendix~\ref{app:trajectory}]. At the largest sizes of the independent $O(2)$ calculation, extrapolations of the vector and spin-one regulators place the charge-one gap between the one-loop value and the leading large-$N$ value of the full twist~\cite{Giombi:2021uae}, in agreement with Ref.~\cite{Tang:2026kzf}.\footnote{This extrapolation depends on the correction law more than the matched endpoints differ from each other [Table~\ref{tab:endpoint}(b)].} A second comparison, $O(4)\to O(3)$, also retains a twisted plane while removing an untwisted spectator, and its low levels match in the same way (Appendix~\ref{app:trajectory}).

\FloatBarrier
\subsection{Removing the twisted fields}\label{sec:memory}\label{sec:core}
On the easy-axis branch the roles are reversed, since only the untwisted field $\phi_3$ survives. The symmetry now allows a trivial infrared line, but it also allows local $\mathbb Z_2$-even interactions on that line, so the endpoint has to be tested. We take the deep easy-axis limit, where the twisted flavours are frozen out, and compare the line spectrum with that of the bulk at the same couplings. We follow the Ising spin field $\sigma$, its first descendant at transverse spins zero and one, and the Ising energy field $\varepsilon$ (not to be confused with the expansion parameter $\epsilon$). With the subtracted Hamiltonian all of these levels coincide with their defect-free values at every size we studied [Fig.~\ref{fig:branches}(c) and the control in Fig.~\ref{fig:flowcontrols}(c)], so the line cannot be distinguished from no defect.

Two properties of the Ising line make this endpoint natural. A $\mathbb Z_2$-even term on the trivial line couples first to $\varepsilon$, whose dimension exceeds one~\cite{Kos:2016ysd}, so the trivial line is stable against such terms. The most familiar ordered alternative, the direct sum of the two pinned magnetic lines, is unstable. The wall between them has $\Delta_{+-}<1$~\cite{Lanzetta:2025xfw}, both on the fuzzy sphere and in a Pad\'e approximant anchored in two dimensions~\cite{Cuomo:2024psk} (Table~\ref{tab:anchors}).

The two branches thus show the monodromy following the light fields, surviving when they carry it and disappearing when they do not.

\section{Spectator dressing and defect entropy}\label{sec:entropy}
Fields that the twist leaves alone can still change a monodromy that survives. A free untwisted spectator cancels between the partition functions with and without the defect, so removing it leaves $\log\mathfrak g$ unchanged, and any change in the interacting theory measures how the spectator dresses the twist. The wavefunction-overlap method for a local line that can end~\cite{Zhou:2023fqu} does not apply to a monodromy line, which bounds a symmetry surface.\footnote{The alternative cap-entanglement calculation has not yet reached a controlled normalization (Appendix~\ref{app:entanglement}).} We therefore study the dressing in a controlled limit, an $O(N)$ model at large $N$ with one twisted complex component and $N-2$ untwisted real spectators.

\subsection{A free reference on the hyperbolic cylinder}
Many properties of a monodromy defect simplify after a Weyl transformation. Writing the flat metric as $\mathrm ds^2=r^2[\mathrm d\theta^2+(\mathrm dr^2+\mathrm dy^2)/r^2]$, the complement of the defect becomes conformal to $S^1\times \mathbb H^{d-1}$, with the defect at the boundary of hyperbolic space and $\theta$ running around the circle. The monodromy then becomes a twisted boundary condition on a thermal circle of length $2\pi$, so that defect data become thermodynamic quantities~\cite{Giombi:2021uae}.\footnote{The same map relates entanglement entropy to thermal entropy on hyperbolic space~\cite{Casini:2011kv}, and for a defect the entanglement entropy combines $\log\mathfrak g$ with the conformal weight introduced below~\cite{Lewkowycz:2013laa,Jensen:2013lxa,Kobayashi:2018lil}.} In three dimensions the expectation value $\mathfrak g$ of a circular defect follows from the free energy on $S^1\times \mathbb H^2$ with the regularized volume $\operatorname{Vol}(\mathbb H^2)=-2\pi$, and its logarithm is the defect entropy. For a complex scalar twisted by $e^{2\pi\mathrm i\alpha}$, a conformally coupled mode with $\mathbb H^2$ spectral parameter $\nu$ has energy $\nu$ and Plancherel density $\nu\tanh(\pi\nu)/2\pi$, so that
\begin{equation}
 \log\mathfrak g_{\rm free}(\alpha)=\int_0^\infty \mathrm d\nu\,\nu\tanh(\pi\nu)\,
 \log\frac{1-2e^{-2\pi\nu}\cos2\pi\alpha+e^{-4\pi\nu}}{(1-e^{-2\pi\nu})^2}.
 \label{eq:loggfree}
\end{equation}
It is positive for noninteger $\alpha$. The conformal weight $\mathfrak h$ of the defect fixes the stress-tensor one-point function, $\langle T_{ab}\rangle=-\mathfrak h\,\delta_{ab}/(2\pi r^d)$ along the defect~\cite{Billo:2016cpy}. In three dimensions it is the response of the free energy $\mathcal F$ to the length $\Theta$ of the thermal circle, $\mathfrak h=-(\pi/\operatorname{Vol}(\mathbb H^2))\,\partial_\Theta\mathcal F|_{\Theta=2\pi}$~\cite{Giombi:2021uae}, and for the free twisted complex scalar
\begin{equation}
 \mathfrak h_{\rm free}(\alpha)=\frac{\alpha(1-\alpha)(1-2\alpha)\tan\pi\alpha}{12},\qquad
 \mathfrak h_{\rm free}(\tfrac12)=\frac1{24\pi}.
 \label{eq:hfree}
\end{equation}

\subsection{How the spectators respond}
At large $N$ we write the critical $O(N)$ model with the auxiliary field $\varsigma$, whose equation of motion fixes $\langle\phi_a\phi_a\rangle$~\cite{Moshe:2003xn},
\begin{equation}
 S=\int \mathrm d^3x\left[\tfrac12(\partial\phi_a)^2+\tfrac12\varsigma\,\phi_a\phi_a\right].
 \label{eq:largeMaction}
\end{equation}
In the presence of the defect, conformal symmetry fixes the saddle to $\langle\varsigma\rangle=\varsigma_*/r^2$, which becomes a constant mass $\varsigma_*$ on $S^1\times \mathbb H^2$. The constraint then requires the defect to leave $\langle\phi_a\phi_a\rangle$ unchanged at every point, which in the $\mathbb H^2$ frame reads
\begin{equation}
 2\,\delta g_{\rm tw}+(N-2)\,\delta g_{\rm sp}=0,
 \label{eq:gap}
\end{equation}
where $\delta g_{\rm tw}$ and $\delta g_{\rm sp}$ are the coincident-point changes of the propagator of one twisted and one spectator component. The source is the deficit left by the twisted pair, whose fluctuations near the line are suppressed, $\delta g_{\rm tw}=-\mathcal A_\alpha/(4\pi^2)$ with
\begin{equation}
 \mathcal A_\alpha=\frac\pi2(1-2\alpha)\tan\pi\alpha,\qquad \mathcal A_{1/2}=1,
 \label{eq:Aalpha}
\end{equation}
as the exact flat-space correlator~\cite{Dowker:2022mex} also shows at coincident points. The spectators must compensate with enhanced fluctuations, but only one of their modes can respond at leading order. On $S^1\times \mathbb H^2$ each spectator mode of transverse spin $s$ is a field on $\mathbb H^2$, which is Euclidean AdS$_2$, of boundary dimension $\Dh_s=\tfrac12+\sqrt{s^2+\varsigma_*}$, and only the $s=0$ mode responds at order $\sqrt{\varsigma_*}$. With the regular boundary condition its response is $-\sqrt{\varsigma_*}/8$, of the same sign as the source, and $\varsigma_*<0$ would make the mode unstable, so Eq.~\eqref{eq:gap} has no solution. The solution instead uses the singular branch of the spectators' lowest line mode $\hat\phi_{\rm sp}$~\cite{Breitenlohner:1982jf,Klebanov:1999tb}, with $\Dh=\tfrac12-\sqrt{\varsigma_*}$ and response $+\sqrt{\varsigma_*}/8$, and the gap equation then gives
\begin{equation}
 \sqrt{\varsigma_*}=\frac{4\mathcal A_\alpha}{\pi^2N}+O(N^{-2}),\qquad
 \Dh_{\rm sp}=\frac12-\frac{4\mathcal A_\alpha}{\pi^2N}+\cdots,
 \label{eq:sqrta}
\end{equation}
where the dots stand for the other $O(1/N)$ corrections, such as the anomalous dimension of $\phi$, which we have not computed.\footnote{The contemporaneous analysis of Feng et al.~\cite{Feng:2026rrk} also identifies this singular branch, which they call the alternate boundary condition, and the resulting enhancement of the spectator correlations. They study partial monodromies with a fixed fraction of twisted components at large $N$ and follow the defect fixed-point branches between three and four dimensions.}

The saddle thus lowers the spectators' lowest line dimension, a property of the interacting field rather than of the microscopic orbitals of Sec.~\ref{sec:method}. In $O(3)$ the corresponding lattice mode of the subtracted Hamiltonian lies well below the order-parameter dimension $\Delta_\phi$, close to the large-$N$ value at $N=3$ (Table~\ref{tab:spectator}). Since the shift is linear in the source at large $N$, a single twisted real component, as in the reflection $\operatorname{diag}(1,1,-1)$, should lower the spectator mode half as much as a twisted plane. This is close to what our lattice and Refs.~\cite{Tang:2026kzf,Feng:2026rrk} find for these two $O(3)$ defects [Fig.~\ref{fig:spectators}(a)].

\begin{table}[t]\centering\small
\begin{tabular}{lcccc}\toprule
$O(3)$ half-twist & large $N$ at $N=3$ & lattice & Ref.~\cite{Tang:2026kzf} & Ref.~\cite{Feng:2026rrk}\\\midrule
spectator line mode $\Dh_{\rm sp}$ & 0.37 & 0.41 & --- & 0.409(11)\\
lowest neutral line operator & 1.28 & 1.23 & 1.26(4) & ---\\
shift ratio, component$\,/\,$plane & $1/2$ & $0.049/0.110$ & $0.032/0.059$ & $0.047/0.110$\\
\bottomrule\end{tabular}
\caption{The spectator channel of the $O(3)$ half-twist, in bulk units: the leading large-$N$ saddle of Eq.~\eqref{eq:sqrta} evaluated at $N=3$, the lattice with the local potential subtracted, Ref.~\cite{Tang:2026kzf}, and the Monte Carlo of Ref.~\cite{Feng:2026rrk} without an added line coupling; the order-parameter dimension is $\Delta_\phi=0.519$. The neutral entry of the saddle is the lowest zero of the singlet-channel kernel of Appendix~\ref{app:singlet}; the naive spectator composite at $1-8\mathcal A_\alpha/(\pi^2N)$ is not a state of the large-$N$ theory. The last row compares the predicted ratio of the spectator shifts for a single twisted real component and for a twisted plane with the shifts found on the lattice and in Refs.~\cite{Tang:2026kzf,Feng:2026rrk}. The large-$N$ derivation is in Appendix~\ref{app:largeM}.}
\label{tab:spectator}
\end{table}

This is also the channel in which the $\epsilon$ expansion fails. At one loop the tadpole gives every component a potential $\mathfrak v/r^2$, and with $n_{\rm tw}$ twisted real components the untwisted strength is $n_{\rm tw}/(n_{\rm tw}+2)$ times the twisted one, so both are attractive. For $s\ne0$ the spectator modes are then lowered by that fraction of the twisted shift at equal $|s|$. At $s=0$, however, the effective spin $\tilde s=\sqrt{\mathfrak v}$ of an attractive potential is imaginary, and the leading expansion cannot place this mode, as observed in Ref.~\cite{Tang:2026kzf}. A defect coupling of $\hat\phi_{\rm sp}^2$ is then marginal at this order and has to be included~\cite{Giombi:2021uae}, while the large-$N$ saddle settles the same channel on the singular branch.

The same saddle predicts the neutral singlet channel that tests stability under symmetric line perturbations. The spectator double trace $\hat\phi_{\rm sp}^2$ would have dimension $1-8\mathcal A_\alpha/(\pi^2N)$, below marginality, but at large $N$ it is not a state of the theory. In the singlet channel the auxiliary field resums the spectator bubbles, so the double trace drops out, just as $\phi^2$ is replaced by $\varsigma$ in the bulk. The neutral line operators are instead the zeros of the $\varsigma$ kernel (Appendix~\ref{app:singlet}). The lowest of them lies between the composite and the lowest twisted double trace, which it approaches as $N\to\infty$. For $N\ge4$ the lowest zero exceeds one throughout the angle scan, consistent with the small-angle expansion, so this channel contains no relevant perturbation within the retained large-$N$ approximation.\footnote{This is a statement about the neutral singlet kernel at this order. At $N=3$ its lowest zero falls marginally below one at very small twist, by a margin the omitted corrections can change (Appendix~\ref{app:singlet}).} At $N=3$ and half twist it agrees with the subtracted lattice estimate and with Ref.~\cite{Tang:2026kzf} to a few percent (Table~\ref{tab:spectator}).

The saddle also dresses the defect entropy classically. Integrating the response of the free energy to $\varsigma_*$ from zero to the saddle gives
\begin{equation}
 \delta_{\rm cl}\log\mathfrak g=-\frac{16\mathcal A_\alpha^3}{3\pi^4N^2},
 \label{eq:classical}
\end{equation}
which is only of second order (Appendix~\ref{app:largeM}), so the leading effect comes from fluctuations.

\subsection{The fluctuation correction and an entropy comparison}
\begin{figure}[t]\centering
\includegraphics[width=\textwidth]{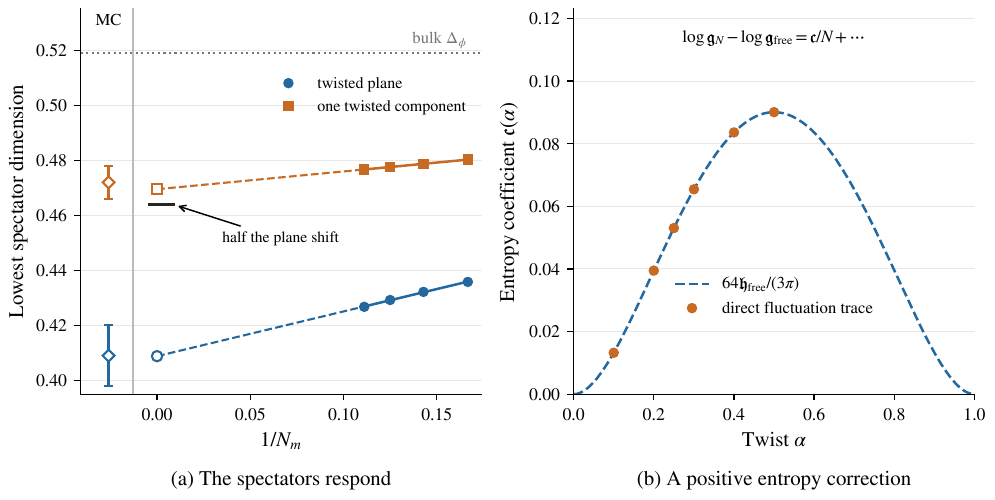}
\caption{The spectators respond to the twist. (a) The lowest spectator dimension in the $O(3)$ model with a half-twisted plane or one reflected component. Filled markers are fuzzy-sphere measurements with the local potential subtracted; solid lines are linear fits in $1/\Nm$ within the measured window, and dashed lines extend to the open extrapolated markers. The separate MC column shows Ref.~\cite{Feng:2026rrk} without an added line coupling. The short black bar is the bulk dimension lowered by half the extrapolated plane shift, as the leading large-$N$ response predicts. (b) The coefficient $\mathfrak c(\alpha)$ of the entropy correction $\mathfrak c/N$, with direct fluctuation traces and Eq.~\eqref{eq:cresult}. Its positive sign gives the leading endpoint entropy increase of Eq.~\eqref{eq:decouple}, conditional on the bulk flow connecting the saddles.}
\label{fig:spectators}
\end{figure}

The first correction to the free value comes from the fluctuations of $\varsigma$,
\begin{equation}
 \log\mathfrak g=\log\mathfrak g_{\rm free}(\alpha)+\frac{\mathfrak c(\alpha)}{N}+O(N^{-2}),\qquad
 \mathfrak c=-\Tr\!\left[\Pi_0^{-1}\delta\Pi\right],
 \label{eq:ctrace}
\end{equation}
where $G_0$ is the propagator of one real component without the defect and $G_\Dc=\langle\phi_{\mathbb C}(x)\phi_{\mathbb C}^\dagger(x')\rangle_\Dc$ for $\phi_{\mathbb C}=(\phi_1+\mathrm i\phi_2)/\sqrt2$, so that $G_\Dc=G_0$ without the twist, $\Pi_0=G_0^2$ is the $\varsigma$ self-energy kernel without the defect, up to a factor $N/2$, and
\begin{equation}
 \delta\Pi=|G_\Dc|^2-G_0^2+N\,G_0\,\delta G_{\rm sp}
 \label{eq:deltaPi}
\end{equation}
is its change, from the twisted pair and from the spectators' singular branch. Here $\delta G_{\rm sp}(x,x')$ is the two-point shift and $\delta g_{\rm sp}$ its coincident limit, and replacing $N-2$ by $N$ in this kernel changes $\log\mathfrak g$ only at $O(N^{-2})$. On $S^1\times \mathbb H^2$ every kernel depends only on the angle and the $\mathbb H^2$ distance between its arguments, so the trace is diagonal in the $S^1$ mode $\kappa$ and the $\mathbb H^2$ spectral parameter $\nu$. Each spectral function is a sum, over pairs of line modes with $s+s'=\kappa$, of bubbles of two AdS$_2$ propagators~\cite{Fitzpatrick:2011dm,Carmi:2018qzm,Giombi:2020rmc}, which we evaluate in closed form level by level (Appendix~\ref{app:largeM}). The gap equation cancels the leading large-momentum term, and a direct evaluation with rotationally symmetric smooth cutoffs gives the coefficients of Table~\ref{tab:entropy}(a) at six angles. Since the sums are symmetric under $\alpha\to1-\alpha$, these angles sample the whole range. Every coefficient is positive, and with nothing fitted they agree, within the cutoff and truncation variations, with
\begin{equation}
 \mathfrak c(\alpha)=\frac{64}{3\pi}\,\mathfrak h_{\rm free}(\alpha)
 =\frac{16}{9\pi}\,\alpha(1-\alpha)(1-2\alpha)\tan\pi\alpha,
 \label{eq:cresult}
\end{equation}
whose half-twist value is $8/(9\pi^2)$ [Fig.~\ref{fig:spectators}(b)].

The conformal weight appears for a physical reason. The short-distance expansion of the kernel follows the bulk operator product expansion, and once the gap equation removes the scalar term associated with the auxiliary-field constraint, a traceless stress-tensor contribution proportional to $\mathfrak h_{\rm free}$ remains. Its momentum-space contribution is only conditionally integrable, so its value depends on how the cutoff treats the circle and $\mathbb H^2$ directions. A sum performed one circle mode at a time gives zero, because the twisted and spectator contributions cancel in every fixed circle sector, for every $\kappa$ and $\alpha$ (Appendix~\ref{app:sector}). A cutoff in $\kappa^2+\nu^2$ therefore differs from zero only through the stress-tensor term, and this difference is the right-hand side of Eq.~\eqref{eq:cresult}.\footnote{This step assumes that the large-momentum expansion holds uniformly in the circle and $\mathbb H^2$ directions, which we check on the tabulated integrand in Appendix~\ref{app:largeM}, and it identifies the isotropic cutoff with the covariant circular-defect entropy, which dimensional regularization supports near four dimensions (Appendix~\ref{app:analytic}).} Near four dimensions, where dimensional regularization evaluates the trace covariantly, the sum one circle mode at a time again vanishes, and the covariant value is this stress-tensor term (Appendix~\ref{app:analytic}).

The positive coefficients have a consequence that does not rely on the general formula. If removing one spectator connects the corresponding $O(N)$ and $O(N-1)$ saddles, their entropies differ by
\begin{equation}
 \log\mathfrak g_{N-1}-\log\mathfrak g_N
 =\frac{\mathfrak c(\alpha)}{N(N-1)}+O(N^{-3}).
 \label{eq:decouple}
\end{equation}
The leading difference is positive at every angle evaluated, so the infrared endpoint has the larger entropy. This is allowed, because the line $\mathfrak g$-theorem assumes a fixed conformal bulk~\cite{Cuomo:2021rkm,Jensen:2015swa}, and Eq.~\eqref{eq:decouple} compares endpoints without implying anything about monotonicity between them. An untwisted spectator can therefore affect the defect even though it carries none of its monodromy.

Near four dimensions perturbation theory supplies a separate comparison at fixed $N$. For one twisted complex component the first-order change of the continued defect entropy $\widetilde F$, which equals $\log\mathfrak g$ at $d=3$, is
\begin{equation}
 \widetilde F_{N-1}-\widetilde F_N=\frac{\pi\alpha^2(1-\alpha)^2}{2}\,\frac{\epsilon}{(N+7)(N+8)}+\cdots,
 \label{eq:decoupleeps}
\end{equation}
which is positive at every angle. At this order a free spectator contributes only through the bulk fixed-point coupling. At the next order its lowest mode responds to the twist. The real part of the continued entropy difference is again positive at the angles evaluated, but the second-order term is larger than the first at $\epsilon=1$ [Table~\ref{tab:entropy}(b)].\footnote{At fixed $N$ near four dimensions, the attractive tadpole pushes the spectator mode below the Breitenlohner--Freedman value. Its two boundary conditions give complex-conjugate branches, with imaginary terms starting at order $\epsilon^{3/2}$. The comparison retains their common real part through order $\epsilon^2$. It is not a controlled entropy measurement of a real three-dimensional defect. Appendix~\ref{app:entropyeps} gives the self-consistent spectator channel and the remaining second-order terms.}

The first-order, second-order and large-$N$ estimates for $O(3)\to O(2)$ and $O(4)\to O(3)$ are all positive, while their magnitudes differ several-fold [Table~\ref{tab:entropy}(b)]. These are different limits of spectator dressing.\footnote{Taking large $N$ before expanding in $\epsilon$ allows the spectators to screen the twisted pair's deficit. For one twisted plane this halves the leading coefficient relative to the large-$N$ limit of Eq.~\eqref{eq:decoupleeps}. Appendix~\ref{app:entropyeps} derives this noncommutativity, and Table~\ref{tab:entropy}(a) compares the resulting coefficient with the three-dimensional trace.} The sign and magnitude of the entropy change along the three-dimensional small-$N$ flow still require an independent measurement.

\section{Pinning and the fate of ordered lines}\label{sec:pinmono}
Having followed the monodromy through the bulk flow, we now perturb the line itself. The resulting pinned endpoints let us connect the two kinds of flow and ask whether the line can support order.

\subsection{The pinned monodromy line}\label{sec:monopin}
A charged defect mode $\hat\psi_s$ of the $O(2)$ monodromy with $\Dh_s<1$ is a relevant line perturbation, which we add to the defect action as
\begin{equation}
 \delta S=-h_{\rm pin}\int \mathrm d\tau\,\big(\hat\psi_s+\hat\psi_s^\dagger\big).
 \label{eq:monopinaction}
\end{equation}
Here $\tau$ is the coordinate along the line and $h_{\rm pin}$ is the pinning strength. This coupling breaks the $U(1)$ symmetry and the transverse rotations separately, but it preserves $L_z-sQ$, where $L_z$ generates transverse rotations and $Q$ is the $U(1)$ charge. Ref.~\cite{Kravchuk:2025evf} identifies the endpoint of this flow as a monodromy pinning defect. Since $Q$ is broken there, the line carries a tilt at $\Dh=1$, and junctions measure the cost of changing the pinning direction. A phase difference $\eta$ between the pins at the two poles is a rotation of the pinning direction, because global $U(1)$ and transverse rotations shift both phases equally. At $\eta=\pi$ it joins the lines pinned with opposite sign.

We test the identification at both ends of the flow (Appendix~\ref{app:monopin}). At weak coupling the second-order shifts of the charged doublet approach the ratio three predicted for a single complex coupling (Table~\ref{tab:monopin}). At strong coupling the lowest neutral level approaches the protected tilt dimension at the smaller twists, and the low levels are insensitive to the microscopic form of the pin. At half twist, where the extrapolated tilt level stays above one, the identification rests on continuity in $\alpha$.\footnote{Appendix~\ref{app:monopin} gives the tilt and stiffness tests and the candidate displacement level, which has not converged at the sizes reached.}

\begin{figure}[t]\centering
\includegraphics[width=\textwidth]{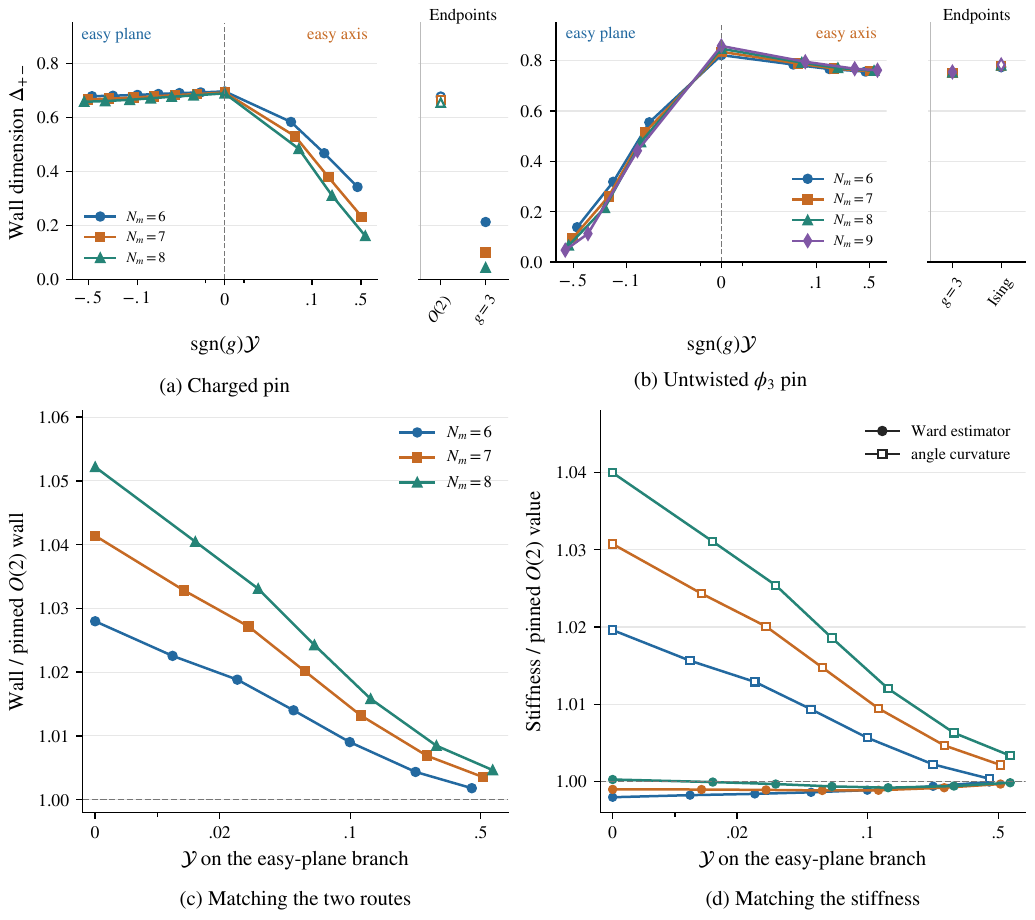}
\caption{Pins follow the light fields. (a,b) Wall dimensions for charged and untwisted-component pins along both branches, with negative $\operatorname{sgn}(g)\mathcal Y$ denoting easy plane and positive values easy axis. The axis is linear near zero and logarithmic farther away. Separate endpoint columns show the frozen $g=3$ values (filled) and independent $O(2)$ or Ising references (open); they have no flow-coordinate position. The charged pin has $h_{\rm pin}=1000$, the $\phi_3$ pin is projected to infinite strength, and the local potential is subtracted. The charged wall retains a strong-pin remnant at $g=3$ that decreases with size; the weaker-pin control and the Ising calibration offset are discussed in Appendix~\ref{app:pinbranches}. (c,d) Along the easy-plane branch, the wall and both stiffness estimators divided by their values on the independently constructed pinned $O(2)$ half-twist at the same size. Unity marks matching junction data for pinning before or after the bulk flow. Colours denote the same sizes throughout; in (d), filled circles give the Ward estimator and open squares the angle curvature.}
\label{fig:pins}
\end{figure}

The pinned $O(2)$ half-twist is also the endpoint of a bulk flow. We compare the wall dimension $\Delta_{+-}$ and two estimates of the tilt stiffness, the Ward estimator $\Ct^{HQ}$ and the small-angle curvature $\Ct^{\angle}$, defined in Sec.~\ref{sec:junction}. We pin the lowest charged mode of the $O(3)$ half-twist with the same coupling and follow the pinned line along the easy-plane flow of Sec.~\ref{sec:flows}. At every size the wall and both stiffness estimators approach those of the pinned $O(2)$ monodromy line as the scaling variable grows [Fig.~\ref{fig:pins}(c,d)]. Pinning before or after the easy-plane flow thus gives matching infrared junction data. The untwisted spectator leaves the Ward stiffness unchanged but raises the small-angle curvature and the wall, by an amount that grows with $\Nm$. Removing the spectator lowers the measured wall of the pinned line. Its effect on the entropy is a separate question, addressed for the unpinned partial twist in the large-$N$ comparison of Sec.~\ref{sec:entropy}.

\FloatBarrier

The charged pin is not the only relevant coupling on the $O(3)$ half-twist. The untwisted component $\phi_3$ has its own line mode below dimension one (Table~\ref{tab:spectator}), and pinning it gives a line on which the twist and the pin act on different fields. Along the two branches of the bulk flow the two pins behave in opposite ways (Figs.~\ref{fig:story} and~\ref{fig:pins}). On the easy-axis branch the twisted pair becomes massive, and the $\phi_3$-pinned half-twist loses its twist but keeps its pin. In the frozen easy-axis limit its wall and low levels match those of the pinned line of the Ising Hamiltonian~\cite{Zhu:2022gjc,Hu:2023ghk}, up to the difference between the two lattice models (Appendix~\ref{app:pinbranches}). The charged pin instead decouples there. Its stiffness estimators vanish as the twisted pair becomes heavy, while the strong-pin wall remnant in Fig.~\ref{fig:pins}(a) decreases with size (Appendix~\ref{app:pinbranches}). On the easy-plane branch the roles are reversed. Pins, like the monodromy, follow the light fields.

\subsection{Junctions on ordinary pinned lines}\label{sec:junction}
To interpret these junction observables, we use the ordinary pinned line as the zero-twist reference. It is obtained by a localized field,
\begin{equation}
 S_{\rm pin}=S-h_{\rm pin}\int \mathrm d\tau\,\phi_1(\tau,\mathbf x_\perp=0),
 \label{eq:pinaction}
\end{equation}
where $\mathbf x_\perp$ is the transverse position. It is relevant because $\Delta_\phi<1$, and it flows to the ordinary pinned line~\cite{Cuomo:2021kfm}. The two signs of $h_{\rm pin}$ give two conformal lines $\Dc_+$ and $\Dc_-$, pinned in opposite directions, and their sum $\Dc_+\oplus\Dc_-$ is a candidate $\mathbb Z_2$-ordered line. It is stable against a dilute fugacity of domain walls only if the operator that changes one pinning into the other has $\Delta_{+-}>1$, while for $\Delta_{+-}<1$ the walls proliferate~\cite{Lanzetta:2025xfw}.

For these lines~\cite{Assaad:2013xua,ParisenToldin:2016szc,Cuomo:2021kfm,Cuomo:2022xgw,Gimenez-Grau:2022czc,Bianchi:2022sbz,Pannell:2023pwz,Bianchi:2023gkk,Hu:2023ghk,Yang:2026qtk,Barrat:2026ozj}, a Ward identity, a two-loop calculation and a large-$N$ saddle provide independent comparisons for the junction observables.\footnote{With the subtracted Hamiltonian, pinning an untwisted spectator on a partially twisted line gives the same junction data as the ordinary pinned line at the sizes tested (Appendix~\ref{app:pinning}).}

\subsubsection{Separating the stiffness from the angular shape}
\begin{figure}[t]\centering
\includegraphics[width=0.97\textwidth]{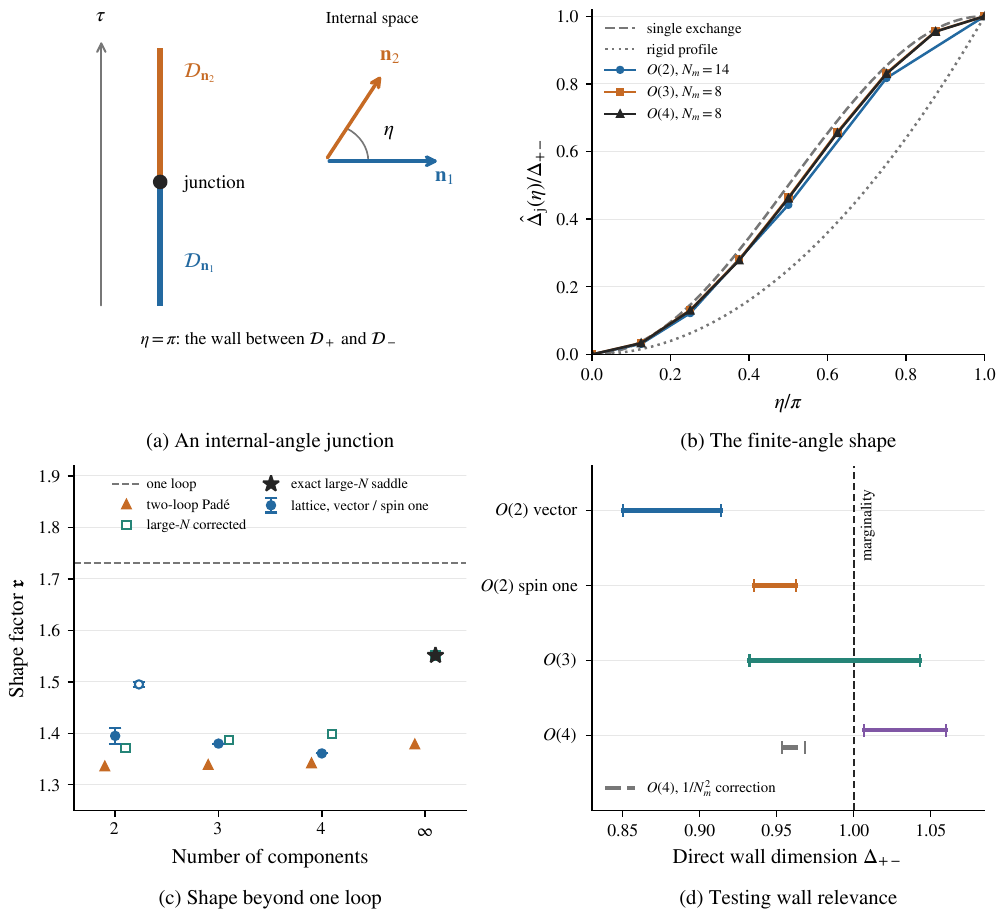}
\caption{From a junction to the ordering question. (a) The spatial line stays straight while its internal pinning direction changes by $\eta$; opposite pinnings give the wall at $\eta=\pi$. (b) Finite-size angular profiles normalized by the wall dimension, with the single-exchange and rigid reference shapes. (c) Shape-factor estimates from Table~\ref{tab:shape}. Filled lattice circles are vector models and the open circle at $N=2$ is the spin-one regulator. The star is the exact large-$N$ saddle; the two-loop entries use the stated resummations. (d) Direct-wall extrapolations from the saved junction energies. Coloured segments span the linear and quadratic fits in $1/\Nm$ or the tilt excess; the separate grey segment shows the faster $\Nm^{-2}$ correction for $O(4)$. The dashed line is marginality. Segments in (c,d) describe only the stated fit choices, not statistical confidence intervals. Panel (d) omits the slow line and bulk corrections examined in Appendix~\ref{app:wallroutes}; allowing both leaves marginality unresolved even for $O(2)$. Stiffness and reconstruction diagnostics are in Fig.~\ref{fig:junctioncontrols}, with further convergence tests in Fig.~\ref{fig:junctionconvergence}.}
\label{fig:junction}
\end{figure}

Rotating the pinning direction at one pole by an internal angle $\eta$ creates a junction on the original line, whose spatial geometry stays straight [Fig.~\ref{fig:junction}(a)]. The junction has dimension $\Dh_{\rm j}(\eta)$, and at $\eta=\pi$ it is the wall, $\Dh_{\rm j}(\pi)=\Delta_{+-}$. The Ward identity for the broken internal rotation relates its small-angle curvature to the tilt normalization~\cite{Padayasi:2021sik,Drukker:2022pxk,Herzog:2023dop,Girault:2025kzt},
\begin{equation}
 \Dh_{\rm j}(\eta)=\Ct\frac{\eta^2}{2}+O(\eta^4),\qquad
 \langle\hat t_a(\tau)\hat t_b(0)\rangle=\frac{\Ct\delta_{ab}}{\tau^2}.
 \label{eq:junction}
\end{equation}
Here $\hat t_a$, $a=2,\ldots,N$, are the tilt operators, the defect operators generated by the broken internal rotations, and the tilt normalization $\Ct$ is the stiffness. It is useful to separate this stiffness from the finite-angle shape through
\begin{equation}
 \mathfrak r=\frac{\Dh_{\rm j}(\pi)}{2\Ct}.
 \label{eq:rdef}
\end{equation}
A single exchange gives $\Dh_{\rm j}=\Ct(1-\cos\eta)$ and $\mathfrak r=1$, while the rigid quadratic profile gives $\mathfrak r=\pi^2/4$. The interacting junction lies between these two reference shapes.

Expanding the one-loop junction of Ref.~\cite{Cuomo:2024psk} gives
\begin{equation}
 \mathfrak r=1+\frac{6\zeta(3)}{\pi^2}\epsilon+O(\epsilon^2),
 \label{eq:roneloop}
\end{equation}
independent of $N$ at this order (Appendix~\ref{app:pinning}). Setting $\epsilon=1$ overshoots the lattice estimates substantially (Table~\ref{tab:shape}). In Appendix~\ref{app:twoloop} we find
\begin{equation}
 \mathfrak r=1+\frac{6\zeta(3)}{\pi^2}\epsilon+\mathfrak r^{(2)}(N)\,\epsilon^2+O(\epsilon^3),
 \label{eq:rtwoloop}
\end{equation}
where the second-order coefficient $\mathfrak r^{(2)}(N)$, listed in Table~\ref{tab:shape}, is large and negative and depends only weakly on $N$. We resum the series in two ways, by a $[1/1]$ Pad\'e approximant at $\epsilon=1$ and by correcting the exact large-$N$ value $\mathfrak r_\infty$ of Appendix~\ref{app:junctionlargeN} by the $N$-dependence of $\mathfrak r^{(2)}$. The two resummations give estimates spanning a narrow band that overlaps the lattice estimates for $O(3)$ and $O(4)$ and lies below them for $O(2)$ [Table~\ref{tab:shape} and Fig.~\ref{fig:junction}(c)].\footnote{At $N=\infty$, where the exact curve is known, the same Pad\'e approximant falls short of it at $d=3$, which suggests a downward bias but does not establish a bound at finite $N$ (Appendix~\ref{app:twoloop}).}

On the lattice the angular profiles lie between the two reference shapes [Fig.~\ref{fig:junction}(b)]. At each size we extract the small-angle curvature $\Ct^{\angle}$ from an angle scan and form $\mathfrak r_{\Nm}=\Dh_{{\rm j},\Nm}(\pi)/(2\Ct^{\angle}(\Nm))$, whose growth tracks the approach of the lattice tilt level to its protected value one. Fits in the tilt excess give the lattice entries of Table~\ref{tab:shape} for the vector models and for the independent spin-one $O(2)$ regulator~\cite{Dey:2026cso} [Fig.~\ref{fig:junctionconvergence}(a) in Appendix~\ref{app:pinning}]. The shape varies across $N$ by less than it differs between the two $O(2)$ regulators. Since at second order $\mathfrak r$ depends on $N$ only weakly [Eq.~\eqref{eq:rtwoloop}], the data do not yet resolve a dependence of that size.\footnote{Normalized by the Ward stiffness of the next subsection instead of the curvature, the $O(2)$ shape factor agrees between the two regulators (Appendix~\ref{app:wallroutes}).}

\subsubsection{Measuring the stiffness with a Ward identity}
The shape alone cannot decide whether the wall is relevant, so we determine $\Ct$ separately. Let $Q_{12}$ generate the broken internal rotation. Since it commutes with the bulk Hamiltonian, $[H,Q_{12}]$ is the variation of the pinning term, and for a finite-size state chosen to approach the tilt primary
\begin{equation}
 \Ct^{HQ}(\Nm)\equiv\frac{|\langle\hat t_{\Nm}|[H,Q_{12}]|0_\Dc\rangle|^2}{4\gamma_B^2}\ \longrightarrow\ \Ct.
 \label{eq:tiltHQ}
\end{equation}
Here $|0_\Dc\rangle$ is the ground state of the pinned line, $|\hat t_{\Nm}\rangle$ is the finite-size state selected as the tilt, and $\gamma_B$ is the bulk energy unit of Eq.~\eqref{eq:flowtune}. The operator identity holds at finite size, so the extrapolation enters only through the approach of the selected state to the tilt primary.

We take the Ward estimator as the stiffness. It uses the tilt primary alone, whereas the small-angle curvature reaches $\Ct$ only when the whole lattice tilt tower has become conformal. The blind finite-size checks support this choice. In $O(2)$ the two continuations agree within a few percent, although their finite-size values are separated [Table~\ref{tab:wallestimators} and Fig.~\ref{fig:junctioncontrols}(a), Appendix~\ref{app:wallroutes}].\footnote{Their continuations agree closely in both $O(2)$ regulators, while in $O(3)$ and $O(4)$ a larger difference remains at the sizes reached (Table~\ref{tab:wallestimators}). The exact relation between the curvature, the Ward stiffness and the lattice tilt tower, and the blind tests, are in Appendix~\ref{app:tiltnorm}. The curvature has to be taken at small angle, since a global fit to the whole angular profile overstates it by an amount that grows with $\Nm$ (Appendix~\ref{app:pinning}).}

The wall itself is the energy of the ground state with opposite pinnings at the two poles. Figure~\ref{fig:junction}(d) displays its continuum extrapolations against marginality, and Table~\ref{tab:walls} gives the values. The stiffness and the shape factor are not a second measurement of it, since $\Dh_{\rm j}(\pi)=2\Ct^{\angle}\mathfrak r$ holds at each size by definition. Their value lies in explaining the wall, by comparing the stiffness with the Ward identity and the shape with the two-loop and large-$N$ calculations. The factorized estimate in Table~\ref{tab:wallestimators} of Appendix~\ref{app:wallroutes}, the Ward stiffness times the shape factor, therefore differs from the measured wall only through the ratio of the two stiffness estimates and through the separate extrapolation of the shape factor.

\begin{table}[t]\centering\small
\begin{tabular}{lcc}\toprule
model & Ward stiffness $\Ct^{HQ}$ & direct-wall extrapolation $\Delta_{+-}$\\\midrule
$O(2)$ vector & 0.326--0.331 & 0.85--0.91\\
$O(2)$ spin one & 0.329--0.336 & 0.94--0.96\\
$O(3)$ vector & 0.335--0.341 & 0.93--1.04\\
$O(4)$ vector & 0.341--0.348 & 1.01--1.06\\
\bottomrule\end{tabular}
\caption{Stiffness and direct-wall extrapolations for ordinary pinned lines. The stiffness ranges span the fits of Appendix~\ref{app:tiltnorm}. The wall ranges cover only the linear and quadratic fits in $1/\Nm$ or the tilt excess, not the slow-correction fits of Appendix~\ref{app:wallroutes}. None is a confidence interval; the faster $\Nm^{-2}$ correction gives $0.95$--$0.97$ for $O(4)$ instead. The $O(2)$ walls favour relevance, while the larger-$N$ verdict depends on the correction law. Table~\ref{tab:wallestimators} compares the curvature and factorized estimates, and Table~\ref{tab:wallfits} gives the direct fits.}
\label{tab:walls}
\end{table}

The verdict follows from the measured wall. In $O(2)$ the measured walls extrapolate below one in both regulators, so wall proliferation, and with it the instability of the candidate ordered line, is favoured with a small margin.\footnote{The vector wall sequence has not reached its asymptotic convergence at the largest sizes, and the spin-one factorized estimate reaches marginality. In the spin-one regulator, whose wall sequence converges best, the wall extrapolates below one with or without a single slow correction from the leading irrelevant line operator, which disfavours a limit at one. A fit with both slow corrections is too poorly conditioned to exclude it (Appendix~\ref{app:wallroutes}).} In $O(3)$ and $O(4)$ the measured walls are still rising at the largest sizes. The wall is close to marginal for $N=3$ and $4$, and the lattice does not decide its relevance at these sizes.\footnote{Their extrapolations reach marginality or stay just below it depending on the correction law, which the leading irrelevant line operator controls (Appendix~\ref{app:wallroutes}). The factorized estimates lie lower, because their shape factor is defined with the curvature (Appendix~\ref{app:tiltnorm}).}

These lattice values can be set against the analytic ones collected in Table~\ref{tab:anchors}. At $\epsilon=1$ the measured stiffnesses and walls lie between the leading and first-order values of the one-loop junction (Appendix~\ref{app:pinning}), so the expansion does not yet decide relevance in three dimensions. The second-order coefficient of the shape factor is known, but its value in three dimensions requires resummation. The wall also needs the two-loop tilt normalization, and with it the three-loop running of the pinning coupling, which is not available. Combining the resummed shape factor with the lattice stiffness instead gives an $O(2)$ wall below one, in line with the measured walls.

What the analytic results do control is the growth of the wall with $N$. At large $N$ the junction is a classical saddle (Appendix~\ref{app:junctionlargeN}), and both the stiffness and the wall grow linearly in $N$. Walls are therefore irrelevant at large $N$, so an ordering threshold $N_c$ exists.\footnote{If the $O(2)$ wall is relevant, as the data favour, then a continuous crossing has $N_c>2$, with $N=3$ and $4$ close to marginal, which is compatible with the above-one extrapolation for $O(3)$ of Ref.~\cite{Yang:2026qtk}. At $N=1$ the wall between opposite Ising magnetic lines lies below our $O(2)$ estimates, as the growth suggests. For $N=2$ a $[1/1]$ Pad\'e approximant between $d=4$ and the Kosterlitz--Thouless point places the wall above one (Table~\ref{tab:anchors}), but such interpolations are unreliable for $O(2)$, whose approach to two dimensions has an essential singularity. The analytic routes to the threshold are collected in Appendix~\ref{app:wallroutes}.} The leading slope does not locate it, because the corrections of relative order $1/N$ are large at small $N$, as the one-loop factor $N+8$ already shows.

\subsection{How the twist changes the wall}\label{sec:twistwalls}
\begin{table}[t]\centering\small
\begin{tabular}{lccc}\toprule
 & & \multicolumn{2}{c}{$\Delta_{+-}(\alpha)/\Delta_{+-}(0)$}\\
$\alpha$ & $O(2)$ wall $\Delta_{+-}(\alpha)$ & $O(2)$ & $N=\infty$\\\midrule
0.2 & 0.75--0.78 & 0.94--0.96 & 0.927\\
0.3 & 0.70--0.71 & 0.87--0.89 & 0.831\\
0.5 & 0.53--0.56 & 0.65--0.69 & 0.476\\
\bottomrule\end{tabular}
\caption{Walls on pinned monodromy lines. The absolute $O(2)$ dimensions show relevance; the ratios show softening relative to the same pin at zero twist [Fig.~\ref{fig:walls}(b)]. The lattice ratios are formed at equal $\Nm$ before extrapolation. Ranges span fits linear in $1/\Nm$ over the last three sizes, linear and quadratic in $1/\Nm$ over all sizes, and linear in the tilt excess; they are not confidence intervals. Sizes are $\Nm=6$--$12$ at $\alpha=0.3,0.5$ and $\Nm=6$--$10$ at $\alpha=0.2$. Table~\ref{tab:monojunctiondetails} separates the curvature-based stiffness and shape ratios, and Table~\ref{tab:monopinlargeN} gives the large-$N$ saddle values.}
\label{tab:monojunction}
\end{table}

For comparison with the pinned $O(2)$ monodromy line, we now use a large-$N$ model in which every component is twisted in complex pairs. This differs from the partial twist with $N-2$ spectators in Sec.~\ref{sec:entropy}, but the same constraint balances the field fluctuations and the condensate. The straight pinned line carries a classical field $\sqrt N\phi_*e^{\mathrm i\alpha\theta}$, where $\phi_*$ is its amplitude in the hyperbolic frame. The order parameter winds by a fraction $\alpha$ of a turn around the defect. The angular gradient costs energy, and the saddle reduces that cost by weakening the condensate. Its squared amplitude $\phi_*^2$, and with it the stiffness $\Ct=2N\phi_*^2$, decrease monotonically with $\alpha$ [Fig.~\ref{fig:walls}(a), Appendix~\ref{app:monopin}]. At large $N$ this weakening accounts for most of the reduction of the wall dimension.

For the pinned $O(3)$ line of Sec.~\ref{sec:monopin}, the relevant large-$N$ comparison instead retains the untwisted spectators. At half twist their soft response supports a pinned saddle with $\phi_*^2\propto N^{-1/2}$ even though the leading condensate vanishes (Appendix~\ref{app:monopin}). Connecting this mechanism to the small-$N$ lattice endpoint remains an open question.

The lattice establishes the reduction of the wall more clearly than its division between stiffness and shape. The absolute wall dimensions in Table~\ref{tab:monojunction} lie below one at every twist studied, and their ratios show the wall becoming smaller as the twist grows [Fig.~\ref{fig:walls}(b)]. The angle curvature softens more than the Ward stiffness at the sizes reached (Tables~\ref{tab:monojunctiondetails} and~\ref{tab:softening}), and in the spin-one regulator the Ward stiffness stays close to its untwisted value (Appendix~\ref{app:monopin}). Ratios at equal size to the untwisted line built with the same pin cancel most of the dependence on the pin. The reduction of the curvature-based shape factor agrees at moderate twist with the fully twisted large-$N$ saddle. The measured walls therefore destabilize the candidate ordered $O(2)$ monodromy line, with a larger twist moving it further from marginality.

\subsection{Where an ordered line can go}\label{sec:order}
The wall dimension tests the stability of the candidate ordered line against domain walls. The entropy supplies a necessary condition for a flow from that line to the unpinned one. The ordered line has expectation value $2\mathfrak g_{\rm pin}$, where $\mathfrak g_{\rm pin}$ is that of a single pinned line. Where its walls are relevant it flows on the line, and by the $g$-theorem~\cite{Cuomo:2021rkm} it can end on the unpinned line, of expectation value $\mathfrak g_{\rm u}$, only if $\log\mathfrak g_{\rm u}-\log\mathfrak g_{\rm pin}<\log2$.\footnote{The reverse flow would need a relevant line operator that preserves the exchange of the two pinnings but breaks the continuous symmetry. The lines studied here have none. On the trivial line the lowest such operator is the traceless symmetric tensor, whose dimension exceeds one, and on the $O(2)$ monodromy line the lowest charge-two level lies above one at every twist computed (Appendix~\ref{app:pinentropy}).}

\begin{figure}[t]\centering
\includegraphics[width=\textwidth]{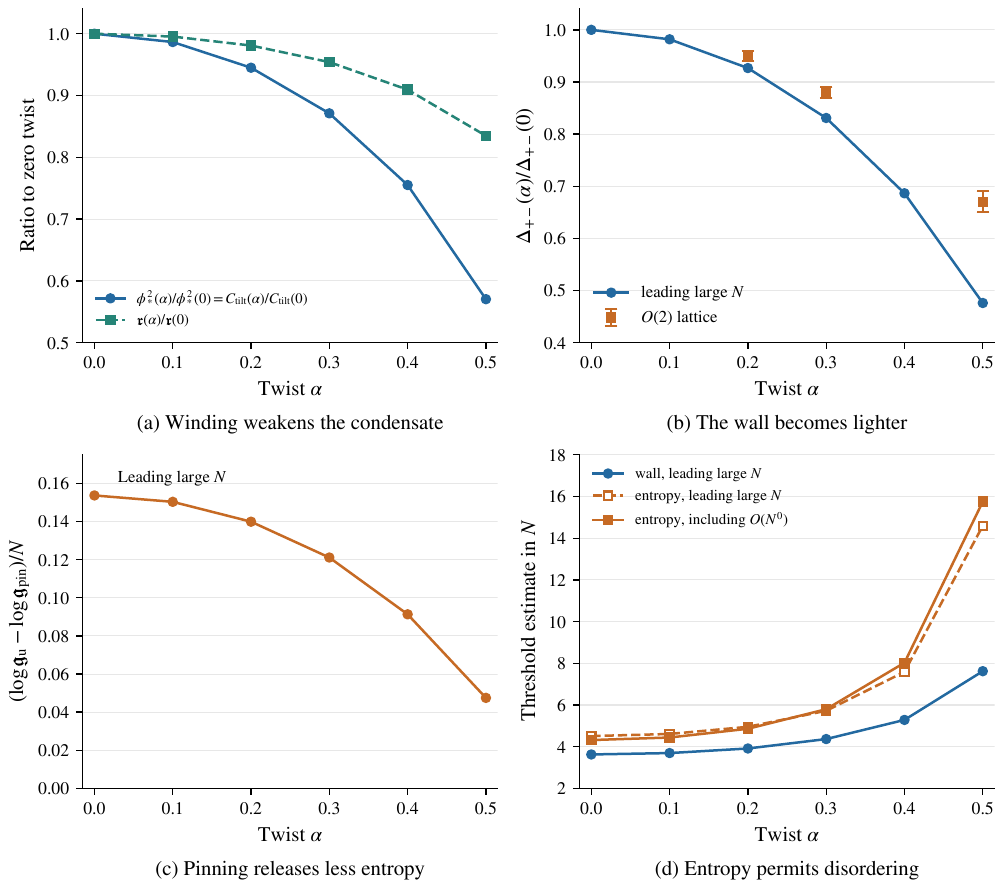}
\caption{Winding, walls and the entropy constraint in the fully twisted large-$N$ model. (a) Increasing the twist reduces the squared condensate amplitude and hence the stiffness, $\Ct=2N\phi_*^2$. The angular shape changes less. Values are normalized to zero twist. (b) The resulting wall reduction compared with the pinned $O(2)$ lattice line; bars span the fit choices of Table~\ref{tab:monojunction}. (c) The leading entropy released by pinning, per component, decreases with the twist. (d) Threshold estimates from the leading wall dimension and from the entropy at leading and next order (Table~\ref{tab:order}). Below the wall estimate the wall is relevant at that order; below the entropy estimate the $g$-theorem permits disordering into the unpinned line. The entropy thresholds lie above the leading wall threshold at every computed twist. These are large-$N$ estimates, with the next wall correction unknown; they do not locate the finite-$N$ phase boundary. Lines join computed twists.}
\label{fig:walls}
\end{figure}

For the fully twisted large-$N$ model introduced above, the wall and entropy follow from the same saddle equations. On $S^1\times\mathbb H^2$, with every component twisted in complex pairs, the expectation value of a line whose fields see the background $\varsigma$ obeys
\begin{equation}
 \frac{\mathrm d\log\mathfrak g}{\mathrm d\varsigma}=-\frac N2\,\mathcal A_\alpha(\varsigma),
 \label{eq:dloggdsigma}
\end{equation}
where $\mathcal A_\alpha(\varsigma)$ is the coincident deficit of one twisted complex component at background $\varsigma$, which reduces to Eq.~\eqref{eq:Aalpha} at $\varsigma=0$ (Appendix~\ref{app:pinentropy}). The unpinned line sits at the zero $\varsigma_{\rm u}$ of $\mathcal A_\alpha$, the pinned line at $\varsigma_{\rm pin}=\tfrac14-\alpha^2$, and the classical action of the condensate vanishes on its own saddle. Hence
\begin{equation}
 \log\mathfrak g_{\rm u}-\log\mathfrak g_{\rm pin}=\frac N2\int_{\varsigma_{\rm u}}^{\varsigma_{\rm pin}}\mathcal A_\alpha(\varsigma)\,\mathrm d\varsigma+O(N^0),
 \label{eq:entropygap}
\end{equation}
which is positive at every twist, as the $g$-theorem requires for the pinning flow, because $\log\mathfrak g$ is largest where $\mathcal A_\alpha$ vanishes. The same function fixes the condensate, since Eq.~\eqref{eq:phistaralpha} of Appendix~\ref{app:monopin} reads $\phi_*^2=\mathcal A_\alpha(\varsigma_{\rm pin})/4\pi^2$, and hence the stiffness $\Ct=2N\phi_*^2$, which together with the shape factor gives the wall. The entropy released by pinning decreases with the twist [Fig.~\ref{fig:walls}(c)]. At zero twist Eq.~\eqref{eq:entropygap} reproduces the large-$N$ entropy of the ordinary pinned line~\cite{Cuomo:2021kfm}.

Figure~\ref{fig:walls}(d), with values in Table~\ref{tab:order} of Appendix~\ref{app:pinentropy}, compares the thresholds inferred from these expansions. The leading entropic threshold lies above the leading wall threshold at every twist, so the entropy permits disordering into the unpinned line wherever the leading wall estimate is relevant. This compatibility supplies a candidate endpoint, whose dynamical realization still has to be tested. The twist raises both thresholds, because a condensate that winds both stiffens the walls less and releases less entropy when it forms. The term of order $N^0$, which we compute in three dimensions, lowers the entropic threshold slightly at small twist and raises it at large twist, and keeps it above the leading wall threshold at every twist evaluated. At zero twist the one-loop entropy near four dimensions places it lower still, in the window where the lattice walls approach marginality. In $O(2)$, where the lattice walls are favoured to be relevant, the entropy released by pinning at zero twist lies well below $\log2$ even with the term of order $N^0$ included (Appendix~\ref{app:pinentropy}), so the disordering flow is allowed there as well, although the expansion is not controlled at $N=2$.\footnote{Yang et al.\ measure the expectation value of the ordinary pinned line on the fuzzy sphere and read $\mathfrak g_{\rm pin}<\tfrac12$ as a condition for the ordered line to be stable~\cite{Yang:2026qtk}. In the present reading this inequality forbids the ordered line from disordering into the trivial line, so relevant walls would have to drive it elsewhere, which supports their conclusion as a consistency argument rather than a proof of stability. Their values place the entropic threshold below both expansions, and slightly below the one-loop value (Appendix~\ref{app:pinentropy}).} The comparison concerns the plain lines of our lattice and large-$N$ setups, which carry no additional degrees of freedom on the line.

\section{A construction in five dimensions}\label{sec:five}\label{sec:fiveconstruction}
Bulk flow, spectator dressing and local order have counterparts for codimension-two defects in higher dimensions. On the fuzzy four-sphere they become questions about five-dimensional conformal theories,\footnote{Free-scalar and Yang--Lee~\cite{Fisher:1978pf} theories in five dimensions have been regulated on the fuzzy four-sphere in Ref.~\cite{Qian:2026fbx}.} and the first step is to construct the twisted defect itself. The defect worldvolume is now three-dimensional and intersects the radial $S^4$ on a great $S^2$. Its ordering problem and entropy constraints must therefore be formulated for a three-dimensional defect. The line-wall criterion used above applies specifically to a one-dimensional defect. States carry a transverse spin and a parallel $SO(3)$ spin, which give additional descendant tests of any proposed conformal limit. The geometric construction carries over directly. An odd flavour is a section of the Yang-monopole bundle tensored with a flat line bundle of holonomy $-1$ around the great sphere, while an even flavour is untwisted.

We prescribe the regular domain, with local behaviour $r^{|s|}e^{\mathrm is\theta}$. Finite trial sections then follow by multiplying monopole sections by $w^{1/2}$ or $\bar w^{1/2}$ and by polynomials in the coordinates of $S^4$, the base of the monopole bundle, where $w$ is a complex coordinate transverse to the great sphere. The square-root factors depend only on the base and preserve the transformation law of the monopole fibre. Their first derivatives have finite norm, so the Gram matrix and the covariant angular quadratic form define a Hermitian variational problem. Selecting a separated finite band and projecting local interactions then gives a finite Hermitian many-body Hamiltonian with the specified holonomy, whose sections, forms and projector are given in Appendix~\ref{app:s4}.

The harmonic weights of the untwisted projected density approach their continuum values at fixed angular momentum, giving a direct check that the regulator resolves functions on $S^4$ as the cutoff grows. Recovering a five-dimensional scalar theory also requires the additional fibre modes to separate from its low-energy scalar sector.\footnote{Appendix~\ref{app:s4locality} gives the density form factors and explains this dynamical requirement. The extra modes associated with the $\CP^1$ fibre of $\CP^3\to S^4$ are a familiar feature of the fuzzy four-sphere~\cite{Medina:2002pc}.} For a free five-dimensional scalar the regular half-twist has $\Dh_{1/2}=2$, $\Dh_{3/2}=3$ and a displacement at four. Exact correlators and matrix elements of the parallel conformal generators then test whether the states form the intended defect theory.\footnote{The finite-cutoff tests in Appendices~\ref{app:s4tests} and~\ref{app:s4generator} have not yet established the many-body free-defect limit. The generator test also requires improving the microscopic energy density, a freedom that leaves the spectrum unchanged. Following a five-dimensional defect along a bulk flow remains a future application of the construction.}

\section{Discussion}\label{sec:discussion}
A monodromy line can change because the bulk flows underneath it or because the line flows on its own, and the two leave different marks on the defect entropy. Along the bulk flow of the anisotropic $O(3)$ model the monodromy follows the light fields. The easy-plane branch retains the twisted pair and approaches the $O(2)$ half-twist, while the easy-axis branch makes that pair massive and approaches a trivial line. The spectral and coupling comparisons on the easy-plane branch, together with the defect-free spectrum recovered after removing the local potential on the easy-axis branch, give a quantitative realization of this picture.

The fields that carry no twist nevertheless affect the defect, because they must compensate the deficit of fluctuations left by the twisted pair. At large $N$ their dressing gives a positive correction to the entropy, proportional to the conformal weight of the free defect. This correction grows as $N$ decreases, so a bulk flow connecting neighbouring spectator-dressed fixed points ends on a defect of larger entropy, which a flow on the line could not do. The symmetry identifies the surviving monodromy, while the entropy records the influence of the fields removed along the way.

The endpoint of the easy-plane flow carries a relevant charged mode, and a coupling to it takes the defect further, to a pinned monodromy line. Pinning before or after the bulk flow gives matching infrared junction data. A pin of the untwisted component survives only the easy-axis flow, so pins follow the light fields as the monodromy does. The walls between opposite pinnings test the stability of the candidate ordered line. They are relevant on the pinned $O(2)$ monodromy lines studied and favoured to be relevant, with a smaller margin, on the ordinary $O(2)$ line. For ordinary $O(3)$ and $O(4)$ pinned lines the walls lie close to marginality, while at sufficiently large $N$ they become irrelevant.

The entropy connects this stability test to the possible endpoint of a line flow. In the fully twisted large-$N$ model, the constraint relates the condensate and the entropy to the deficit of field fluctuations, just as it governs spectator dressing for the partial twist. Winding weakens this condensate and lowers both the wall dimension and the entropy released by pinning. The resulting entropy permits disordering into the unpinned line wherever the leading wall estimate is relevant. The next-order entropy correction preserves this compatibility with the leading wall estimate at every twist evaluated. Establishing the disordering flow itself requires following the wall perturbation.

Whether the entropy increases in the three-dimensional $O(3)\to O(2)$ flow requires an independent determination at small $N$. A measurement combining cap entanglement with the stress-tensor one-point function could test whether the spectator effect found at large $N$ persists in this flow.\footnote{Appendix~\ref{app:entanglement} develops the cap-entanglement calculation and tests it against the free defect, identifying the normalization problem that must be resolved before extracting the entropy.}

The spectator sector also raises a question about the existence of the defect. Near four dimensions, the perturbative branch considered here has complex-conjugate fixed points, whereas the three-dimensional lattice supports a real defect. Feng et al.~\cite{Feng:2026rrk} have now analyzed the coupled defect beta functions at one loop and followed the fixed-point branches at large $N$, finding a stable three-dimensional saddle on the singular branch of the spectators. Connecting this branch structure to the finite-$N$ defects realized on the lattice remains an open question.\footnote{Appendix~\ref{app:entropyeps} examines the spectator channel near four dimensions, including its resummation and the limits of the resulting continuation.}

For ordinary pinned lines, determining the ordering threshold $N_c$ requires resolving the relevance of the $O(3)$ and $O(4)$ walls. The slow corrections associated with the leading irrelevant line operator make this difficult at the accessible sizes. Controlling these corrections would determine where the candidate ordered line becomes stable against domain walls. The first fluctuation correction to the large-$N$ junction saddle would provide an independent analytic handle on the wall dimension and its approach to this threshold.\footnote{Appendices~\ref{app:tiltnorm} and~\ref{app:wallroutes} compare the stiffness estimators and wall extrapolations and examine their finite-size corrections. Appendix~\ref{app:junctionlargeN} gives the large-$N$ junction saddle, while Appendix~\ref{app:monopin} studies the effect of the twist on the pinned line.}

The pinned partial twist presents a further small-$N$ question. Its half-twist condensate survives through the soft spectator response described in Sec.~\ref{sec:twistwalls}, with squared amplitude $\phi_*^2\propto N^{-1/2}$. The region in which this response must be resummed has width $O(1/N)$, so it is not narrow for $O(3)$. The finite Ward stiffness of the pinned $O(3)$ half-twist on the lattice (Sec.~\ref{sec:monopin}) suggests that its condensate survives as well. An effective theory of the spectator and condensate fluctuations at small $N$ could determine whether the same mechanism connects the two descriptions.\footnote{Appendix~\ref{app:monopin} derives the near-half-twist saddle and gives the lattice tests.}

The four-sphere construction provides a starting point for asking these questions in five dimensions. Its immediate purpose is to define the monodromy and its regular domain in a finite Hamiltonian. The longer-term task is to establish a many-body conformal limit that retains the prescribed defect domain, using the free correlator, descendant spectrum and displacement Ward identity as independent tests. This would allow bulk flows and flows localized on the defect to be studied together in a higher-dimensional setting.\footnote{Appendix~\ref{app:s4} gives the construction, and Appendix~\ref{app:s4tests} reports the current tests of the free-defect spectrum, correlator and Ward normalization.}

\noindent\emph{Note added.} While this manuscript was being finalized, related works by Feng et al.~\cite{Feng:2026rrk}, Barrat and Giombi~\cite{Barrat:2026ozj}, and Tang and Zhu~\cite{TangZhu:2611.08102} appeared. Feng et al.\ study partial monodromy defects using renormalization group methods, the large-$N$ expansion and Monte Carlo simulations. Their identification of the spectators' singular branch and their analysis of defect fixed-point branches overlap with the spectator mechanism and branch analysis discussed here. Barrat and Giombi study the pinning-field line defect in the critical $O(N)$ model at finite temperature, including the tilt and longitudinal singlet correlators. Tang and Zhu study the continuous family of $U(1)$ monodromy defects of the $O(2)$ model on the fuzzy sphere, dual to fractional Wilson lines of the critical noncompact Abelian Higgs model. They find no relevant symmetry-preserving defect scalar and identify the charge-one defect operators, monopole operators in the dual description, as relevant symmetry-breaking deformations, in line with Appendix~\ref{app:three} and the charged pin of Sec.~\ref{sec:monopin}.

\acknowledgments
We thank Wei Zhu for comments on the draft. Claude Opus 5.5 is used for generating numerical scripts, doing analysis and for drafting the manuscript. GPT Astra 6.0 was used for technical audits. The authors have guided the AI at major steps and take responsibility for the results presented in the paper. K.L. and J.N. are supported by the Beijing Institute of Mathematical Sciences and Applications.

\appendix

\section{Three-dimensional construction and fixed-point benchmarks}\label{app:three}
Before the endpoints of the bulk flows can be trusted, the twisted orbital space, the Hamiltonians, the bulk energy unit and the normalization of matrix elements have to be tested on the free scalar and on interacting fixed points whose defect data are known, including those of Refs.~\cite{Qian:2026lrp,Tang:2026kzf} where they overlap with ours.

\subsection{Fractional flux and the regular orbital space}
Radial quantization maps a line through the origin to its two intersections with each spatial $S^2$, so on the fuzzy sphere the defect sits at the two poles. The untwisted lowest Landau level has monopole spin $j$ and $\Nm=2j+1$ orbitals. On a sphere of radius $R$ its coordinate matrices are $\mathcal X_i=RL_i/\sqrt{j(j+1)}$, with $L_i$ the spin-$j$ generators, so that $\sum_i\mathcal X_i^2=R^2$. A fractional flux $\alpha$ through the poles shifts the allowed angular momenta, and in a singular gauge the regular orbitals are
\begin{equation}
 \begin{gathered}
 \Phi_m^{(\alpha)}(\vartheta,\theta)=\mathfrak n_m e^{\mathrm im\theta}
 \cos^{j+m}\!\frac\vartheta2\,\sin^{j-m}\!\frac\vartheta2,
 \qquad m\in(-j+\alpha+\mathbb Z)\cap[-j,j],\\
 \mathfrak n_m^2=\frac{\Gamma(2j+2)}{4\pi\Gamma(j+m+1)\Gamma(j-m+1)},
 \end{gathered}
 \label{eq:orb}
\end{equation}
where $\vartheta$ is the polar angle and $\theta$ the azimuth. A flavour of charge $q_a$ sees the flux $q_a\alpha$. The common monopole phase cancels from $\mathsf f_a^\dagger\mathsf f_b$, which is left with the monodromy $e^{2\pi\mathrm i(q_b-q_a)\alpha}$, so every symmetry-invariant local interaction stays single-valued. The orbital space thus imposes the fractional transverse spins, and what remains to be tested is dynamical, namely the conformal energies and the matrix elements.

For noninteger flux the regular band has $\Nm-1$ orbitals per twisted flavour, and the missing endpoint orbital returns only at integer flux. Taking $\alpha\to0$ at fixed $\Nm$ through noninteger fluxes is therefore not the same finite problem as setting $\alpha=0$, a difference that matters whenever a small-angle defect is compared with the untwisted theory (Appendix~\ref{app:entanglement}).

\subsection{Local Hamiltonians and the local potential}
Each model comes with two contact prescriptions for its Hamiltonian $H_\alpha$ at twist $\alpha$, which differ by a one-body potential,
\begin{equation}
 H_\alpha(h,g;\lambda)=H_\alpha(h,g;0)+\lambda V_{\rm core},
 \qquad V_{\rm core}=\sum_{a,m}\delta\mathcal U_a(m)\mathsf n_{ma}.
 \label{eq:coreknob}
\end{equation}
The original prescription has $\lambda=1$, and $\lambda=0$ removes the potential $\delta\mathcal U_a(m)$, which we define below Eq.~\eqref{eq:ordering} once the interactions are in place.

To build these interactions we write $\mathsf f_a(\Omega)=\sum_m\Phi_{ma}(\Omega)\mathsf f_{ma}$, where $\Omega=(\vartheta,\theta)$ is a point on the sphere and $\Phi_{ma}$ are the orbitals of flavour $a$. For a local kernel $\mathcal K=v_0\delta+v_1\nabla^2\delta$, projection onto these orbitals gives
\begin{equation}
 V^{(aa'bb')}_{1234}=\int \mathrm d\Omega\,\mathrm d\Omega'\,
 \Phi_{1a}^*(\Omega)\Phi_{4a'}(\Omega)\mathcal K(\Omega,\Omega')
 \Phi_{2b}^*(\Omega')\Phi_{3b'}(\Omega'),
 \label{eq:Vproj}
\end{equation}
where the labels $1,\ldots,4$ stand for orbitals, with coefficients fixed by the untwisted pair pseudopotentials $(V_0,V_1)$. For Ising we rotate to $\mathsf f_\pm=(\mathsf f_\uparrow\pm\mathsf f_\downarrow)/\sqrt2$ and twist only $\mathsf f_-$. The Hamiltonian is
\begin{equation}
 H_{\rm I}=\frac12\sum\mathsf P_{aa'bb'}V^{(aa'bb')}_{1234}
 \mathsf f_{1a}^\dagger\mathsf f_{2b}^\dagger\mathsf f_{3b'}\mathsf f_{4a'}
 -h\sum_m(\mathsf n_{m+}-\mathsf n_{m-}),\qquad
 \mathsf P=\mathbf1\otimes\mathbf1-\mathsf X\otimes\mathsf X,
 \label{eq:isingH}
\end{equation}
where $\mathsf X$ is the first Pauli matrix, acting on the two flavours $\pm$, $(V_0,V_1,h)=(4.75,1,3.16)$ and the fermion number is $N_e=\Nm$~\cite{Zhu:2022gjc}, with the interaction normal ordered.

The vector construction adds a singlet flavour $0$ to the $N$ vector flavours~\cite{Guo:2025odn},
\begin{align}
 H_{\rm WF}&=\int \mathrm d^2x\left[\mathsf n^2-u\sum_{a=1}^N\phi_a\nabla^2\phi_a+h\mathsf n_0\right],
 &\phi_a&=\mathsf f_0^\dagger\mathsf f_a+\mathsf f_a^\dagger\mathsf f_0,\label{eq:vectorH}\\
 H_{\rm G}&=\int \mathrm d^2x\left[\mathsf n^2+u\sum_{0\le a<b\le N}\mathcal L_{ab}\nabla^2\mathcal L_{ab}
 +\frac h{R^2}\mathsf n_0\right],
 &\mathcal L_{ab}&=\mathrm i(\mathsf f_a^\dagger\mathsf f_b-\mathsf f_b^\dagger\mathsf f_a),
 \label{eq:gaussH}
\end{align}
with $\mathsf n=\sum_{a=0}^N\mathsf f_a^\dagger\mathsf f_a$ and $\mathsf n_0=\mathsf f_0^\dagger\mathsf f_0$ the total and singlet densities, $R^2=\Nm$ and $N_e=\Nm$. We use $H_{\rm WF}$ with $u=0.270,0.264,0.255$ for the Wilson--Fisher models with $N=2,3,4$, and $H_{\rm G}$ with $N=2$ and $u=0.148$ for the free complex-scalar target. A twisted vector plane is diagonalized into flavours of charges $\pm1$. The finite Hamiltonian that targets the Gaussian theory is itself interacting, and only its infrared limit is free.

An independent spin-one realization of $O(2)$~\cite{Dey:2026cso} gives us a second regulator. It uses three flavours of charges $+1,0,-1$, on which the spin-one matrices $\mathsf S^x$ and $\mathsf S^y$ act, and
\begin{equation}
 \begin{split}
 H_{\rm XY}&=\sum\mathsf P_{aa'bb'}V^{(aa'bb')}_{1234}
 (\mathsf f_{1a}^\dagger\mathsf f_{4a'})(\mathsf f_{2b}^\dagger\mathsf f_{3b'})
 +\Lambda\sum_m(\mathsf n_{m,+1}+\mathsf n_{m,-1}),\\
 \mathsf P_{aa'bb'}&=\delta_{aa'}\delta_{bb'}
 -\tfrac12(\mathsf S^x_{aa'}\mathsf S^x_{bb'}+\mathsf S^y_{aa'}\mathsf S^y_{bb'}),\qquad (V_0,V_1)=(4,1),
 \end{split}
 \label{eq:xyH}
\end{equation}
at the same filling $N_e=\Nm$. The vector and spin-one products are kept un-normal-ordered, and their one-body contraction is, in the notation of Eq.~\eqref{eq:Vproj},
\begin{equation}
 \mathcal U_a(m)=\sum_b\mathsf P_{abba}\sum_{m'\in\operatorname{orb}(b)}
 V^{(abba)}_{m\,m'\,m\,m'},
 \label{eq:ordering}
\end{equation}
where $\operatorname{orb}(b)$ is the set of orbitals of flavour $b$. Removing the untwisted constant separately for each flavour defines $\delta\mathcal U_a(m)$, and with it the potential $V_{\rm core}$ of Eq.~\eqref{eq:coreknob}.

\begin{figure}[t]\centering
\includegraphics[width=\textwidth]{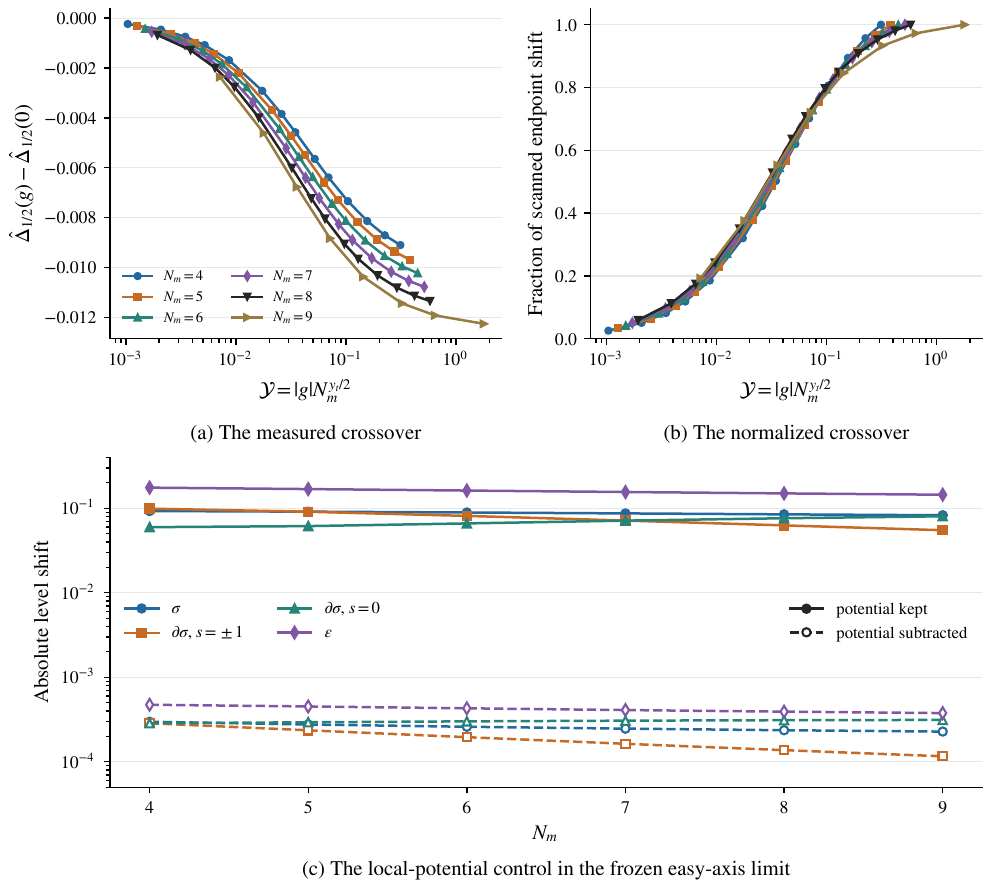}
\caption{Controls on the flow comparison. (a) The measured charge-one gap shift along the easy-plane branch, with the displacement candidate fixed at two. (b) The same shift divided by its value at the most negative coupling in each scan. The common endpoint at one follows from this normalization; the intervening crossover tests the scaling variable with the bulk exponent fixed. (c) Absolute shifts of the four easy-axis line levels from the defect-free values at $g=3$, with the original local potential (filled) and after subtraction (open). These controls accompany the endpoint comparisons of Fig.~\ref{fig:branches}.}
\label{fig:flowcontrols}
\end{figure}

The frozen easy-axis limit of Sec.~\ref{sec:core} shows why the subtraction matters, since only the subtracted Hamiltonian reproduces the defect-free levels there [Fig.~\ref{fig:flowcontrols}(c)].

\subsection{Bulk tuning and matrix-element normalization}\label{app:estimators}

\begin{table}[t]\centering\small
\begin{tabular}{lp{0.68\textwidth}}\toprule
comparison & $\beta_{\phi\hat\psi_{1/2}}$\\\midrule
finite size, $\Nm=6$--$9$ & flowed and independent $O(2)$ agree to $0.08$--$0.13\%$\\
extrapolated, $\Nm=7$--$9$ & $0.865$--$0.876$ for both, differing by at most $1.2\times10^{-3}$\\
ultraviolet $O(3)$ & extrapolates $0.004$--$0.008$ below the independent $O(2)$\\
one loop, $\epsilon=1$ & $0.847$ for every model at $|s|=\tfrac12$\\
two loops, $\epsilon=1$ & $0.8480$ for $O(2)$ and $0.8453$ for ultraviolet $O(3)$\\
references & free field in $d=3$: $0.798$; Ising twist on the lattice: $0.862$\\
\bottomrule\end{tabular}
\caption{Bulk-to-defect coupling comparisons. Finite-size values use normalized equatorial matrix elements. Extrapolations use the common window $\Nm=7$--$9$ with $\Nm^{-1}$ and $\Nm^{-1/2}$ corrections, as shown in Fig.~\ref{fig:branches}(b); the full-precision inputs are in Appendix~\ref{app:trajectory}. Perturbative values follow from Eqs.~\eqref{eq:betaeps} and~\eqref{eq:betaeps2}. The spread describes the correction laws tested, not a confidence interval.}
\label{tab:coupling}
\end{table}

At a conformal defect the levels $E_n^\Dc$ measure dimensions, $E_n^\Dc-E_0^\Dc\propto\Dh_n/R$ up to cutoff corrections, and one bulk energy fixes the unit. For the vector models we take $\gamma_B=E_{\partial\phi}-E_\phi$ and tune the untwisted field so that
\begin{equation}
 \frac{E_\phi-E_0}{E_{\partial\phi}-E_0}
 =\frac{\Delta_\phi}{\Delta_\phi+1},
 \label{eq:bulktune}
\end{equation}
with target dimensions $\Delta_\phi=1/2$ for the Gaussian model and $0.519088$, $0.518942$, $0.51812$ for $O(2)$, $O(3)$, $O(4)$~\cite{Chester:2019ifh,Chester:2020iyt,Hasenbusch:2021rse}. For Ising we use the stress tensor, $\gamma_B=(E_T-E_0)/3$, with $E_T$ its energy. Where explicitly indicated we calibrate instead with the displacement operator $D$, through $\gamma_D=(E_D^\Dc-E_0^\Dc)/2$ with $E_D^\Dc$ its energy, to compare with displacement-calibrated spectra. The spin-one tuning gives $\Lambda=2.9563,2.9075,2.8747,2.8516,2.8347$ at $\Nm=6,\ldots,10$ and $2.8123,2.7985$ at $\Nm=12,14$, and the locally tuned vector $O(2)$ fields are $h=-0.0860,-0.0853,-0.0848$ at $\Nm=6,8,10$ in our normalization.

Matrix elements need a normalization as well. We denote bulk-to-defect coefficients by $\beta_{\mathcal O\hat{\mathcal O}}$, reserving $b_1$ for the bulk-response slopes. For unit-normalized bulk and defect primaries the cylinder profile is
\begin{equation}
 \langle\hat{\mathcal O}_s|\mathcal O(\vartheta,\theta)|0_\Dc\rangle
 \propto \beta_{\mathcal O\hat{\mathcal O}}(\sin\vartheta)^{\Dh_s-\Delta_{\mathcal O}}e^{-\mathrm is\theta}.
 \label{eq:profile3}
\end{equation}
We divide the equatorial overlap by the untwisted overlap of the same density at the same $\Nm$. This removes the leading microscopic normalization and leaves corrections from admixtures of other continuum operators.\footnote{Normalizing a lattice density by a reference matrix element is standard in fuzzy-sphere calculations of operator product coefficients~\cite{Hu:2023xak,Han:2023yyb}. The remaining admixtures can be reduced with improved conformal generators~\cite{Fardelli:2024qla} or treated in conformal perturbation theory~\cite{Lao:2023zis}.} The Ising density is $\mathsf n^z=\mathsf f_+^\dagger\mathsf f_-+\mathrm{h.c.}$, and for the vector model we use the charged combination of $\phi_a$.

Near four dimensions the coefficient has a simple origin. A free field in a potential $\mathfrak v/r^2$ behaves near the defect as $r^{\tilde s}$ in the spin-$s$ channel, with effective spin $\tilde s=\sqrt{s^2+\mathfrak v}$, and its mode sum gives
\begin{equation}
 \beta_s^2=\frac{\Gamma(\tilde s+\frac d2-1)}{\Gamma(\frac d2-1)\Gamma(\tilde s+1)},
 \label{eq:betanu}
\end{equation}
which reduces to the free coefficient at $\mathfrak v=0$. We have checked this in three dimensions against the closed-form mode propagator. The Fourier integral along the defect of the product of the two modified Bessel functions of order $\tilde s$, at transverse distances $r$ and $r'$, equals $\mathrm Q_{\tilde s-1/2}(\cosh\chi)/2\sqrt{rr'}$, where $\mathrm Q$ is the Legendre function of the second kind and $\cosh\chi=(r^2+r'^2+y^2)/2rr'$. At one loop in $d=4-\epsilon$ the interaction enters only through the tadpole, which is a potential of this kind with $\mathfrak v=-2(n_t+1)\alpha(1-\alpha)\epsilon/(N+8)$ for a twisted component, $n_t$ being the number of twisted vector planes. Through $\tilde s\simeq|s|+\mathfrak v/2|s|$ it reproduces Eq.~\eqref{eq:oneloop}. At $d=4$, however, Eq.~\eqref{eq:betanu} equals one for every $\tilde s$, so the shift leaves $\beta$ unchanged at $O(\epsilon)$, and expanding in $d$ gives Eq.~\eqref{eq:betaeps}. Away from four dimensions Eq.~\eqref{eq:betanu} depends on $\tilde s$, and the twisted content of the model enters at $O(\epsilon^2)$. The part of that order induced by the one-loop shift, $\delta\log\beta_s^2=-\tfrac\epsilon2\operatorname{dg}'(|s|+1)\,\delta\tilde s$, with $\operatorname{dg}'$ the derivative of the digamma function, grows with the attraction, in the same order as the couplings of the five defects of Ref.~\cite{Tang:2026kzf}.

The other contribution at order $\epsilon^2$ is the sunset self-energy. At the fixed point it is $\Sigma_a=128\pi^4\epsilon^2(N+8)^{-2}\,G_a\big[\sum_cG_c^2+2G_a^2\big]$, where $G_a$ is the propagator of component $a$ and the sum runs over the $N$ components. We evaluate it at $d=4$ on $S^1\times \mathbb H^3$, where the untwisted propagator is $[8\pi^2(\cosh\chi-\cos\theta)]^{-1}$ and the half twist multiplies it by $\cos(\theta/2)/\cosh(\chi/2)$. In the circle mode $s$ the $\mathbb H^3$ propagator becomes $[\nu^2+s^2-\widetilde\Sigma_s(\nu)]^{-1}$, with $\nu$ the spectral parameter on $\mathbb H^3$ and $\widetilde\Sigma_s(\nu)$ the $\mathbb H^3$ spectral transform of the circle-mode-$s$ part of $\Sigma_a$. The coefficient of the leading defect primary, and hence $\beta_s^2$, is then multiplied by the residue $1+\partial_{\nu^2}\widetilde\Sigma_s$ at $\nu=\mathrm i|s|$. The mode projections are elementary, $\int \mathrm d\theta\,e^{-\mathrm i\kappa\theta}(\cosh\chi-\cos\theta)^{-3}=\pi\,\partial_{\cosh\chi}^2(e^{-|\kappa|\chi}/\sinh\chi)$, so each residue is a single integral over $\chi$ with a logarithmic divergence common to all modes. We remove it, together with the bulk normalization, by repeating the calculation without the defect. That calculation must reproduce the trivial-defect coefficients $\Gamma(\Delta_\phi+|s|)/[\Gamma(\Delta_\phi)\,|s|!]$ with $\Delta_\phi-1+\epsilon/2=(N+2)\epsilon^2/[4(N+8)^2]$, and it does so for $s=0,\ldots,3$ to fifty digits. For the lowest half-twisted mode, the sunset made of three twisted propagators gives $4\pi^2(4-5\log2)$, and the one with a twisted and two spectator propagators gives $4\pi^2(1-\log2)$, both in units of $\epsilon^2/[4\pi^2(N+8)^2]$. Together with the expansion of Eq.~\eqref{eq:betanu}, the result for $n_t$ half-twisted planes is
\begin{multline}
 \log\beta_{1/2}^2=-(1-\log2)\,\epsilon+\Big[\frac{\pi^2}{24}-\frac12+\frac{(n_t+1)(\pi^2/2-4)}{4(N+8)}\\
 +\frac{2(n_t+1)(4-5\log2)+(N-2n_t)(1-\log2)}{(N+8)^2}\Big]\epsilon^2+\cdots.
 \label{eq:betaeps2}
\end{multline}
For the $O(2)$ and $O(3)$ half twists at $\epsilon=1$, expanded strictly to second order, it gives $\beta=0.8480$ and $0.8453$. The second-order terms barely move the absolute value, but they place the ultraviolet $O(3)$ coupling $0.0027$ below the $O(2)$ one, and this difference between the models is the robust prediction.

The spectators of a partial twist add one subtlety. Their $s=0$ mode feels the attractive potential $-2n_t\alpha(1-\alpha)\epsilon/(N+8)$ at the Breitenlohner--Freedman value, so near four dimensions its dimension is complex at order $\sqrt\epsilon$, and the twisted dimensions acquire imaginary parts at order $\epsilon^{3/2}$ (Appendix~\ref{app:entropyeps}). Because Eq.~\eqref{eq:betanu} does not depend on $\tilde s$ at $d=4$, this reaches $\beta$ only at order $\epsilon^{5/2}$, beyond Eq.~\eqref{eq:betaeps2}. In three dimensions the spectator occupies its singular branch instead, as in the large-$N$ saddle of Sec.~\ref{sec:entropy}.

Three-point data on the defect come from a near-pole estimate,
\begin{equation}
 C_{123}^{\rm est}(\vartheta_0)=
 \frac{\langle\hat1|\mathcal O_{\rm lat}(\vartheta_0)|\hat3\rangle}
 {\langle\hat2|\mathcal O_{\rm lat}(\vartheta_0)|0_\Dc\rangle}
 (\cos\vartheta_0)^{-(\Dh_1-\Dh_2-\Dh_3)},
 \label{eq:poleest}
\end{equation}
where $|\hat1\rangle$ and $|\hat3\rangle$ are defect states, $\mathcal O_{\rm lat}(\vartheta_0)$ is a lattice density at polar angle $\vartheta_0$ near a pole, and $\hat2$ is the lowest state with the charge and spin transferred by the insertion. We average over $\vartheta_0=\pi/12,\pi/6,\pi/4$ and test the resulting plateau on the free model~\cite{Tang:2026kzf}.

\subsection{The free scalar as a control}\label{app:freebench}\label{app:freecore}
The exact free half-twist tells us how far to trust the level calibration and the field normalization separately. With the potential subtracted, five low levels lie within five percent of their free values at $\Nm=10$. The median deviation of a normalized correlator, over polar angles, relative azimuths and cylinder times up to one, falls from about $0.08$ to $0.016$, although the largest sampled deviation does not improve.

\begin{figure}[t]\centering
\includegraphics[width=\textwidth]{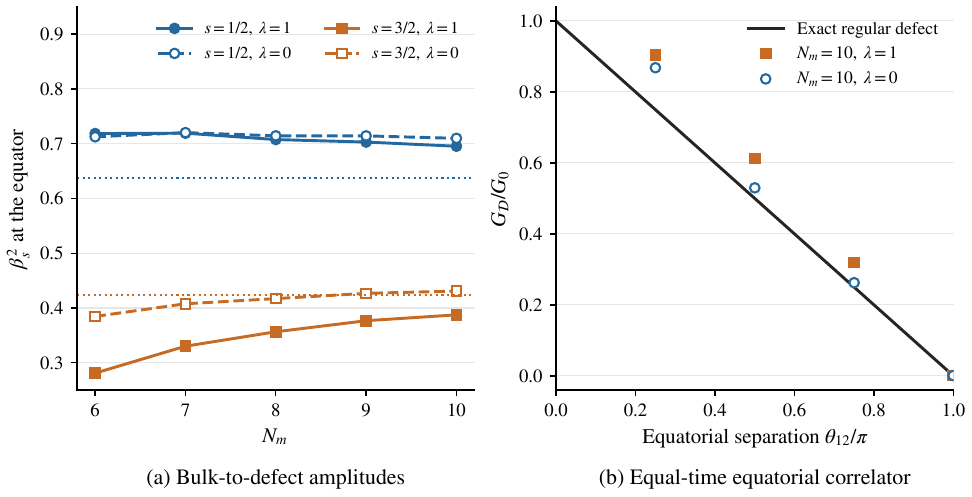}
\caption{Free-theory controls in three dimensions. (a) Squared equatorial bulk-to-defect coefficients with the original ($\lambda=1$, filled) and subtracted ($\lambda=0$, open) potential. Dotted lines are exact values. (b) The normalized equal-time equatorial correlator at $\Nm=10$ with both potentials, compared with $1-\theta_{12}/\pi$, where $\theta_{12}$ is the relative azimuth of the two points; this is Eq.~\eqref{eq:green35} at cross-ratio $\xi=0$ [Eq.~\eqref{eq:crossratios}] on the displayed sheet.}
\label{fig:freebenchmark}
\end{figure}

In three dimensions the free bulk-to-defect coefficients, written $\beta_s=\beta_{\phi\hat\psi_s}$, are
\begin{equation}
 \beta_s^2=\frac{\Gamma(|s|+1/2)}{\sqrt\pi\,\Gamma(|s|+1)},\qquad
 \beta_{1/2}^2=\frac2\pi,\qquad \beta_{3/2}^2=\frac4{3\pi}.
\end{equation}
At $\Nm=10$ with the potential subtracted, the estimates of $\beta_{1/2}^2$ and $\beta_{3/2}^2$ are about $0.710$ and $0.431$ (Fig.~\ref{fig:freebenchmark}). The higher-spin mode is close to its target, but the lowest squared amplitude is still about $11.5\%$ high. Accurate level spacings therefore do not imply equally accurate field normalizations.

\subsection{Selected interacting benchmarks}
Table~\ref{tab:benchmarks} records the fixed-point checks that we found most useful. For the Ising twist, calibrating with the displacement brings the lowest odd mode close to Monte Carlo, but the higher odd mode is closer on the bulk scale, and neither choice removes all cutoff effects. A ratio of dimensions does not depend on this choice. The lowest even scalar over the lowest odd mode is $2.46$ at $\Nm=14$, against about $2.47$ from Monte Carlo~\cite{Billo:2013jda} and from the independent fuzzy-sphere calculation of Ref.~\cite{Tang:2026kzf}. The comparison in Ref.~\cite{Qian:2026lrp} gives a different ratio, whose origin the present tests do not resolve.

\begin{table}[t]\centering\small
\begin{tabular}{llccc}\toprule
model and observable & unit & $\Nm=10$ & larger size & external target\\\midrule
Ising $\hat\psi_{1/2}$ & bulk & 0.9934 & 0.9632 ($20$) & 0.9187(6)\\
 & displacement & 0.9227 & 0.9184 ($20$) & \\
Ising $\hat\psi_{3/2}$ & bulk & 2.0122 & 1.9859 ($20$) & 1.99(5)\\
 & displacement & 1.8689 & 1.8936 ($20$) & \\
Ising $D$ candidate & bulk & 2.1533 & 2.0976 ($20$) & 2\\\midrule
$O(2)$, spin-one $\hat\psi_{1/2}$ & displacement & 0.8789 & 0.8798 ($14$) & ---\\
$O(2)$, vector $\hat\psi_{1/2}$ & displacement & 0.9123 & 0.9065 ($14$) & ---\\
$O(2)$, spin-one $\alpha_c$ & displacement & 0.3896 & 0.3949 ($14$) & 0.417 (1L), 0.383 (large $N$)\\
$O(2)$, vector $\alpha_c$ & displacement & 0.3941 & 0.3936 ($14$) & 0.417 (1L), 0.383 (large $N$)\\\bottomrule
\end{tabular}
\caption{Selected independent fixed-point checks with the original prescription. Here $\hat\psi_s$ is the lowest mode of transverse spin $s$, odd for the Ising twist and of charge one for the $O(2)$ twist, and $D$ is the displacement. Parentheses in the fourth column give $\Nm$, not uncertainties. Ising targets are from Ref.~\cite{Billo:2013jda}, except for the protected displacement. The $O(2)$ crossing is defined by the charged $s=\alpha-1$ mode reaching dimension one, and it concerns symmetry-breaking perturbations. Its references are the one-loop (1L) value and the leading large-$N$ value of the full twist from Eq.~\eqref{eq:largeNgap}.}
\label{tab:benchmarks}
\end{table}

The Ising bulk-to-defect coefficient of the lowest odd mode is about $0.862$ for $\Nm=8$--$14$, close to $0.857(3)$ in Ref.~\cite{Tang:2026kzf} but below the Monte Carlo value $0.968(2)$~\cite{Billo:2013jda}. The defect three-point estimates give $C_{\hat\psi\hat\psi\bar D}=1.350(19)$ and $C_{\hat\psi\hat\psi^\dagger\hat{\mathcal O}_0}=0.98(2)$, against the bootstrap values $1.345$ and $0.95$~\cite{Gaiotto:2013nva}. Here $\bar D$ is the displacement component of transverse spin $-1$, $\hat{\mathcal O}_0$ is the lowest neutral spin-zero defect operator, and the errors describe variations of the fits.

For $O(2)$ the charge-one modes $s=\alpha$ and $s=\alpha-1$ become degenerate at half twist. The twist $\alpha_c$ at which the charged mode becomes marginal, which we call the marginal crossing, is more stable between the two regulators than the half-twist dimensions themselves (Table~\ref{tab:benchmarks}). At fixed nonzero spin, the one-loop comparison in $d=4-\epsilon$ is
\begin{equation}
 \Dh_s=1-\frac\epsilon2+|s|
 -\frac{n_t+1}{N+8}\frac{\alpha(1-\alpha)}{|s|}\epsilon+O(\epsilon^2),
 \label{eq:oneloop}
\end{equation}
for $n_t$ equally twisted vector planes~\cite{Soderberg:2017oaa,Giombi:2021uae}. At $\epsilon=1$ it has uncontrolled higher-order corrections, especially at small $|s|$, so we use it for orderings rather than values. The one-plane $O(3)$ half-twist and the two-plane $O(4)$ half-twist give lowest charged dimensions $0.9259$ and $0.9011$ at $\Nm=10$ in displacement units, with marginal crossings $0.4117$ and $0.3807$, an ordering that agrees with Eq.~\eqref{eq:oneloop}. The $O(2)$ crossing lies between the one-loop value $5/12$ and the leading large-$N$ value for the full twist. We obtain the latter from the full-twist saddle of Ref.~\cite{Giombi:2021uae} in three dimensions, where $\Dh_s=\tfrac12+\sqrt{\varsigma_*+s^2}$ and the vanishing of the $\phi\bar\phi$ term in the bulk operator product expansion reads, after zeta regularization of the sum over $s\in\mathbb Z+\alpha$,
\begin{equation}
 \sum_{s\in\mathbb Z+\alpha}\left[\operatorname{dg}\big(\tfrac12+\sqrt{\varsigma_*+s^2}\big)-\log|s|\right]=-\log(2\sin\pi\alpha).
 \label{eq:largeNgap}
\end{equation}
At half twist this reproduces $\varsigma_*=-0.168$, $\Dh_{1/2}=0.786$ and $\Dh_{3/2}=1.943$ of Ref.~\cite{Giombi:2021uae}, and the mode $s=\alpha-1$ reaches dimension one at $\alpha=0.383$.

\subsection{Two-mode energies and charged instabilities}
\begin{figure}[t]\centering
\includegraphics[width=\textwidth]{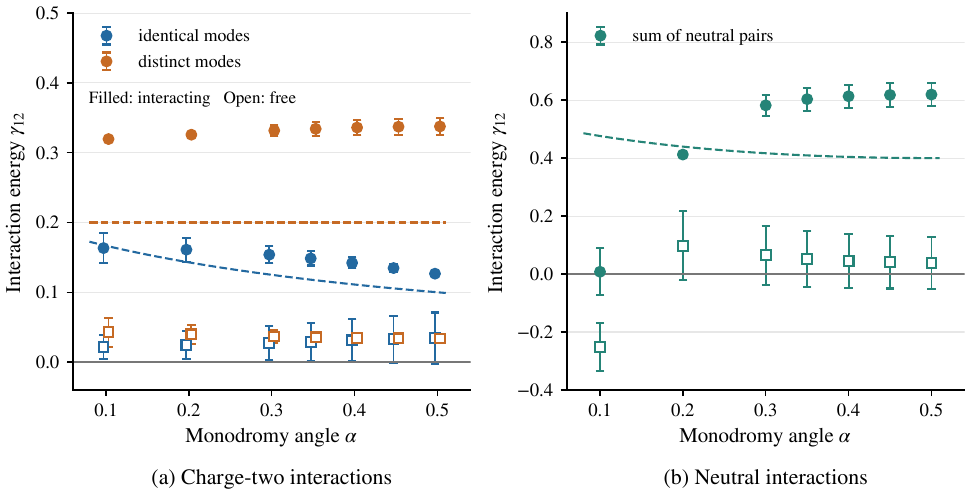}
\caption{Interaction energies with matched free controls. Filled circles are the vector $O(2)$ model and open squares the Gaussian target, extrapolated over $\Nm=7$--$10$. Bars are half the spread between $\Nm^{-1/2}$ and $\Nm^{-1}$ fits. Dashed curves are Eq.~\eqref{eq:contactpred}. In the neutral sector the plotted trace tests the sum of the two shifts, not their individual signs.}
\label{fig:contact}
\end{figure}

The last benchmark asks whether the defect modes interact. For two constituent modes $1$ and $2$, with $\Dh_{[12]}$ the dimension of the lowest two-mode state built from them, we define $\gamma_{12}=\Dh_{[12]}-\Dh_1-\Dh_2$ at the same cutoff and in the same energy unit, so that a free control has $\gamma_{12}=0$. For the two neutral spin-zero pairs we compare the sum of the shifts, which does not depend on their mixing angle. Tang and Zhu evaluate the contact shifts at half twist~\cite{Tang:2026kzf}, and we extend their quartic-overlap calculation to general $0<\alpha<1$, using modes with $|s_1|=\alpha$ and $|s_2|=1-\alpha$ in the same normalized $S^3$ integrals.\footnote{On $S^1\times \mathbb H^{d-1}$ the defect modes are particles in hyperbolic space, and these shifts are the analogue of the binding energies of two-particle states in anti-de Sitter space~\cite{Fitzpatrick:2011dm,Carmi:2018qzm,Giombi:2020rmc}.} For $O(2)$ at one loop and $\epsilon=1$ this gives
\begin{equation}
 \gamma_{\alpha,\alpha}^{(1)}=\frac{0.2}{1+2\alpha},\qquad
 \gamma_{\alpha,\alpha-1}^{(1)}=0.2,\qquad
 \operatorname{tr}\gamma_{\rm neutral}^{(1)}
 =\frac{0.4}{1+2\alpha}+\frac{0.4}{3-2\alpha}.
 \label{eq:contactpred}
\end{equation}
Figure~\ref{fig:contact} compares these predictions with the extrapolated shifts and their free controls. The distinct charge-two pair gives about $0.33$ at the sampled angles, above its one-loop value, while the identical pair gives $0.13$--$0.16$, and the positive neutral trace at $\alpha\ge0.3$ also exceeds its one-loop value. At $\alpha=0.1$ the neutral free control is displaced by about $0.25$, which precludes a quantitative reading there.

At half twist the identical-mode three-point estimator is $1.335$ in $O(2)$ and $1.347$ in the free model at $\Nm=10$, both below $\sqrt2$. The deficit is common to the two models, so it should not be read as an interaction. Finally, the lowest resolved symmetry-preserving neutral scalar stays above one in both $O(2)$ models, whereas the charged mode crosses one near $\alpha=0.39$--$0.40$. The charged crossing allows symmetry-breaking line perturbations~\cite{Kravchuk:2025evf}, which are the pinnings of Sec.~\ref{sec:pinmono} and are distinct from a symmetry-preserving instability.

\section{Conformal perturbation theory of the anisotropy flow}\label{app:cpt}
Near the $O(3)$ point the anisotropy is weak, and conformal perturbation theory (CPT) relates the response of the spectrum to known operator data~\cite{Lao:2023zis,Yang:2025stu}. We use it to ask whether the microscopic Hamiltonian generates the bulk flow with the right operator content.

\subsection{Retuning and first-order slopes}
The light-field condition of Eq.~\eqref{eq:flowtune} requires a thermal retuning, so the continuum deformation on the unit cylinder carries two couplings, $H=H_{\rm CFT}+\mu_t\int_{S^2}t_W+\mu_{\mathcal E}\int_{S^2}\mathcal E$, with both operators integrated over the unit sphere and $\mathcal E$ the $O(3)$ singlet. We choose the tensor component so that
\begin{equation}
 \langle\phi_a|t_W|\phi_a\rangle=c_a c_\phi,\qquad
 (c_1,c_2,c_3)=(1,1,-2),\qquad c_\phi=C_{\phi\phi t}/\sqrt3,
\end{equation}
with light and heavy weights $(c_{\rm L},c_{\rm H})=(-2,1)$ toward Ising and $(1,-2)$ toward $O(2)$. Writing $\varrho_{\mathcal O}=\Delta_{\mathcal O}(3-\Delta_{\mathcal O})/6$ for a bulk primary $\mathcal O$, the light-level condition and the change of the bulk unit give
\begin{equation}
 \mathcal R_{\rm ret}=\frac{1+\varrho_t}{1+\varrho_{\mathcal E}}=0.99085,\qquad
 \mu_{\mathcal E}=-\mathcal R_{\rm ret}
 \frac{c_\phi c_{\rm L}}{C_{\phi\phi\mathcal E}}\mu_t,\qquad
 X=4\pi\mu_t c_\phi(c_{\rm H}-c_{\rm L}),
 \label{eq:retune}
\end{equation}
and the heavy--light splitting is $\bar E_{\phi_{\rm H}}-\bar E_{\phi_{\rm L}}=X+O(X^2)$. Any independently measured primary level $n$, with operator product coefficients $C_{n\mathcal On}$, then has the first-order slope
\begin{equation}
 b_1(n)=\frac{
 C_{nt_Wn}-\mathcal R_{\rm ret}\dfrac{c_\phi c_{\rm L}}{C_{\phi\phi\mathcal E}}C_{n\mathcal E n}
 +\dfrac{\Delta_n}{\Delta_\phi}c_\phi c_{\rm L}(\varrho_t-\mathcal R_{\rm ret}\varrho_{\mathcal E})}
 {c_\phi(c_{\rm H}-c_{\rm L})}.
 \label{eq:b1}
\end{equation}
We use the bulk operator product coefficients $C_{\phi\phi t}=0.87451$, $C_{\phi\phi\mathcal E}=0.524261$, $C_{tt\mathcal E}=0.98348$, $C_{\mathcal E\mathcal E\mathcal E}=0.5055$ and $C_{ttt}=1.49957$ in the conventions of Ref.~\cite{Chester:2020iyt}, which give the first-order predictions in Table~\ref{tab:cptnumbers}. The lattice slopes move toward them as $\Nm$ grows, which tests the bulk operator content at first order.

\begin{table}[t]\centering\small
\begin{tabular}{llccc}\toprule
response & branch & continuum & $\Nm=9$ & finite-size fit\\\midrule
$b_1(t_W)$ & Ising & 0.682 & 0.724 & $0.708\pm0.010$\\
 & $O(2)$ & 1.198 & 1.276 & $1.220\pm0.031$\\
$b_1(\mathcal E)$ & Ising & 0.656 & 0.944 & $0.70\pm0.14$\\
 & $O(2)$ & 0.328 & 0.472 & $0.35\pm0.07$\\\bottomrule
\end{tabular}
\caption{First-order bulk response coefficients. Continuum values follow from the CFT data quoted in the text. The last column varies fitting powers and size windows.}
\label{tab:cptnumbers}
\end{table}

\section{Trajectory and displacement diagnostics}\label{app:open}

\subsection{Controls on the bulk trajectory}\label{app:trajectory}

We begin with the second flow of the main text, $O(4)\to O(3)$. The $(1,2)$ plane is twisted and the untwisted component $\phi_4$ is made heavy. At $\Nm=7$ the flowed gaps $(D,\hat\psi_{1/2},\hat\psi_{3/2},\hat\phi_3)$ are $(2.018,0.941,1.946,0.524)$, against $(2.024,0.943,1.941,0.523)$ in the independently tuned $O(3)$ model, while the heavy mode rises to $15$--$18$ in bulk units and leaves the low-energy window.

Returning to the anisotropy flow from $O(3)$, the flow coordinate scales as $g\Nm^{(3-\Delta_t)/2}$ near the ultraviolet point. Three-size endpoint fits give the light energy operator near $1.415$ on the Ising side and $1.522$ on the $O(2)$ side, against $1.4126$ and $1.5114$, and stress-tensor candidates near $3.025$ and $3.013$. Neither was imposed by the tuning of the light order parameter, so both are checks.

To follow the defect along the crossover we take the charge-one mode in displacement units and normalize its shift by the value at the most negative coupling in each scan, $g=-0.09$ for $\Nm=4$--$8$ and $g=-0.25$ for $\Nm=9$. Linear interpolation at $\mathcal Y=0.02$, $0.05$, $0.1$ and $0.2$ gives fractions $0.35$--$0.40$, $0.61$--$0.64$, $0.77$--$0.80$ and $0.87$--$0.93$, respectively. The denominator is a finite-coupling scan endpoint, not an independently determined continuum infrared limit. In bulk units the charge-two and displacement-candidate levels collapse with root-mean-square spread $0.009$ and $0.010$. They move by the same $+0.026$, however, and share one crossover shape, which identifies that common change as a drift of the bulk unit, and in displacement units the charge-two level moves by only $-0.003$. Fitting the exponent of $\mathcal Y$ to minimize the spread gives $1.53$ and $2.00$ for these two channels, which bracket $y_t=1.79$ without resolving it at these sizes, and the neutral level crossing is complete by $\mathcal Y\simeq0.05$ at every size.

The two tuning checks using known bulk dimensions give consistent crossovers. Replacing the light-level condition by $E_\phi/E_T=\Delta_\phi/3$ gives crossover curves that agree at about the percent level, and replacing the ultraviolet order-parameter target by the infrared value shifts the extrapolated light level by at most $0.002$. A size-crossing prescription that uses no external target lands $0.034\pm0.007$ low instead and has not converged through $\Nm=9$, so locating the critical trajectory from size scaling alone requires larger systems.

The full-precision fits of the coupling use the common window $\Nm=7$--$9$ in all three series. For reference, the values rounded to six decimals are
\begin{center}\small
\begin{tabular}{rccc}\toprule
$\Nm$ & ultraviolet & $g=-0.09$ & independent $O(2)$\\\midrule
6 & 0.850943 & 0.851419 & 0.850329\\
7 & 0.852809 & 0.853644 & 0.852699\\
8 & 0.853886 & 0.855077 & 0.854266\\
9 & 0.854626 & 0.856159 & 0.855465\\
\bottomrule\end{tabular}
\end{center}
The $\Nm^{-1}$ intercepts are $0.861021$, $0.864973$ and $0.865153$, and the $\Nm^{-1/2}$ intercepts are $0.868264$, $0.874990$ and $0.876167$. The endpoint comparison is therefore sensitive to the correction law at the $10^{-3}$ level, even though the finite-size matching is more precise.

\subsection{Displacement and the geometric Ward identity}\label{app:ward}
A line defect breaks the translations transverse to it, and the broken Ward identity defines the displacement operator~\cite{Billo:2016cpy},
\begin{equation}
 \partial_\mu T^{\mu i}(x)=\delta^{(2)}(x_\perp)D_i(\tau),\qquad
 \langle D_i(\tau)D_{i'}(0)\rangle=\frac{C_D\delta_{ii'}}{\tau^4}.
 \label{eq:ward}
\end{equation}
Here $T$ is the stress tensor, $i$ labels the two transverse directions, $x_\perp$ are the transverse coordinates, $D_i$ is the displacement operator and $C_D$ its normalization.\footnote{The same normalization fixes the energy radiated by a moving line and the small-angle limit of the cusp anomalous dimension~\cite{Lewkowycz:2013laa,Cuomo:2024psk}.} A geometric tilt has a definite expansion in the displacement family. We write $\mathcal W(\hat{\mathcal O})$ for the weight of a defect state $\hat{\mathcal O}$ in it, the squared matrix element between $|0_\Dc\rangle$ and $\hat{\mathcal O}$ of the rotation generator that tilts the line. In our normalization the primary weight vanishes, the first allowed descendant has weight $\mathcal W(\partial D)=32C_D/9$, and the next allowed descendant has $\mathcal W(\partial^3D)/\mathcal W(\partial D)=1.8$.

The free complex scalar at half twist gives an exact target. Its displacement is a neutral product of the two lowest modes $\hat\psi_{\pm1/2}$, $D_+=D_x+\mathrm iD_y=C\hat\psi_{1/2}\hat\psi_{-1/2}^\dagger$, and the integrated translation identity fixes $|C|=1/\pi$, hence
\begin{equation}
 C_D^{\rm free}=\frac1{2\pi^2}=0.050661,
 \label{eq:cdfree}
\end{equation}
where the complex derivative and displacement components are paired with conjugate conventions.

The finite calculation uses a projected rotation generator with adjacent-orbital coefficient $\sqrt{(j-m)(j+m+1)}$ and omits transitions to excluded endpoints. With the potential subtracted and the bulk retuned at each size, the resulting estimate of $C_D$ is $0.0059$, $0.0089$, $0.0119$, $0.0149$ and $0.0178$ at $\Nm=6$--$10$ by exact diagonalization, and $0.0231$ and $0.0276$ at $\Nm=12$ and $14$ by DMRG, rising from $12\%$ to $54\%$ of Eq.~\eqref{eq:cdfree}. The growth bends, and an $\Nm^{-1}$ extrapolation of the last three sizes gives $1.02$ times the exact value, while other correction powers give $0.74$--$1.60$. The method is therefore consistent with the exact normalization, but it does not pin the correction law.

\section{The defect entropy at large \texorpdfstring{$N$}{N}}\label{app:largeM}
Here we derive the large-$N$ correction to the defect entropy stated in Sec.~\ref{sec:entropy}. It has a classical part, from the response of the spectators to the saddle, and a fluctuation part, from the determinant of the auxiliary field. The subtleties all lie in the second.

\subsection{Propagators on \texorpdfstring{$S^1\times \mathbb H^2$}{S1 x H2}}
In the $\mathbb H^2$ frame a conformally coupled real scalar obeys $(-\partial_\theta^2-\nabla_{\mathbb H^2}^2-\tfrac14)G=\delta$. Its propagator depends on the angle $\theta$ and the geodesic distance $\chi$ on $\mathbb H^2$,
\begin{equation}
 G_0(\theta,\chi)=\frac1{4\pi\sqrt{2(\cosh\chi-\cos\theta)}},
 \label{eq:G0H2}
\end{equation}
which is the Weyl transform of the flat propagator. The circle decomposes it into fields on Euclidean AdS$_2$. Each $S^1$ mode of transverse spin $s$ has dimension $\Dh_s=\tfrac12+|s|$, so that a field twisted by $e^{2\pi\mathrm i\alpha}$ has
\begin{equation}
 \begin{gathered}
 G_\alpha(\theta,\chi)=\frac1{2\pi}\sum_{s\in\mathbb Z+\alpha}e^{\mathrm is\theta}\,\mathcal G_{\frac12+|s|}(\chi),\\
 \mathcal G_\Delta(\chi)=\frac{\Gamma(\Delta)\,(2\cosh\chi)^{-\Delta}}{2\sqrt\pi\,\Gamma(\Delta+\frac12)}\,
 {}_2F_1\!\left(\tfrac\Delta2,\tfrac{\Delta+1}2;\Delta+\tfrac12;\cosh^{-2}\chi\right),
 \end{gathered}
 \label{eq:AdS2modes}
\end{equation}
where $\mathcal G_\Delta$ is the AdS$_2$ propagator, whose $\mathbb H^2$ transform is $1/[\nu^2+(\Delta-\tfrac12)^2]$. At $\alpha=0$ the mode sum reproduces Eq.~\eqref{eq:G0H2}, and at $\alpha=\tfrac12$ it reproduces $G_\Dc=G_0F_3(Z)$ of Eq.~\eqref{eq:green35}, both to $10^{-14}$. Here $Z$ and $\xi$ are the cross-ratios of Eq.~\eqref{eq:crossratios} and $\cosh\chi=1+2\xi$.

The entropy needs the coincident limit. Near coincident points $\mathcal G_\Delta(\chi)=-[\log(\chi/2)+\gamma_{\rm E}+\operatorname{dg}(\Delta)]/2\pi+O(\chi^2\log\chi)$, where $\operatorname{dg}=(\log\Gamma)'$ is the digamma function, so the defect changes the coincident propagator by a regularized sum of these functions. Point splitting along the circle gives
\begin{equation}
 \operatorname{Re}\delta G_\alpha(\theta,0)=-\frac{\mathcal A_\alpha}{4\pi^2}-\frac{\mathfrak h_{\rm free}(\alpha)}{4\pi}\,\theta^2+O(\theta^4),
 \label{eq:Gexpansion}
\end{equation}
with $\mathcal A_\alpha$ of Eq.~\eqref{eq:Aalpha} and $\mathfrak h_{\rm free}$ of Eq.~\eqref{eq:hfree}, which we have checked to eight digits for $\alpha=0.1$--$0.5$. The second term is the stress-tensor contribution to the coincident expansion, and it returns in the fluctuation determinant.

\subsection{The gap equation and the classical dressing}
With the saddle $\varsigma_*$ the spectator modes have $\Dh_s=\tfrac12+\sqrt{s^2+\varsigma_*}$. For $s\ne0$ the change of the coincident propagator is of order $\varsigma_*$, whereas the $s=0$ mode responds at order $\sqrt{\varsigma_*}$, with $\mp\operatorname{dg}'(\tfrac12)\sqrt{\varsigma_*}/4\pi^2=\mp\sqrt{\varsigma_*}/8$ on the regular and singular branches, since $\operatorname{dg}'(\tfrac12)=\pi^2/2$. The singular branch is an allowed boundary condition in AdS$_2$ for $0<\sqrt{\varsigma_*}<1$, and unitarity on the line requires $\Dh>0$. Both hold at large $N$. Equation~\eqref{eq:gap} then reads $-2\mathcal A_\alpha/4\pi^2+(N-2)\sqrt{\varsigma_*}/8=0$, which is Eq.~\eqref{eq:sqrta}.

The classical dressing follows by integrating the gap equation. The free energy on $S^1\times \mathbb H^2$ obeys
\begin{equation}
 \frac{\mathrm d\mathcal F}{\mathrm d\varsigma_*}=\frac12\operatorname{Vol}_3\left[2\,\delta g_{\rm tw}+(N-2)\,\delta g_{\rm sp}\right],
 \qquad \operatorname{Vol}_3=2\pi\operatorname{Vol}(\mathbb H^2)=-4\pi^2,
\end{equation}
since the constant part of $\varsigma$ cancels against the bulk counterterm. Integrating from zero to the saddle, with $\delta g_{\rm sp}=+\sqrt{\varsigma_*}/8$, gives $\delta\mathcal F=\tfrac13\operatorname{Vol}_3\varsigma_*\delta g_{\rm tw}=16\mathcal A_\alpha^3/(3\pi^4N^2)$ and hence Eq.~\eqref{eq:classical}. The twisted pair has no $s=0$ mode, so its own response is analytic in $\varsigma_*$ and enters only at order $N^{-4}$.

\subsection{The fluctuation determinant}
Integrating out $\phi_a$ leaves the full $\varsigma$ kernel $\Sigma_\varsigma=\tfrac12\sum_aG_a^2$, with $G_a$ the propagator of component $a$, and the one-loop free energy $\tfrac12\Tr\log\Sigma_\varsigma$. Without the defect $\Sigma_\varsigma=(N/2)\Pi_0$ with $\Pi_0=G_0^2$, and at the order we need the defect changes it by $(|G_\Dc|^2-G_0^2)+(N-2)G_0\delta G_{\rm sp}$. The change of $-\tfrac12\Tr\log\Sigma_\varsigma$ is therefore $\mathfrak c/N$, with $\mathfrak c$ of Eq.~\eqref{eq:ctrace}. For a kernel $K(\theta,\chi)$ the trace is $\operatorname{Vol}_3$ times its coincident value, and in modes
\begin{equation}
 \mathfrak c=\sum_{\kappa\in\mathbb Z}\int_0^\infty \mathrm d\nu\,\nu\tanh(\pi\nu)\,
 \frac{\delta\widetilde\Pi_\kappa(\nu)}{\widetilde\Pi_{0,\kappa}(\nu)},
 \label{eq:cmodes}
\end{equation}
where the tildes denote the transform in the $S^1$ mode $\kappa$ and the $\mathbb H^2$ spectral parameter $\nu$.

Both spectral functions are sums of AdS$_2$ bubbles, because the product of two propagators at the same pair of points decomposes as~\cite{Fitzpatrick:2011dm}
\begin{equation}
 \mathcal G_{\Delta_1}\mathcal G_{\Delta_2}=\sum_{\ell \ge0}e_\ell \,\mathcal G_{\Delta_1+\Delta_2+2\ell },\qquad
 e_\ell =\frac{(\frac12)_\ell \,(\Delta_{12}+2\ell )_{1/2}\,(\Delta_{12}+\ell )_\ell }
 {2\sqrt\pi\,\ell !\,(\Delta_1+\ell )_{1/2}(\Delta_2+\ell )_{1/2}(\Delta_{12}+\ell -\frac12)_\ell },
 \label{eq:bubbles}
\end{equation}
where $\Delta_{12}=\Delta_1+\Delta_2$ and $(\Delta)_{1/2}=\Gamma(\Delta+\frac12)/\Gamma(\Delta)$, which we have checked against the product to $10^{-15}$. The bubble is then $\mathcal B_{\Delta_1\Delta_2}(\nu)=\sum_\ell e_\ell/[\nu^2+(\Delta_{12}+2\ell-\tfrac12)^2]$, and
\begin{equation}
 \begin{gathered}
 \widetilde\Pi_{0,\kappa}=\sum_{s\in\mathbb Z}\mathcal B_{\frac12+|s|,\frac12+|\kappa-s|}
 =\frac\pi8\left|\frac{\Gamma(\frac{|\kappa|+1/2+\mathrm i\nu}2)}{\Gamma(\frac{|\kappa|+3/2+\mathrm i\nu}2)}\right|^2,\\
 \delta\widetilde\Pi_\kappa=\sum_{s\in\mathbb Z+\alpha}\mathcal B_{\frac12+|s|,\frac12+|s-\kappa|}-\widetilde\Pi_{0,\kappa}
 -\frac{4\mathcal A_\alpha}{\pi^2}\,\partial_\Delta\mathcal B_{\Delta,\frac12+|\kappa|}\Big|_{\Delta=\frac12},
 \end{gathered}
 \label{eq:spectral}
\end{equation}
in a common normalization in which the $2\pi$ of each $\theta$ transform cancels from the ratio. Grouping the pairs by their total dimension, every level $\Delta_{12}+2\ell$ receives finitely many terms, so the sums over $s$ are exact level by level. The closed form of $\widetilde\Pi_{0,\kappa}$ follows from the transform of $G_0^2$, and the level sums reproduce it. At large $\nu$ the bubble obeys $\nu^2\partial_{\Delta_1}\mathcal B_{\Delta_1\Delta_2}\to\sum_\ell\partial_{\Delta_1}e_\ell=-\operatorname{dg}'(\Delta_1)/2\pi$, which we have checked to seven digits at general dimensions. It makes the twisted and spectator terms of $\nu^2\delta\widetilde\Pi_\kappa$ tend to $-\mathcal A_\alpha/\pi$ and $+\mathcal A_\alpha/\pi$, which is the gap equation in spectral form.

\subsection{Numerical evaluation}
We evaluate Eq.~\eqref{eq:cmodes} with smooth cutoffs $F_{\rm cut}((\kappa^2+\nu^2)/p_{\max}^2)$, for $|\kappa|\le160$, levels up to $8000$ and closed-form tails, and extrapolate in $1/p_{\max}$. Gaussian and Lorentzian $F_{\rm cut}$ give $0.090063$ at $\alpha=\tfrac12$ with an uncertainty of $10^{-5}$. With $|\kappa|\le120$ and levels up to $2000$ they give $0.01324$, $0.03947$, $0.05305$, $0.06542$ and $0.08360$ at $\alpha=0.1$, $0.2$, $\tfrac14$, $0.3$ and $0.4$, against $0.013238$, $0.039469$, $0.053052$, $0.065425$ and $0.083597$ from Eq.~\eqref{eq:cresult}. Both cutoff families converge from below to the same values. The integrand of the sector $-\kappa$ at angle $\alpha$ equals that of the sector $\kappa$ at $1-\alpha$, so the sums are symmetric under $\alpha\to1-\alpha$.

One feature of the integrand stands out. In every tested sector the integral over $\nu$ vanishes, which Appendix~\ref{app:sector} proves for all $\kappa$. Numerically it vanishes to at most $5\times10^{-7}$ for $|\kappa|\le40$ at $\alpha=\tfrac12$, against positive and negative parts of order $0.07$ at $\kappa=0$, to $10^{-9}$ for $|\kappa|\le2$ at $\alpha=\tfrac14$, and to $3\times10^{-8}$ for $|\kappa|\le2$ at the other four angles, against integrals of the absolute value between $0.004$ and $0.13$. The linearized full twist, in which the twisted fields respond to $\varsigma$ in the bulk, behaves differently, and its sectors give $O(0.1)$.

\subsection{The fixed-sector identity}\label{app:sector}
Write $\mathcal I_\kappa$ for the $\nu$ integral of the sector $\kappa$ in Eq.~\eqref{eq:cmodes}, and take $\kappa\ge0$, since the sector $-\kappa$ at angle $\alpha$ equals the sector $\kappa$ at $1-\alpha$. We show that $\mathcal I_\kappa=0$ for every $\kappa$ and $\alpha$. The proof first turns each sector into a single integral over the $\mathbb H^2$ distance, and then shows that every term of that integral is a boundary term.

In position space the circle mode of $\Pi_0$ is
\begin{equation}
 \Pi_{0,\kappa}(\chi)=\sum_{s\in\mathbb Z}\mathcal G_{\frac12+|s|}\mathcal G_{\frac12+|\kappa-s|}=\frac{e^{-\kappa\chi}}{8\sinh\chi},
\end{equation}
the Fourier coefficient of $G_0^2\propto(\cosh\chi-\cos\theta)^{-1}$, in the normalization of Eq.~\eqref{eq:spectral}. Since $\Pi_0$ is a conformal kernel of dimension one in three dimensions, its inverse is the kernel of dimension two, proportional to $\partial_{\cosh\chi}(\cosh\chi-\cos\theta)^{-1}$. Its circle mode is $W_\kappa=(2/\pi^2)(\sinh\chi)^{-1}\partial_\chi(e^{-\kappa\chi}/\sinh\chi)$, and the sector integral becomes
\begin{equation}
 \mathcal I_\kappa=8\int_0^\infty \mathrm d\chi\,\partial_\chi\!\left(\frac{e^{-\kappa\chi}}{\sinh\chi}\right)\delta\Pi_\kappa(\chi),
 \label{eq:Iposition}
\end{equation}
where $\delta\Pi_\kappa(\chi)$ is Eq.~\eqref{eq:spectral} with every bubble $\mathcal B_{\Delta_1\Delta_2}$ replaced by the product $\mathcal G_{\Delta_1}\mathcal G_{\Delta_2}$. We define the lower limit by multiplying the integrand by a power of $\chi$ and continuing analytically to the zeroth power. This is unambiguous, because the integrand contains powers of $\log\chi$ times $\chi^{-2}$ but no $\chi^{-1}$. Equation~\eqref{eq:Iposition} agrees with the spectral integral to $10^{-9}$ for $\kappa\le2$.

To see that these integrals are boundary terms, let $\varphi_1$ and $\varphi_2$ solve the radial equations $(\sinh\chi)^{-1}(\sinh\chi\,\varphi_{1,2}')'=\Delta_{1,2}(\Delta_{1,2}-1)\varphi_{1,2}$, with $\Delta_1+\Delta_2=1+\kappa$. Then
\begin{multline}
 \Xi=e^{-\kappa\chi}\Big\{\big[2(\Delta_1-1)(\Delta_2-1)\sinh\chi+e^{\chi}(\kappa-1+\coth\chi)\big]\varphi_1\varphi_2
 +\big[\Delta_2e^{\chi}-(\Delta_2-1)e^{-\chi}\big]\varphi_1'\varphi_2\\
 +\big[\Delta_1e^{\chi}-(\Delta_1-1)e^{-\chi}\big]\varphi_1\varphi_2'+2\sinh\chi\,\varphi_1'\varphi_2'\Big\}
 \label{eq:antider}
\end{multline}
obeys $\Xi'=\varphi_1\varphi_2\,\partial_\chi(e^{-\kappa\chi}/\sinh\chi)$, as one checks by differentiating and eliminating $\varphi_{1,2}''$ with the radial equations. The radial equation depends on $\Delta_1$ only through $\Delta_1(\Delta_1-1)$. Replacing $\Delta_1$ by its shadow $1-\Delta_1$ in the coefficients therefore gives a second antiderivative, valid when $\Delta_2-\Delta_1=\kappa$. These are the two lines on which every pair in $\delta\Pi_\kappa$ lies. The pairs $(\tfrac12+|s|,\tfrac12+|\kappa-s|)$ with $0\le s\le\kappa$ have $\Delta_1+\Delta_2=1+\kappa$, and all others have $|\Delta_1-\Delta_2|=\kappa$. At large $\chi$, $\Xi$ vanishes. At small $\chi$ its four coefficients are $1/\chi$, $1+(\Delta_2-\Delta_1)\chi$, $1+(\Delta_1-\Delta_2)\chi$ and $2\chi-2\kappa\chi^2$, up to higher orders. Each integral is therefore minus the $\chi^0$ term of $\Xi$, which the coincident form of $\mathcal G_\Delta$ below Eq.~\eqref{eq:AdS2modes} determines.

Differentiating the propagator equation at $\Delta=\tfrac12$ removes its source. So $\partial_\Delta\mathcal G_\Delta|_{1/2}=-(\pi/4)\mathrm P_{-1/2}(\cosh\chi)$, with $\mathrm P_{-1/2}$ the Legendre function, is the regular solution, with coincident value $-\operatorname{dg}'(\tfrac12)/2\pi=-\pi/4$. Only the term in $\varphi_1\varphi_2'$ then leaves a constant, and
\begin{equation}
 8\int_0^\infty \mathrm d\chi\,\partial_\chi\!\left(\frac{e^{-\kappa\chi}}{\sinh\chi}\right)\partial_\Delta\mathcal G_\Delta\big|_{\frac12}\,\mathcal G_{\frac12+\kappa}=\kappa,
\end{equation}
so the spectator contributes $-4\kappa\mathcal A_\alpha/\pi^2$ to $\mathcal I_\kappa$. For two propagators the logarithms cancel between the middle terms of Eq.~\eqref{eq:antider}, and on both lines
\begin{equation}
 4\pi^2\int_0^\infty \mathrm d\chi\,\partial_\chi\!\left(\frac{e^{-\kappa\chi}}{\sinh\chi}\right)\mathcal G_{\frac12+|s|}\mathcal G_{\frac12+|\kappa-s|}
 =\mathfrak b_\kappa(s)\equiv2\kappa+(\kappa-2s)\big[\operatorname{dg}(\tfrac12+|s|)-\operatorname{dg}(\tfrac12+|\kappa-s|)\big].
 \label{eq:bkappa}
\end{equation}
For $\kappa=1$ and $0<s<1$ this is $(1-2s)\pi\tan\pi s=2\mathcal A_s$. We have checked Eq.~\eqref{eq:bkappa} against direct quadrature, and against the spectral evaluation through Eq.~\eqref{eq:bubbles} to $10^{-10}$.

The twisted part of $\mathcal I_\kappa$ is $(2/\pi^2)[\mathfrak s_\kappa(\alpha)-\mathfrak s_\kappa(0)]$, with $\mathfrak s_\kappa(\alpha)=\sum_{s\in\mathbb Z+\alpha}\mathfrak b_\kappa(s)$. For $\kappa=0$ every term vanishes. For integer $\kappa\ge1$ and $s<0$,
\begin{equation}
 \mathfrak b_\kappa(s)=\mathfrak b_\kappa(\kappa-s)=\sum_{\ell=0}^{\kappa-1}\frac{2\ell+1-\kappa}{\frac12-s+\ell},
\end{equation}
whose weights sum to zero, so the two tails sum to $-\sum_\ell(2\ell+1-\kappa)[\operatorname{dg}(\ell+\tfrac32-\alpha)+\operatorname{dg}(\ell+\tfrac12+\alpha)]$. Together with the $\kappa$ terms with $0\le s<\kappa$, every digamma function in $\mathfrak s_\kappa(\alpha)$ has the argument $\tfrac12\pm\alpha+\ell$, with $0\le\ell\le\kappa$. We reduce each to $\operatorname{dg}(\tfrac12\pm\alpha)$ with the recurrence of the digamma function. The coefficients of $\operatorname{dg}(\tfrac12+\alpha)$ and $\operatorname{dg}(\tfrac12-\alpha)$ are then $\pm\kappa(1-2\alpha)$. The poles of the rational remainder cancel term by term, leaving $-2\kappa^2$, which cancels the constants $2\kappa$ of the $\kappa$ inner terms. The reflection formula $\operatorname{dg}(\tfrac12+\alpha)-\operatorname{dg}(\tfrac12-\alpha)=\pi\tan\pi\alpha$ then gives
\begin{equation}
 \mathfrak s_\kappa(\alpha)=\kappa\pi(1-2\alpha)\tan\pi\alpha=2\kappa\mathcal A_\alpha,\qquad \mathfrak s_\kappa(0)=0,
\end{equation}
which we have also confirmed numerically to fifteen digits. Hence $\mathcal I_\kappa=4\kappa\mathcal A_\alpha/\pi^2-4\kappa\mathcal A_\alpha/\pi^2=0$ for every $\kappa$ and $\alpha$. The spectator coefficient $4\mathcal A_\alpha/\pi^2$ in Eq.~\eqref{eq:spectral} is fixed by the gap equation, so the cancellation is the gap equation acting within each circle sector.

\subsection{The analytic formula}\label{app:analytic}
Since every sector integral vanishes, the sum taken one circle mode at a time is zero, with each sector integrated over all $\nu$ before the sum over $\kappa$ is cut off. The rotationally symmetric sum is therefore the difference between a spherical cutoff and this slab limit. We assume that the large-$(\kappa,\nu)$ expansion of the ratio holds uniformly, so that the difference depends only on its leading term. Near coincident points $\delta\Pi\simeq G_0\,\mathcal T$, where $\mathcal T$ is the quadratic part of $2\operatorname{Re}\delta G_\Dc+N\delta G_{\rm sp}$. By Eq.~\eqref{eq:Gexpansion} and the equation of motion of the shifts, $\mathcal T=-(\mathfrak h_{\rm free}/2\pi)(\theta^2-\tfrac12\chi_\perp^2)$, with $\chi_\perp$ the displacement on $\mathbb H^2$, which is traceless. In momentum space, with $p=(\kappa^2+\nu^2)^{1/2}$ and $\hat p$ the unit vector along $(\kappa,\nu)$, the ratio then behaves as $\mathcal C_{ii'}\hat p_i\hat p_{i'}/p^3$, where the indices take the values $\kappa$ (circle) and $\perp$ ($\mathbb H^2$), with $\mathcal C_{\kappa\kappa}=-2\mathcal C_{\perp\perp}$ and $\mathcal C_{\perp\perp}=-16\mathfrak h_{\rm free}/\pi$. A sharp spherical cutoff minus the slab sum gives $-[\tfrac29\mathcal C_{\kappa\kappa}+\tfrac{16}9\mathcal C_{\perp\perp}]=-\tfrac43\mathcal C_{\perp\perp}$, which is Eq.~\eqref{eq:cresult}.

The sector identity is exact, while this evaluation of the simultaneous cutoff limit assumes uniformity of the tail. We have tested that assumption on the tabulated integrand. At all six angles the leading coefficient reproduces $\mathcal C_{\perp\perp}=-16\mathfrak h_{\rm free}/\pi$ to $4\times10^{-5}$, and the largest sampled remainder over the momentum directions in a shell falls as $p^{-4}$. A uniform bound of this form on the spectral ratio would make the omitted tail integrable, since the asymptotic measure is proportional to $p^2\,\mathrm dp$ and $\int_{p_{\max}}^\infty p^{-2}\,\mathrm dp=1/p_{\max}$. This bound suffices for the cutoff difference to depend only on the leading stress-tensor term. The sampled directions and cutoffs support it but do not prove it uniformly. Both smooth-cutoff families converge to Eq.~\eqref{eq:cresult} at all six angles.

The identification with the physical entropy also requires a prescription for the determinant. We use a cutoff depending on the spectrum only through $\kappa^2+\nu^2$, which treats the local circle and hyperbolic momenta isotropically, and normalize by the defect-free partition function. The gap equation removes the scalar ultraviolet term, leaving the traceless stress-tensor contribution. Under the uniform-tail assumption, changing the radial cutoff profile leaves the finite term unchanged. Integrating one circle mode at a time instead uses an anisotropic ultraviolet limit and shifts the answer by a term proportional to $\mathfrak h_{\rm free}$. We identify the isotropic value with the circular-defect entropy.\footnote{Local isotropy motivates this prescription, but the sector identity alone does not establish its equivalence to a covariant regulator of the circular defect in flat space, including possible finite terms in the Weyl transformation. The dimensional-regularization comparison below supplies an independent check near four dimensions, not a proof of that equivalence in three dimensions.}

Near four dimensions this choice can be tested against dimensional regularization, which is covariant and in which the trace can be evaluated in closed form. In $d=4-\epsilon$ the bubble has a pole, $\widetilde\Pi_0\simeq1/8\pi^2\epsilon$, so $\Pi_0^{-1}\simeq8\pi^2\epsilon\,\delta$ is local and the trace reduces to a coincident limit of $\delta\Pi$. On $S^1\times\mathbb H^3$ every propagator is elementary, and with the spectator mode at $s=0$ screening the twisted pair, as at large $N$ in Appendix~\ref{app:entropyeps}, the kernel near coincident points is
\begin{equation}
 \delta\Pi=\big(\delta g^{(4)}_{\rm tw}\big)^2-\frac{\alpha^2(1-\alpha)^2}{64\pi^4}\,\frac{\theta^2-\chi^2/3}{\theta^2+\chi^2}+o(1),\qquad
 \delta g^{(4)}_{\rm tw}=-\frac{\alpha(1-\alpha)}{8\pi^2}.
 \label{eq:dPi4d}
\end{equation}
Here $\delta g^{(4)}_{\rm tw}$ is the coincident shift of one twisted component, $\chi$ is now the geodesic distance on $\mathbb H^3$, and the second term comes from the $\theta^2$ term of $\operatorname{Re}\delta G_\alpha(\theta,0)$, made traceless by the equation of motion of the shifts. Dimensional regularization, like any rotationally symmetric cutoff, keeps the isotropic part and gives $\mathfrak c=\tfrac\pi4\alpha^2(1-\alpha)^2\epsilon$, the large-$N$ coefficient of Appendix~\ref{app:entropyeps}. The sum one circle mode at a time approaches the coincident point along the circle, where the two terms of Eq.~\eqref{eq:dPi4d} cancel identically in $\alpha$. Near four dimensions this sum therefore vanishes again, and the covariant value is the stress-tensor term alone. The same term in general dimension, with the difference between the spherical and slab limits of the traceless kernel evaluated in $d$ dimensions, reduces to Eq.~\eqref{eq:cresult} at $d=3$ and to this value as $d\to4$. At the intermediate dimension $d=7/2$ we have evaluated the sector integrals numerically, with the general-dimension forms of Eqs.~\eqref{eq:bubbles} and~\eqref{eq:spectral}. They vanish to about $10^{-6}$ for $\kappa=0,1$ and to a few times $10^{-5}$ for $\kappa=2,3$, against integrals of the absolute value between $0.07$ and $0.005$. This is the accuracy with which the same evaluation reproduces the three-dimensional identity, and with a different smooth cutoff the residuals decrease with the cutoff in the same way in both dimensions.

\subsection{The full twist and comparison with Ref.~\texorpdfstring{\cite{Giombi:2021uae}}{GHJK}}
When all components are half-twisted there are no spectators, and the twisted fields acquire the mass themselves. This case was solved in Ref.~\cite{Giombi:2021uae}, and it tests our normalizations. The gap equation $\mathcal A_{1/2}(\varsigma_*)=0$, with
\begin{equation}
 \mathcal A_{1/2}(\varsigma)=\Big[\sum_{s\in\mathbb Z+\frac12}\operatorname{dg}\big(\tfrac12+\sqrt{s^2+\varsigma}\big)-\sum_{s\in\mathbb Z}\operatorname{dg}\big(\tfrac12+|s|\big)\Big]_{\rm reg}
\end{equation}
and $\mathcal A_{1/2}(0)=1$, gives $\varsigma_*=-0.168374$, hence $\Dh_{1/2}=0.785703$ and $\Dh_{3/2}=1.942784$. The defect expectation value per complex component is $\log\mathfrak g_{\rm free}(\tfrac12)+\int_{\varsigma_*}^0\mathcal A_{1/2}(\varsigma)\,\mathrm d\varsigma=0.1617$. These agree with the saddle $-0.168$, the dimensions $0.786$ and $1.943$ and the defect expectation value of Ref.~\cite{Giombi:2021uae}, which checks the normalization of the gap function, the coincident limit and the trace measure independently. The general-dimension bubbles of Eq.~\eqref{eq:bubbles} also reproduce $\nu^2\partial_\varsigma\widetilde\Pi\to-\partial_\varsigma\mathcal A_{1/2}(0)/\pi$ with $\partial_\varsigma\mathcal A_{1/2}(0)=4.2072$.

\subsection{The neutral line spectrum of a partial twist}\label{app:singlet}
The same kernel decides whether a partial twist is stable against symmetric perturbations on the line. At large $N$ the singlet sector of the line is governed by the propagator of the auxiliary field, $\Sigma_\varsigma^{-1}$, with $\Sigma_\varsigma=\tfrac12\sum_aG_a^2$ the kernel of the fluctuation determinant above. In the circle mode $\kappa=0$ it is a sum of the AdS$_2$ bubbles of Eq.~\eqref{eq:bubbles},
\begin{equation}
 K_0(\nu)=\tfrac12(N-2)\sum_{s\in\mathbb Z}\mathcal B_{\Dh^{\rm sp}_s\Dh^{\rm sp}_{-s}}(\nu)+\sum_{s\in\mathbb Z+\alpha}\mathcal B_{\Dh^{\rm tw}_s\Dh^{\rm tw}_s}(\nu),
 \label{eq:K0singlet}
\end{equation}
where the spectator modes have $\Dh^{\rm sp}_0=\tfrac12-\sqrt{\varsigma_*}$ on the singular branch and $\Dh^{\rm sp}_s=\tfrac12+\sqrt{s^2+\varsigma_*}$ for $s\ne0$, the twisted modes have $\Dh^{\rm tw}_s=\tfrac12+\sqrt{s^2+\varsigma_*}$, and $\sqrt{\varsigma_*}$ is given by Eq.~\eqref{eq:sqrta}. A neutral line operator of dimension $\Dh$ is a pole of $\Sigma_\varsigma^{-1}$ at $\nu^2=-(\Dh-\tfrac12)^2$, and hence a zero of $K_0$.

The poles of $K_0$ sit at the double-trace dimensions, among them the spectator composite at $2\Dh^{\rm sp}_0=1-8\mathcal A_\alpha/(\pi^2N)$. They are zeros of $\Sigma_\varsigma^{-1}$, and they are absent from the spectrum. In the four-point function of the spectators, the exchange of $\varsigma$ cancels the disconnected double-trace pole of the singlet channel at leading order, just as it removes $\phi^2$ in the bulk.\footnote{The same mechanism, with the spectrum given by the zeros of the bubble and the double traces removed, governs the large-$N$ $O(N)$ model in anti-de Sitter space~\cite{Carmi:2018qzm} and its boundary flows~\cite{Giombi:2020rmc}.} Every coefficient $e_\ell$ in Eq.~\eqref{eq:bubbles} is positive, so $K_0$ increases with $(\Dh-\tfrac12)^2$ between consecutive poles and has exactly one zero between each pair. Without the twist the zeros are at $\Dh=2,4,\ldots$, the restriction of $\varsigma$ to a trivial line, and with $|s|\le600$ and $\ell\le400$ we find $2.0005$ and $4.0011$. With the twist the lowest zero lies between the spectator composite and the lowest twisted double trace at $1+2\sqrt{\min(\alpha,1-\alpha)^2+\varsigma_*}$, and Table~\ref{tab:singlet} lists it. As $\alpha\to0$ the two poles close in on one from opposite sides, at $1-4\alpha/N$ and $1+2\alpha\sqrt{1+4/N^2}$, with residues in the ratio $\tfrac12(N-2):1$. The zero is pinned between them at the residue-weighted point,
\begin{equation}
 \Dh-1\simeq\frac{2\alpha}{N^2}\Big[(N-2)\sqrt{N^2+4}-4\Big],
 \label{eq:singletsmall}
\end{equation}
which reproduces the solver to $6\%$ at $N=3$ and $1\%$ at $N=4$ for $\alpha=10^{-3}$. The bracket changes sign at $N\simeq3.09$. Together with the numerical angle scan, the small-angle expansion gives a lowest neutral singlet above one for $N\ge4$ within this kernel approximation. The conclusion concerns this singlet channel at the retained order; other channels and higher orders require separate tests. At $N=3$ it falls below one for $\alpha<0.016$, by at most $3.5\times10^{-4}$ near $\alpha=0.008$, far inside the neglected $1/N$ corrections, so this order does not decide the stability of an $O(3)$ partial twist at small angle.

\begin{table}[t]\centering\small
\begin{tabular}{lcccccc}\toprule
$N$ & 3 & 4 & 6 & 10 & 30 & $\infty$\\\midrule
$\alpha=\tfrac12$ & 1.283 & 1.435 & 1.572 & 1.686 & 1.828 & 2\\
$\alpha=\tfrac14$ & 1.159 & 1.268 & 1.355 & 1.416 & 1.473 & 1.5\\
$\alpha=0.1$ & 1.035 & 1.092 & 1.136 & 1.164 & 1.189 & 1.2\\
\bottomrule\end{tabular}
\caption{Lowest neutral line operator of a partial twist with one twisted complex component, from the zeros of Eq.~\eqref{eq:K0singlet} at leading order in $1/N$. The truncation error is below $5\times10^{-4}$. Table~\ref{tab:spectator} compares the $O(3)$ half-twist entry with the lattice.}
\label{tab:singlet}
\end{table}

\subsection{The entropy near four dimensions}\label{app:entropyeps}

\begin{table}[t]\centering\small
\begin{tabular}{lcccccc}\toprule
\multicolumn{7}{l}{(a) Free defect entropy and large-$N$ coefficient}\\\midrule
$\alpha$ & $0.1$ & $0.2$ & $\tfrac14$ & $0.3$ & $0.4$ & $\tfrac12$\\\midrule
$\log\mathfrak g_{\rm free}$, Eq.~\eqref{eq:loggfree} & 0.009441 & 0.028660 & 0.038772 & 0.048057 & 0.061800 & 0.066717\\\midrule
direct trace & 0.01324 & 0.03947 & 0.05305 & 0.06542 & 0.08360 & 0.090063\\
Eq.~\eqref{eq:cresult} & 0.013238 & 0.039469 & 0.053052 & 0.065425 & 0.083597 & 0.090063\\
near $d=4$, $O(\epsilon)$ & 0.006362 & 0.020106 & 0.027612 & 0.034636 & 0.045239 & 0.049087\\
\midrule\multicolumn{7}{l}{(b) Increase $\log\mathfrak g_{N-1}-\log\mathfrak g_N$ at $\epsilon=1$, one twisted plane}\\\midrule
flow & $\alpha$ & \multicolumn{2}{c}{$O(\epsilon)$} & \multicolumn{2}{c}{through $O(\epsilon^2)$} & large $N$\\\midrule
$O(3)\to O(2)$ & $\tfrac14$ & \multicolumn{2}{c}{$5.0\times10^{-4}$} & \multicolumn{2}{c}{$2.7\times10^{-3}$} & $8.8\times10^{-3}$\\
$O(3)\to O(2)$ & $\tfrac12$ & \multicolumn{2}{c}{$8.9\times10^{-4}$} & \multicolumn{2}{c}{$4.5\times10^{-3}$} & $1.5\times10^{-2}$\\
$O(4)\to O(3)$ & $\tfrac14$ & \multicolumn{2}{c}{$4.2\times10^{-4}$} & \multicolumn{2}{c}{$3.1\times10^{-3}$} & $4.4\times10^{-3}$\\
$O(4)\to O(3)$ & $\tfrac12$ & \multicolumn{2}{c}{$7.4\times10^{-4}$} & \multicolumn{2}{c}{$4.8\times10^{-3}$} & $7.5\times10^{-3}$\\
\bottomrule\end{tabular}
\caption{The defect entropy order by order. (a) The free defect entropy and the large-$N$ coefficient $\mathfrak c(\alpha)$ of Eq.~\eqref{eq:ctrace}. The direct traces use rotationally symmetric smooth cutoffs, with cutoff and truncation variations of about $10^{-5}$ (Appendix~\ref{app:largeM}), and the next row is Eq.~\eqref{eq:cresult}. The last row gives the leading term of $\mathfrak c$ near four dimensions, $\tfrac\pi4\alpha^2(1-\alpha)^2\epsilon$ at $\epsilon=1$. Its $O(\epsilon^2)$ term would need the second order of the large-$N$ theory itself and has not been computed (Appendix~\ref{app:entropyeps}). (b) The entropy increase when one spectator is removed, from the $\epsilon$ expansion at fixed $N$ truncated at $O(\epsilon)$ [Eq.~\eqref{eq:decoupleeps}] and at $O(\epsilon^2)$ (Appendix~\ref{app:entropyeps} and Table~\ref{tab:entropyeps2}), both at $\epsilon=1$, and from the large-$N$ result Eq.~\eqref{eq:decouple} evaluated at the same $N$. The $\epsilon$-expansion entries in (b) retain the real part of the continued branches. All three approximations are evaluated outside their controlled limits and are compared without assigning any one of them a reference status.}
\label{tab:entropy}
\end{table}

Near four dimensions the defect becomes a surface, and its expectation value has a logarithmic anomaly. Following Ref.~\cite{Kobayashi:2018lil} we continue
\begin{equation}
 \widetilde F=\sin\!\big(\tfrac\pi2(d-2)\big)\log\langle\Dc\rangle,
\end{equation}
which is finite at $d=4$ and equals $\log\mathfrak g$ at $d=3$. At first order in the quartic interaction of Eq.~\eqref{eq:bulkaction} at its fixed point, $(2\pi^2\epsilon/(N+8))(\phi_a\phi_a)^2$, the change is $\delta\log\langle\Dc\rangle=-(2\pi^2\epsilon/(N+8))\int\langle(\phi_a\phi_a)^2\rangle_\Dc$, with the defect-free value subtracted. In dimensional regularization the coincident propagator vanishes without the defect, so only the twisted components, whose coincident propagator the defect shifts, contribute. A free spectator drops out, as in Sec.~\ref{sec:entropy}. For $n_t$ twisted planes, Wick's theorem gives $\langle(\phi_a\phi_a)^2\rangle_\Dc=4n_t(n_t+1)\,(\delta g^{(4)}_{\rm tw})^2$. Here $\delta g^{(4)}_{\rm tw}$ of Eq.~\eqref{eq:dPi4d}, the four-dimensional counterpart of $\delta g_{\rm tw}$ in Eq.~\eqref{eq:gap}, is constant on $S^1\times \mathbb H^3$. The regularized volume $\operatorname{Vol}(\mathbb H^{3-\epsilon})=\pi^{1-\epsilon/2}\Gamma(-1+\epsilon/2)\simeq-2\pi/\epsilon$ supplies the pole of the anomaly, and $\sin(\tfrac\pi2(d-2))\simeq\pi\epsilon/2$ removes it, leaving
\begin{equation}
 \widetilde F=\widetilde F_{\rm free}+\frac{\pi n_t(n_t+1)\alpha^2(1-\alpha)^2}{4}\,\frac{\epsilon}{N+8}+\cdots.
\end{equation}
For $n_t=1$, the difference between $N-1$ and $N$ components is Eq.~\eqref{eq:decoupleeps}.

Keeping the regularized volume of $\mathbb H^{d-1}$ at $d=3$ instead of expanding $\widetilde F$ gives a value smaller by a factor $2/\pi$. Within the $\epsilon$ expansion the continuation is fixed once the scheme is chosen, so this factor only shows how large the higher orders are at $\epsilon=1$.

The next order brings in the spectators, through their mode at $s=0$. On $S^1\times\mathbb H^3$ each circle mode is a field on $\mathbb H^3$. One with boundary dimension $\Dh$ has the dimensionally regularized coincident value $-(\Dh-1)/4\pi$ and, relative to $\Dh=1$, the free energy $-(\Dh-1)^3\operatorname{Vol}(\mathbb H^3)/12\pi$. In general dimension the untwisted mode at $s=0$ sits at the Breitenlohner--Freedman value $\Dh_{\rm BF}=(d-2)/2$. We write $u_{\rm sp}=\Dh_{\rm sp}-(d-2)/2$ for the shift from this value and evaluate the leading response coefficients at $d=4$, where both expressions above depend nonanalytically on the potential. This mode must therefore be treated self-consistently. Its coincident contribution, $-u_{\rm sp}/8\pi^2$ per spectator, feeds back into the spectators' own potential. Adding it to the tadpole of the twisted pair, $\mathfrak v_{\rm sp}=-2n_t\alpha(1-\alpha)\epsilon/(N+8)$, gives
\begin{equation}
 u_{\rm sp}^2+\mathfrak w\,u_{\rm sp}-\mathfrak v_{\rm sp}=0,\qquad
 \mathfrak w=\frac{(N_{\rm sp}+2)\,\epsilon}{N+8},
 \label{eq:spgap}
\end{equation}
where $N_{\rm sp}=N-2n_t$ counts the spectators. This is the Hartree resummation of the cactus diagrams in this channel. With $\epsilon N$ held fixed its three terms are of the same order, $\epsilon^2$, and every neglected effect is smaller by a factor of $\epsilon$. The same equation is the vanishing of the $\beta$ function of the marginal defect coupling of $\hat\phi_{\rm sp}^2$~\cite{Giombi:2021uae}, with the tadpole as its source. On shell, with the first-order piece removed, the spectators contribute
\begin{equation}
 \widetilde F_{\rm sp}
 =-\pi N_{\rm sp}\Big[\frac{u_{\rm sp}^3}{12}+\frac{\mathfrak w\,u_{\rm sp}^2}{16}\Big]
 =\frac{\pi N_{\rm sp}}{96}\Big[\mathfrak w^3+6\mathfrak w\,\mathfrak v_{\rm sp}\mp\big(\mathfrak w^2+4\mathfrak v_{\rm sp}\big)^{3/2}\Big],
 \label{eq:Fsp}
\end{equation}
with the upper sign for the root $u_{\rm sp}=\tfrac12\big[-\mathfrak w+(\mathfrak w^2+4\mathfrak v_{\rm sp})^{1/2}\big]$.

Equation~\eqref{eq:Fsp} has three consequences. First, the tadpole is attractive, so for
\begin{equation}
 \epsilon<\epsilon_{\rm c}=\frac{8n_t\alpha(1-\alpha)(N+8)}{(N_{\rm sp}+2)^2}
\end{equation}
the two roots of Eq.~\eqref{eq:spgap} are complex conjugates. The two boundary conditions of the mode~\cite{Klebanov:1999tb} have merged, as for the unstable Wilson lines of Ref.~\cite{Aharony:2022ntz}, and the defect is a pair of complex-conjugate fixed points. Their entropies differ by $\pm\mathrm i(\pi N_{\rm sp}/12)|\mathfrak v_{\rm sp}|^{3/2}$. The term of order $\epsilon^{3/2}$ is therefore purely imaginary for either boundary condition, the real part has no such term, and its leading spectator term is $-(\pi/8)n_tN_{\rm sp}(N_{\rm sp}+2)\alpha(1-\alpha)\epsilon^2/(N+8)^2$.

Second, for $\epsilon\ge\epsilon_{\rm c}$ both roots are real and lie on the singular side, $u_{\rm sp}<0$. The root with the upper sign has the lower $\widetilde F$ and is the infrared end of the defect flow between the two. Taken to three dimensions at infinite coupling, the same channel reproduces Eq.~\eqref{eq:sqrta}, so this is the branch of the large-$N$ saddle.\footnote{At fixed $N$, $\epsilon_{\rm c}$ does not locate the merger. It is the zero of the discriminant $\mathfrak w^2+4\mathfrak v_{\rm sp}$ of Eq.~\eqref{eq:spgap}, where its order-$\epsilon$ part balances its order-$\epsilon^2$ part, and the latter also receives second-order corrections to $\mathfrak v_{\rm sp}$ and to the quadratic term that the equation does not contain. The second-order fixed-point coupling alone multiplies $\mathfrak v_{\rm sp}$ by $1+3(3N+14)\epsilon/(N+8)^2$, which moves the zero for the $O(4)$ half twist from $\epsilon=1.5$ to $8$ and removes it for $O(3)$. The value of $\epsilon_{\rm c}$ is controlled only at large $N$, where it is of order $1/N$. At infinite coupling the quadratic term, which comes from the free propagator, drops out and the channel is real. This order therefore cannot decide whether the three-dimensional defect is real, as the lattice and Ref.~\cite{Tang:2026kzf} indicate.}

Third, at large $N$ with $\epsilon$ fixed the upper root approaches $\mathfrak v_{\rm sp}/\mathfrak w$. The spectators then cancel the twisted pair's shift of $\sum_a\langle\phi_a\phi_a\rangle$, as the large-$N$ constraint requires. This screens the part of the first-order coefficient proportional to $n_t^2$ and leaves the exchange part, $(\pi/4)n_t\alpha^2(1-\alpha)^2\epsilon/N$. The large-$N$ limit of the fixed-$N$ result is larger by a factor $n_t+1$, because the screening sets in only for $\epsilon\gtrsim\epsilon_{\rm c}\propto1/N$. The two limits do not commute. The comparison in Table~\ref{tab:entropy}(a) uses the large-$N$ one, which the trace of Eq.~\eqref{eq:ctrace} also gives at leading order in $\epsilon$ (Appendix~\ref{app:analytic}). Its second-order term would need the large-$N$ theory itself at the next order in $\epsilon$, which the fixed-$N$ calculation does not supply.

The remaining contribution can be specified by its Laurent coefficients before substituting the fixed point. Use $\bar\lambda=6\lambda_4/(16\pi^2)$ and the $\overline{\rm MS}$ convention
\begin{equation}
 \lambda_{{\rm q},B}=16\pi^2\mathcal M^\epsilon
 \left[\bar\lambda+\frac{N+8}{3\epsilon}\bar\lambda^2+\cdots\right],
 \qquad \mathcal M^\epsilon=\left(\frac{e^{\gamma_{\rm E}}}{4\pi}\right)^{\epsilon/2},
\end{equation}
at unit renormalization scale and unit hyperbolic radius. The defect-subtracted free energy, excluding the resummed $s=0$ channel, is
\begin{equation}
 \delta\mathcal F_{\rm an}=\bar\lambda\left(\frac{f_{-1}}\epsilon+f_0+\cdots\right)
 +\bar\lambda^2\left(\frac{\mathscr C_{-2}}{\epsilon^2}
 +\frac{\mathscr C_{-1}}\epsilon+\cdots\right).
 \label{eq:entropyLaurent}
\end{equation}
Put $z_\alpha=\alpha(1-\alpha)$ and $N_{\rm sp}=N-2n_t$. The first graph gives
\begin{equation}
 \begin{gathered}
 f_{-1}=-\frac{n_t(n_t+1)z_\alpha^2}{6},\qquad f_0=f_{-1}K_\alpha,\\
 K_\alpha=\frac72-\gamma_{\rm E}-\log2-\operatorname{dg}(1+\alpha)-\operatorname{dg}(2-\alpha).
 \end{gathered}
\end{equation}
The counterterm, twisted cactus and nonzero-mode spectator cactus give respectively
\begin{equation}
 \begin{gathered}
 q_{\rm ct}=-\frac{(N+8)n_t(n_t+1)z_\alpha^2}{18},\qquad
 q_{\rm tw}=\frac{n_t(n_t+1)^2z_\alpha^2}{9},\qquad
 q_{\rm sp}=\frac{N_{\rm sp}n_t^2z_\alpha^2}{18},\\
 K_{\rm tw}=\frac72-\gamma_{\rm E}-2\log2
 -\frac32[\operatorname{dg}(\alpha)+\operatorname{dg}(1-\alpha)]
 -\frac1\alpha-\frac1{1-\alpha},\\
 K_{\rm sp}=\frac72-2\log2-\operatorname{dg}(1+\alpha)-\operatorname{dg}(2-\alpha).
 \end{gathered}
 \label{eq:entropyresidues}
\end{equation}
Here $q_X$ is the coefficient of $\bar\lambda^2/\epsilon^2$, and its simple-pole coefficient is $q_XK_X$, with $K_{\rm ct}=K_\alpha$.

The melon requires two integrals,
\begin{equation}
 J_3=\int\!\mathrm d\mu\,G_0^2\delta P,\qquad
 J_2=\int\!\mathrm d\mu\,(\delta P)^2,\qquad
 \delta P=|G_\alpha|^2-G_0^2,
 \label{eq:entropymelon}
\end{equation}
with one point fixed and $\mathrm d\mu=4\pi\sinh^2\chi\,\mathrm d\chi\,\mathrm d\theta$ on $S^1\times\mathbb H^3$, $-\pi<\theta<\pi$. In this formula
\begin{equation}
 G_0=\frac1{8\pi^2(\cosh\chi-\cos\theta)},\qquad
 G_\alpha=\frac1{8\pi^2\sinh\chi}
 \left[\frac{e^{\alpha(\mathrm i\theta-\chi)}}{1-e^{\mathrm i\theta-\chi}}
 +\frac{e^{(1-\alpha)(-\mathrm i\theta-\chi)}}{1-e^{-\mathrm i\theta-\chi}}\right].
\end{equation}
To fix the finite parts, let $\rho^2=\theta^2+\chi^2$ and define
\begin{equation}
 \operatorname{FP}_{a_{-3},a_{-1}}\int f
 =\lim_{\rho_0\to0}\left[\int_{\rho>\rho_0}\!\mathrm d\mu\,f
 -\frac{a_{-3}}{2\rho_0^2}+a_{-1}\log\rho_0\right].
\end{equation}
Write $\omega_d=2\pi^{d/2}/\Gamma(d/2)$, $C_d=\Gamma(d/2-1)/(4\pi^{d/2})$, and
\begin{equation}
 v_d=\frac{(d-1)\Gamma(d/2-\alpha)\Gamma(d/2-1+\alpha)\sin(\pi\alpha)\Gamma(d/2)}
 {2\pi^{d/2+1}\Gamma(d)(2-d)},\qquad v_4=-\frac{z_\alpha}{8\pi^2},
\end{equation}
for the coincident shift of one real twisted component. Let $\mathscr H_3$ be the finite part of $G_0^2\delta P$ with
$(a_{-3},a_{-1})=(2\omega_4C_4^3v_4,\omega_4C_4^2v_4^2)$, and $\mathscr H_2$ that of $(\delta P)^2$ with
$(a_{-3},a_{-1})=(0,4\omega_4C_4^2v_4^2)$. Restoring the local terms in dimensional regularization gives, through the finite part,
\begin{equation}
 \begin{split}
 J_3&=\mathscr H_3+\frac{\omega_dC_d^2v_d^2}{\epsilon}
 +\frac{\omega_dC_d^3v_d(d-2)}{12d}+O(\epsilon),\\
 J_2&=\mathscr H_2+\frac{4\omega_dC_d^2v_d^2}{\epsilon}+O(\epsilon).
 \end{split}
 \label{eq:entropymelonfinite}
\end{equation}
The last term of $J_3$ is the finite local remainder from the sunset-type singularity. It is retained even though its logarithmic coefficient vanishes at $d=4$.

With $\operatorname{Vol}_d=2\pi^{d/2}\Gamma(1-d/2)$, define $m_{-2}$ and $m_{-1}$ by
\begin{equation}
 -\frac{(16\pi^2)^2\mathcal M^{2\epsilon}n_t\operatorname{Vol}_d}{36}
 \left[(N+2)J_3+(n_t+1)J_2\right]
 =\frac{m_{-2}}{\epsilon^2}+\frac{m_{-1}}\epsilon+O(1).
\end{equation}
Then
\begin{equation}
 \begin{gathered}
 m_{-2}=\frac{[N+2+4(n_t+1)]n_tz_\alpha^2}{18},\qquad
 \mathscr C_{-2}=q_{\rm ct}+q_{\rm tw}+q_{\rm sp}+m_{-2}=0,\\
 \mathscr C_{-1}=q_{\rm ct}K_\alpha+q_{\rm tw}K_{\rm tw}+q_{\rm sp}K_{\rm sp}+m_{-1}.
 \end{gathered}
\end{equation}
Finally, using $\widetilde F=-\sin(\pi\epsilon/2)\mathcal F$ and
$\bar\lambda_* = L_1\epsilon+L_2\epsilon^2$ with
$L_1=3/(N+8)$ and $L_2=9(3N+14)/(N+8)^3$, the common real part is
\begin{equation}
 \begin{split}
 \operatorname{Re}(\widetilde F-\widetilde F_{\rm free})
 &=F_1\epsilon+F_2\epsilon^2+\cdots,\qquad F_1=-\frac\pi2L_1f_{-1},\\
 F_2&=-\frac\pi2\left[L_1f_0+L_2f_{-1}+L_1^2\mathscr C_{-1}\right]
 -\frac{\pi n_tN_{\rm sp}(N_{\rm sp}+2)z_\alpha}{8(N+8)^2}.
 \end{split}
 \label{eq:entropyF2assembly}
\end{equation}
The last term is the real order-$\epsilon^2$ part of Eq.~\eqref{eq:Fsp}. Taking $F_i(N-1)-F_i(N)$ at fixed $n_t$ gives the entries in Table~\ref{tab:entropyeps2}.

These expressions make the pole cancellation explicit and leave finite angular integrals for the melon. The finite-part prescription reproduces two integrals known in closed form to $10^{-14}$, the assembled answer is independent of the renormalization scale, and every contribution vanishes without the twist. Table~\ref{tab:entropyeps2} lists the second-order coefficient, and Table~\ref{tab:entropy}(b) compares the continued difference order by order. Both real coefficients are positive, preserving the perturbative sign of Eq.~\eqref{eq:decoupleeps}. The second order is four to six times the first at the angles shown, so the continuation to $\epsilon=1$ determines neither the magnitude nor the sign for a real three-dimensional defect.

\begin{table}[t]\centering\small
\begin{tabular}{llcccc}\toprule
 & & \multicolumn{2}{c}{order $\epsilon^2$} & sum at $\epsilon=1$, & \\
flow & $\alpha$ & analytic & spectator, $s=0$ & $s=0$ resummed & $\epsilon_{\rm c}$\\\midrule
$O(3)\to O(2)$ & $\tfrac14$ & 0.41 & 1.83 & 2.07 & 1.83\\
$O(3)\to O(2)$ & $\tfrac12$ & 1.20 & 2.43 & 3.86 & 2.44\\
$O(4)\to O(3)$ & $\tfrac14$ & 0.40 & 2.27 & 1.32 & 1.13\\
$O(4)\to O(3)$ & $\tfrac12$ & 1.06 & 3.02 & 3.07 & 1.50\\
\bottomrule\end{tabular}
\caption{The second-order coefficient of the entropy increase $\log\mathfrak g_{N-1}-\log\mathfrak g_N$ near four dimensions for one twisted complex component. The first three numerical columns are in units of $10^{-3}$. The analytic column collects the fixed point, the continuation, the counterterm, and the cactus and melon diagrams with minimal subtraction; the next column is the spectator channel of Eq.~\eqref{eq:Fsp} expanded to $\epsilon^2$. The truncated orders are compared with large $N$ in Table~\ref{tab:entropy}(b). The resummed sum adds the first order to the analytic second order and keeps the real part of Eq.~\eqref{eq:Fsp} at $\epsilon=1$, where the roots of Eq.~\eqref{eq:spgap} are complex, so it only indicates the size of the higher orders. The last column is $\epsilon_{\rm c}$ for the larger model, the zero of the discriminant of Eq.~\eqref{eq:spgap}, which locates the merger only at large $N$.}
\label{tab:entropyeps2}
\end{table}

\subsection{A lattice entanglement test}\label{app:entanglement}
Since the wavefunction overlaps that measure the entropy of a line that can end do not apply here, we also tried to reach the defect entropy through entanglement. The entropy and the conformal weight together fix the entanglement entropy that the line adds to a disk centred on it. The modular Hamiltonian of the disk is the integral of the energy density $T_{00}$ weighted by a conformal Killing vector~\cite{Casini:2011kv}, so the stress tensor shifts this entropy away from $\log\mathfrak g$ by a term fixed by $\mathfrak h$~\cite{Lewkowycz:2013laa,Jensen:2013lxa,Kobayashi:2018lil},
\begin{equation}
 \mathfrak S_\Dc=\log\mathfrak g-2\pi\mathfrak h,
 \label{eq:Sdefect}
\end{equation}
which is $-0.0166$ for the free half-twist.

For the cap $\vartheta<\vartheta_c$ around the north pole, each orbital of Eq.~\eqref{eq:orb} splits into an inside and an outside mode with weight $\mathsf p_m=\operatorname{BetaReg}_{\sin^2(\vartheta_c/2)}(j-m+1,j+m+1)$, the regularized incomplete beta function, also for the shifted grid of a twisted flavour. Expanding each many-body basis state over which occupied orbitals lie inside gives the full Schmidt matrix, block diagonal in the inside particle number, angular momentum and charge. A filled level reproduces $-\sum_m[\mathsf p_m\log\mathsf p_m+(1-\mathsf p_m)\log(1-\mathsf p_m)]$ to $10^{-10}$, the half-twist is north--south symmetric, and the fermionic signs agree with a direct construction in a doubled Fock space to $4\times10^{-16}$.

The free complex scalar shows where the test stands. Its half-twisted cap entropy minus the untwisted one is about $-0.31$, flat in $\Nm=6$--$9$ and in the cap size, but it is already $-0.29$ at $\alpha=0.1$, since the regular band removes one orbital per twisted flavour for every $\alpha\ne0$. Measured from the $\alpha\to0^+$ limit instead, the twist dependence fits $A_{\rm cap}+B_{\rm cap}\mathfrak S_\Dc^{\rm free}(\alpha)$ to $3\times10^{-4}$ at every $\Nm$, with $B_{\rm cap}=1.59$, $1.55$, $1.59$ and $1.65$ at $\Nm=6$--$9$. The shape is therefore right but the normalization is not, and five twist angles cannot separate $\log\mathfrak g$ from $\mathfrak h$.

\section{Pinned lines and junctions}\label{app:pinning}
Section~\ref{sec:junction} rests on three ingredients, the lattice junctions, the stiffness that normalizes them and two analytic limits that test their shape.

\subsection{Pinning and decoupling}
On the fuzzy sphere a pinned line is a local field at the two polar orbitals, the lattice form of Eq.~\eqref{eq:pinaction}. In the vector Hamiltonian of Eq.~\eqref{eq:vectorH}, pinning along the internal unit vector $\mathbf e_1$ at the north pole and along $\cos\eta\,\mathbf e_1+\sin\eta\,\mathbf e_2$ at the south pole adds
\begin{equation}
 -h_{\rm pin}\left[(\mathsf f_{{\rm N}0}^\dagger\mathsf f_{{\rm N}1}+{\rm h.c.})+\cos\eta\,(\mathsf f_{{\rm S}0}^\dagger\mathsf f_{{\rm S}1}+{\rm h.c.})+\sin\eta\,(\mathsf f_{{\rm S}0}^\dagger\mathsf f_{{\rm S}2}+{\rm h.c.})\right],
 \label{eq:pin}
\end{equation}
where N and S label the polar orbitals, and we use $h_{\rm pin}=8$. Changing $h_{\rm pin}$ from $4$ to $32$ at $\Nm=7$ in $O(2)$ moves $\mathfrak r$ from $1.093$ to $1.085$ and $\Ct$ from $0.360$ to $0.364$. In the spin-one regulator of Eq.~\eqref{eq:xyH} the order parameter is the in-plane spin, which is odd under the model's antiunitary symmetry. The pin is therefore $\mathsf S^x$ at the north pole and $\cos\eta\,\mathsf S^x+\sin\eta\,\mathsf S^y$ at the south pole, which makes the Hamiltonian complex. This pin saturates slowly, and we use $h_{\rm pin}=1024$, within about $0.002$ of saturation. The junction dimension is $\Dh_{\rm j}(\eta)=[E_0(\eta)-E_0(0)]/\gamma_B$, and the tilt appears as the lowest excitation at $\eta=0$.

A monodromy in other components barely affects a pinned spectator. In $O(4)$ with the $(1,2)$ plane twisted and the $(3,4)$ plane pinned at the poles with a strong field, the lowest pinned levels at $\Nm=6,7,8$ are $1.301,1.262,1.227$, compared with $1.302,1.263,1.228$ without the twist, and the next level is $1.460,1.464,1.469$ against $1.459,1.463,1.466$. After the local potential is subtracted, the twisted and untwisted $O(3)$ spectator-wall estimates differ by at most $0.003$ over $\Nm=5$--$8$, below $0.5\%$ of the wall dimension $0.80$--$0.84$. Within this accuracy, and for the tested sizes and channels, the ordinary pinned-line junction also describes the pinned spectator in the partially twisted theory. Pinning a charged mode of the $O(2)$ monodromy instead starts a different line flow, whose endpoint and junction data are given in Appendix~\ref{app:monopin}.

\subsection{Shape of the junction}

\begin{figure}[t]\centering
\includegraphics[width=0.95\textwidth]{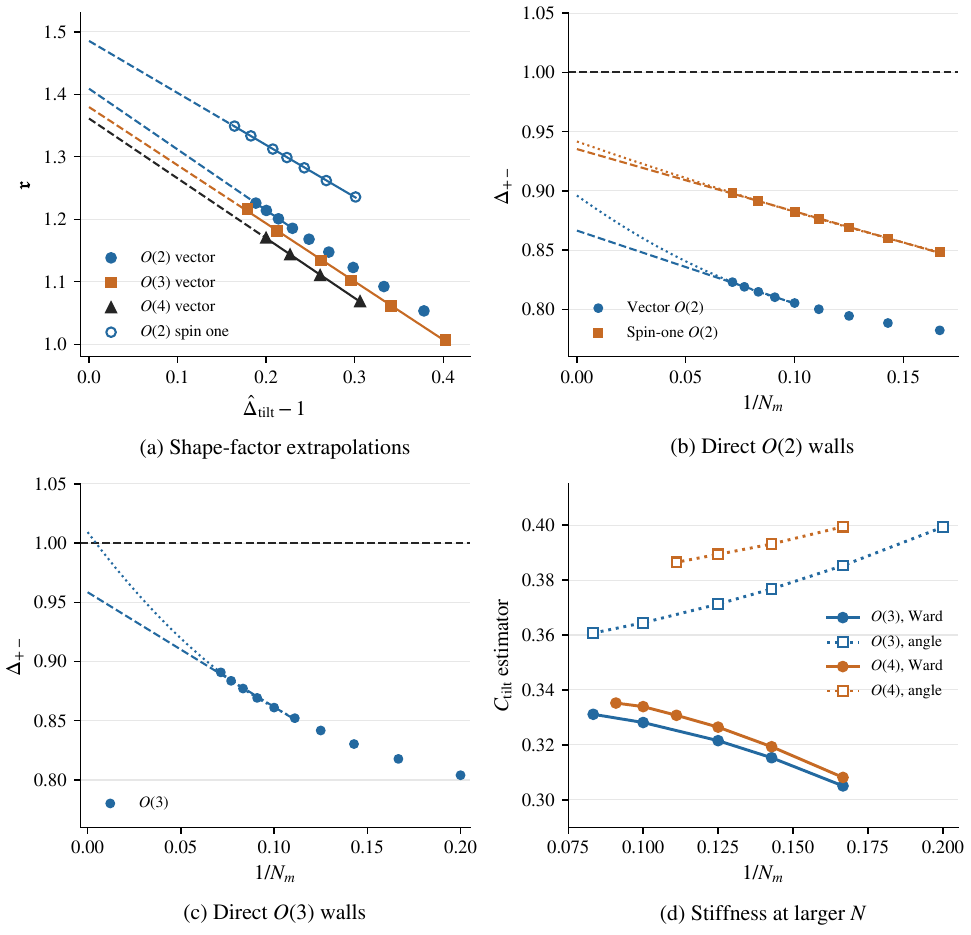}
\caption{Convergence checks for the junction analysis. (a) Shape factors fitted linearly in the tilt excess, using vector $O(2)$ sizes $11$--$14$, $O(3)$ sizes $5$--$8$, $10$ and $12$, $O(4)$ sizes $6$--$9$, and spin-one $O(2)$ sizes $6$--$10$, $12$ and $14$; dashed lines extend the fits toward the continuum. (b) Direct $O(2)$ walls and representative linear (dashed) and quadratic (dotted) $1/\Nm$ extrapolations, fitted at $\Nm=10$--$14$ (vector) and $6$--$14$ (spin one). (c) The direct $O(3)$ walls with the same two ansatzes fitted at $\Nm=9$--$14$; horizontal dashed lines mark marginality. (d) The angle-curvature and Ward stiffness sequences for $O(3)$ and $O(4)$, including the DMRG Ward points at $\Nm=10$ and $12$ for $O(3)$ and at $\Nm=9$--$11$ for $O(4)$, and the DMRG angle points at $\Nm=10$ and $12$ for $O(3)$ and $\Nm=9$ for $O(4)$. Table~\ref{tab:wallfits} gives the complete direct-fit comparison.}
\label{fig:junctionconvergence}
\end{figure}

The angle scans use nine angles in vector-model exact diagonalization and six in vector-model DMRG and spin-one exact diagonalization, always including $\eta=\pi/8$ and $\pi/4$. We take $\Ct^{\angle}$ from the quadratic $\Dh_{\rm j}=\Ct^{\angle}(1-\cos\eta)+\upsilon_2(1-\cos\eta)^2$ through these two smallest angles, so that by Eq.~\eqref{eq:junction} it estimates $\Ct$, and $\mathfrak r=\Dh_{\rm j}(\pi)/2\Ct^{\angle}$ at each size. The second-order curvature $E''(0)$ can also be computed exactly from the ground state and the resolvent at $\eta=0$, and at $\Nm\le9$ the two-angle estimate reproduces it to better than $10^{-4}$ in $O(2)$, $O(3)$ and $O(4)$. A global cubic fit in $1-\cos\eta$ to the whole profile instead overstates the curvature by an amount that grows with $\Nm$, from $7\times10^{-4}$ at $\Nm=6$ to $6\times10^{-3}$ at $\Nm=14$ in vector $O(2)$ and to $1.1\times10^{-2}$ at $\Nm=10$ in the spin-one regulator, whose global-fit curvature appears flat for this reason.

Table~\ref{tab:junction} lists the lattice tilt level, $\Ct^{\angle}$ and $\mathfrak r$, and Fig.~\ref{fig:junctionconvergence} shows the shape fits and the measured wall sequences. The shape factor is close to linear in the excess of the tilt level over one, with slopes $-0.97$, $-0.93$ and $-0.96$ for $O(2)$, $O(3)$ and $O(4)$. Linear extrapolations use $\Nm=11$--$14$, $\Nm=5$--$12$ and $\Nm=6$--$9$, respectively, and give $\mathfrak r=1.409$, $1.380$ and $1.361$. The $O(3)$ fit adds the DMRG scans at $\Nm=10$ and $12$ to exact diagonalization at $\Nm=5$--$8$, which alone gives $1.369$, and the $O(4)$ fit combines exact diagonalization at $\Nm=6$--$8$ with DMRG at $\Nm=9$. In vector $O(2)$ the extrapolation still rises with the fit window, from $1.384$ for $\Nm=6$--$9$ through $1.390$, $1.396$ and $1.401$ to $1.409$ for $\Nm=11$--$14$. The spin-one regulator gives $\mathfrak r=1.2357$, $1.2620$, $1.2824$, $1.2989$, $1.3124$, $1.3335$ and $1.3492$ at $\Nm=6$--$10$, $12$ and $14$, with tilt levels $1.301$, $1.268$, $1.243$, $1.223$, $1.207$, $1.183$ and $1.164$, and $1.486$ for the linear and $1.500$ for the quadratic shape fit.

\begin{table}[t]\centering\footnotesize
\begin{tabular}{llcccc}\toprule
model & $\Nm$ & tilt level & $\Ct^{\angle}$ & $\mathfrak r$ & $\Dh_{\rm j}(\pi)$\\\midrule
$O(2)$ vector & 6 & 1.3782 & 0.3713 & 1.0533 & 0.7823\\
 & 7 & 1.3330 & 0.3609 & 1.0923 & 0.7884\\
 & 8 & 1.2982 & 0.3538 & 1.1228 & 0.7944\\
 & 9 & 1.2707 & 0.3486 & 1.1475 & 0.8001\\
 & 10 & 1.2484 & 0.3447 & 1.1681 & 0.8053\\
 & 11 & 1.2297 & 0.3417 & 1.1856 & 0.8102\\
 & 12 & 1.2139 & 0.3393 & 1.2008 & 0.8148\\
 & 13 & 1.2002 & 0.3373 & 1.2141 & 0.8190\\
 & 14 & 1.1883 & 0.3356 & 1.2259 & 0.8229\\
\midrule
$O(2)$ spin one & 6 & 1.3009 & 0.3432 & 1.2357 & 0.8483\\
 & 7 & 1.2679 & 0.3407 & 1.2620 & 0.8598\\
 & 8 & 1.2430 & 0.3388 & 1.2824 & 0.8690\\
 & 9 & 1.2234 & 0.3374 & 1.2989 & 0.8764\\
 & 10 & 1.2075 & 0.3362 & 1.3124 & 0.8824\\
 & 12 & 1.1827 & 0.3343 & 1.3335 & 0.8916\\
 & 14 & 1.1642 & 0.3329 & 1.3492 & 0.8982\\
\midrule
$O(3)$ vector & 5 & 1.4025 & 0.3993 & 1.0069 & 0.8041\\
 & 6 & 1.3412 & 0.3853 & 1.0614 & 0.8178\\
 & 7 & 1.2960 & 0.3768 & 1.1020 & 0.8304\\
 & 8 & 1.2616 & 0.3712 & 1.1338 & 0.8418\\
 & 10 & 1.2122 & 0.3645 & 1.1814 & 0.8611\\
 & 12 & 1.1788 & 0.3607 & 1.2160 & 0.8772\\
\midrule
$O(4)$ vector & 6 & 1.3061 & 0.3994 & 1.0685 & 0.8535\\
 & 7 & 1.2610 & 0.3931 & 1.1104 & 0.8730\\
 & 8 & 1.2270 & 0.3894 & 1.1434 & 0.8904\\
 & 9 & 1.1999 & 0.3865 & 1.1703 & 0.9045\\
 & 10 & 1.1782 & --- & --- & 0.9174\\
 & 11 & 1.1585 & --- & --- & 0.9275\\
\bottomrule\end{tabular}
\caption{Junction scans in bulk units. Vector $O(2)$ uses ED for $\Nm=6$--$9$ and DMRG for $\Nm=10$--$14$, $O(3)$ uses ED for $\Nm=5$--$8$ and DMRG at $\Nm=10$ and $12$, spin-one $O(2)$ uses ED for $\Nm=6$--$10$ and DMRG at $\Nm=12$ and $14$, and $O(4)$ uses ED for $\Nm=6$--$8$ and DMRG for $\Nm=9$--$11$; the other scans shown here use ED. Vector ED uses nine angles, vector DMRG and spin-one ED use six. The $O(3)$ DMRG at $\Nm=10$ and $12$ and the spin-one DMRG at $\Nm=12$ and $14$ have the energies at $\eta=0$, $\pi/8$, $\pi/4$ and $\pi$, and the $O(4)$ DMRG at $\Nm=10$ and $11$ only those at $\eta=0$ and $\pi$. Every size entering the shape fits is listed. The tilt level approaches one and the angle curvature estimates the continuum stiffness, with finite-size corrections.}
\label{tab:junction}
\end{table}

In the continuum, the one-loop junction of Ref.~\cite{Cuomo:2024psk} is, in our notation,
\begin{equation}
 \begin{aligned}
 \Dh_{\rm j}(\eta)={}&\frac{(N+8)(1-\cos\eta)}{4\pi^2}\\
 &+\epsilon\,\frac{(1-\cos\eta)\left[\pi^2(3N^2+29N+106)+6(N+8)^2(1-\cos\eta)\zeta(3)\right]}{8\pi^4(N+8)}+O(\epsilon^2).
 \end{aligned}
\end{equation}
The coefficient of $1-\cos\eta$ is $\Ct$ at this order~\cite{Girault:2025kzt}, and the $(1-\cos\eta)^2$ term comes from the classical quartic interaction of the pinned profile, whose angular integral is $\int_0^\pi \mathrm d\omega\,\omega^2(\pi-\omega)^2/\sin^2\omega=6\pi\zeta(3)$. In the ratio of Eq.~\eqref{eq:rdef} the terms with $\pi^2$ cancel, which gives Eq.~\eqref{eq:roneloop}.

\subsection{The tilt normalization}\label{app:tiltnorm}
The small-angle curvature $\Ct^{\angle}$ of the junction decreases with $\Nm$ in every model, including the spin-one regulator once the bias of a global fit is removed. Whether it converges to the stiffness is a separate question, and we answer it with an independent estimator built on the exact broken generator. The charge $Q_{12}$ commutes with the bulk Hamiltonian, so $[H,Q_{12}]$ is the pinning term in the tilt direction at the two poles. In the continuum $|\langle k|[H,Q_{12}]|0_\Dc\rangle|=2\sqrt{\Ct(k+1)}\,\gamma_B$ for even parallel levels $k$ of the tilt family and zero for odd ones, so the estimate $|\langle\hat t|[H,Q_{12}]|0_\Dc\rangle|^2/4\gamma_B^2$ is $\Ct$, obtained from the primary matrix element of an exact Ward operator. At finite size the selected tilt state can mix with other states, even though the operator identity is exact. This matters because the lattice tilt family is not yet conformal. Its level-two descendant carries about $0.02$ of the tilt's weight in $Q_{12}|0_\Dc\rangle$ instead of $1/3$, and estimators that replace the measured tilt gap by its protected value, or impose continuum descendant weights, inherit this mismatch.

The vector $O(2)$ generator estimates are $0.3023$, $0.3118$, $0.3172$, $0.3204$ and $0.3224$ for $\Nm=6$--$10$, and DMRG gives $0.3241$ and $0.3244$ at $\Nm=12$ and $14$. The spin-one values at the same sizes are $0.2923$, $0.3037$, $0.3114$, $0.3167$ and $0.3205$, and DMRG gives $0.3252$ and $0.3278$ at $\Nm=12$ and $14$. In $O(3)$ the ED values $0.3051$, $0.3153$ and $0.3216$ at $\Nm=6$--$8$ are supplemented by the DMRG values $0.328164$ and $0.331157$ at $\Nm=10$ and $12$. The $O(4)$ sequence is $0.3082$, $0.3193$, $0.3265$ at $\Nm=6$--$8$, followed by the DMRG values $0.3308$, $0.3339$ and $0.3353$ at $\Nm=9$--$11$.

We compare stiffness fits $C_*+A_{\rm fit}\Nm^{-2}$ and $C_*+A_{\rm fit}\Nm^{-3}$, and a power law with a fitted exponent, using all the listed sizes. They give $\Ct^{HQ}=0.326$--$0.331$ for vector $O(2)$, $0.329$--$0.336$ for spin-one $O(2)$, $0.335$--$0.341$ for $O(3)$ and $0.341$--$0.348$ for $O(4)$. Free-power fits to the small-angle curvatures give instead about $0.323$ in both $O(2)$ regulators, and $0.352$ and $0.377$ for $O(3)$ and $O(4)$. These ranges quantify the sensitivity to the ansatz and leave further continuum corrections unbounded. In $O(2)$ the two continuations agree to $1$--$3\%$ in the vector regulator and $2$--$4\%$ in the spin-one regulator. In vector $O(2)$ the last two increments of the Ward estimator, $0.0017$ and $0.0003$, show that it has nearly converged, and the difference between the curvature and the Ward estimator closes roughly as $\Nm^{-2}$, from $0.069$ at $\Nm=6$ to $0.011$ at $\Nm=14$. In $O(3)$ and $O(4)$ the curvature continuations stay $4\%$ and $9\%$ above the Ward fits.

We take the Ward estimator as the stiffness for every $N$. In $O(2)$ the choice changes the stiffness by a few percent at most, and for $N\ge3$ two blind tests in $O(3)$ support it, with a third size in $O(4)$ consistent with it. Fits of the Ward sequence at $\Nm\le10$ predicted $0.3314$ at $\Nm=12$, against the subsequently measured $0.33116$, whereas a Ward sequence converging to the curvature continuation would have reached $0.3342$. Equation~\eqref{eq:towerS} below predicted the curvature at $\Nm=10$ and $12$, which a joint single-power fit of the curvature and Ward sequences missed at $\Nm=12$ by about $2\times10^{-3}$. In $O(4)$ the three Ward fits at $\Nm\le10$ forecast $0.3352$--$0.3368$ at $\Nm=11$, and the measured $0.33525$ lies at the lower end of this band, below the $0.3366$ reached by a sequence converging to the curvature continuation. These tests favour the Ward continuation over the fitted curvature limit at the measured sizes, without fixing either asymptotic correction law. For $N\ge3$ we therefore treat the curvature continuations as finite-window diagnostics, and the difference between the factorized estimate and the measured wall comes from the shape factor, which is defined with the curvature (Appendix~\ref{app:wallroutes}).

Why the curvature converges so slowly becomes clear from its exact spectral form. Rotating both pins together costs no energy and the reflection exchanges them, so the contact term cancels and $E''(0)=2\langle0_\Dc|V'_{\rm N}(H-E_0)^{-1}V'_{\rm S}|0_\Dc\rangle$, where $V'_{\rm N,S}$ are the rotations of the north and south pins. The curvature is therefore the tilt propagator between the two poles, $\Ct^{\angle}=2\sum_nP_nw_n$, with $w_n=|\langle n|V'_{\rm S}|0_\Dc\rangle|^2/[(E_n-E_0)\gamma_B]$ and $P_n=\pm1$ the reflection parity. In the continuum every member of the tilt family has $w_n=\Ct$ and $P_n=(-1)^k$, and the alternating sum returns $\Ct$ only because all these weights are equal, whereas the Ward stiffness $w_0\Dh_{\rm tilt}$, with $\Dh_{\rm tilt}$ the lattice tilt level in bulk units, uses the primary alone. On the lattice the first two descendants carry $w_1/w_0=0.20$--$0.39$ and $w_2/w_0=0.015$--$0.086$ at $\Nm=6$--$9$, the same in $O(2)$, $O(3)$ and $O(4)$ to within $3\%$, and these three levels reproduce the curvature to about $1\%$ for $\Nm\ge8$.

Separating the primary turns this sum into an exact relation at each size,
\begin{equation}
 \Ct^{\angle}=\Ct^{HQ}+w_0\,(2-\Dh_{\rm tilt}-\mathcal Q),\qquad \mathcal Q=-\frac{2}{w_0}\sum_{n\ge1}P_nw_n,
 \label{eq:towerS}
\end{equation}
where $w_0=\Ct^{HQ}/\Dh_{\rm tilt}$ is the single-pole weight of the tilt, and $\mathcal Q$ collects the descendants. In the continuum the alternating sum gives $\mathcal Q=1$, and the two estimators coincide. On the lattice $\mathcal Q$ depends on the size but hardly on the model. It is $0.307$, $0.457$ and $0.552$ at $\Nm=6$, $7$ and $8$ in vector $O(2)$, and the $O(3)$ and $O(4)$ values agree with these to within $3\%$, with an offset that grows with $\Nm$ and with $N$. In vector $O(2)$ it reaches $0.665$ at $\Nm=10$ and $0.771$ at $\Nm=14$, while the spin-one regulator has its own, larger $\mathcal Q$.

This near-universality is predictive. Taking $\mathcal Q$ from vector $O(2)$ at $\Nm=10$, with the $O(3)$ offset extrapolated from $\Nm=6$--$8$ and the $O(3)$ Ward data as input, Eq.~\eqref{eq:towerS} gives $\Ct^{\angle}=0.3648\pm0.0014$ for $O(3)$ at $\Nm=10$ and $0.3611\pm0.0008$ at $\Nm=12$, and the DMRG angle scans, evaluated afterwards, give $0.36445$ and $0.3607$. The same construction reproduces the $O(4)$ DMRG curvature at $\Nm=9$ to $10^{-4}$. The curvature therefore approaches $\Ct$ only as $\mathcal Q$ approaches one, at a rate set by the lattice pin and common to the three models, while the Ward estimator needs only the primary. In vector $O(2)$, $1-\mathcal Q$ decreases about as $\Nm^{-1.2}$, while the tilt excess, which enters Eq.~\eqref{eq:towerS} with the opposite sign, decreases as $\Nm^{-0.8}$ in $O(2)$ and $\Nm^{-0.9}$ and $\Nm^{-1.1}$ in $O(3)$ and $O(4)$. If these effective powers persist, the tilt excess will eventually dominate and the curvature will approach $\Ct$ non-monotonically. The measured sizes do not establish that asymptotic ordering, but the competing terms explain why a single-power continuation of the curvature is sensitive to the fitting window.

\subsection{The measured wall and its factorization}\label{app:wallroutes}

\begin{table}[t]\centering\small
\begin{tabular}{lcccc}\toprule
 & $\Ct^{HQ}$ (Ward) & $\Ct^{\angle}$ (angle) & $\Delta_{+-}$, measured & $2\Ct^{HQ}\mathfrak r$, factorized\\\midrule
$O(2)$ vector & 0.326--0.331 & $\approx0.323$ & 0.85--0.91 & 0.92--0.93\\
$O(2)$ spin one & 0.329--0.336 & $\approx0.323$ & 0.94--0.96 & 0.98--1.01\\
$O(3)$ vector & 0.335--0.341 & $\approx0.352$ & 0.93--1.04 & 0.92--0.94\\
$O(4)$ vector & 0.341--0.348 & $\approx0.377$ & 1.01--1.06 & 0.93--0.95\\
\bottomrule\end{tabular}
\caption{Lattice stiffness and wall dimension. The measured wall is the junction energy at $\eta=\pi$, extrapolated linearly and quadratically in $1/\Nm$ or in the tilt excess. The stiffness is the Ward estimate, from continuum fits of Eq.~\eqref{eq:tiltHQ} with $\Nm^{-2}$, $\Nm^{-3}$ and fitted-power corrections. The angle column lists free-power continuations of the small-angle curvature of Appendix~\ref{app:pinning}; for $N\ge3$ these are finite-window numbers, not stiffness estimates (Appendix~\ref{app:tiltnorm}). The factorized column multiplies the Ward stiffness by the shape factor of Table~\ref{tab:shape}. Ranges span the fit choices of Appendix~\ref{app:pinning} (Tables~\ref{tab:junction} and~\ref{tab:wallfits}) and are not confidence intervals. The measured walls reach $0.891$ for $O(3)$ at $\Nm=14$ and $0.928$ for $O(4)$ at $\Nm=11$, and both are still rising; for $O(4)$ a fit linear in $\Nm^{-2}$ gives $0.95$--$0.97$.}
\label{tab:wallestimators}
\end{table}

\begin{figure}[t]\centering
\includegraphics[width=\textwidth]{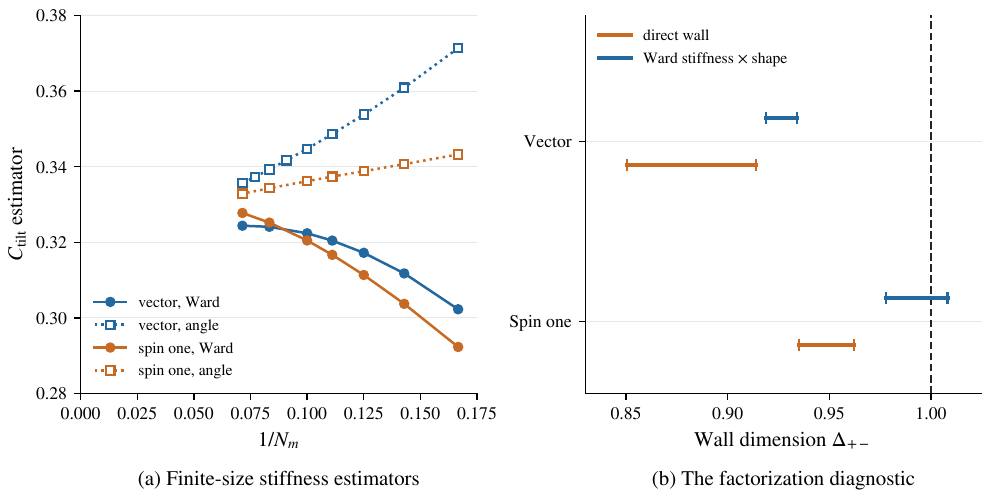}
\caption{Stiffness and reconstruction controls for ordinary $O(2)$ pinned lines. (a) The Ward estimator and the small-angle curvature at finite size in the vector and spin-one regulators. Their continuum continuations are listed in Table~\ref{tab:wallestimators}. (b) The direct-wall fit spread and the Ward stiffness times the separately extrapolated shape factor. The latter is a diagnostic of the stiffness estimates and their extrapolation, not an independent measurement of the wall. Segments span the stated fits and are not confidence intervals; the dashed line marks marginality.}
\label{fig:junctioncontrols}
\end{figure}

Figure~\ref{fig:junctioncontrols} compares the stiffness sequences and their consequences for the factorized wall estimate.

We now turn to the wall itself. We fit the measured wall dimension as a polynomial in $1/\Nm$ or in the tilt excess $x_t=\Dh_{\rm tilt}-1$, and Table~\ref{tab:wallfits} lists the resulting intercepts. Linear and quadratic fits use all sizes and the indicated terminal windows, and quadratic fits require at least four points. We make no statistical interpretation of their spread. For $O(4)$ the linear fits give $1.01$--$1.02$ and the quadratic fits $1.02$--$1.06$, while a fit linear in $\Nm^{-2}$ gives $0.954$ over $\Nm=6$--$11$ and $0.968$ over $\Nm=8$--$11$, which shows how strongly the conclusion depends on the correction law. This dependence belongs to the measured wall itself, and it is set by the leading irrelevant operator on the pinned line.

The neutral spectrum of the straight pinned line identifies this operator. Its lowest neutral level lies at $\Dh\simeq1.52$, $1.49$ and $1.47$ for $O(2)$, $O(3)$ and $O(4)$ at $\Nm=8$. This level is $\hat\phi_1$, the restriction of the order parameter along the pinning direction, whose coupling the pinning flow drives to its fixed point, with $\Dh=1.55(14)$ from the $\epsilon$ expansion and $1.542$ at large $N$~\cite{Cuomo:2021kfm}, and $1.609(11)$, $1.635(58)$ and $1.666(38)$ for $N=2$, $3$ and $4$ from fuzzy-sphere extrapolations~\cite{Yang:2026qtk}, so the slow exponent is $(\Dh-1)/2=0.27\pm0.07$. It is even under the rotation by $\pi$ that exchanges the two poles and reverses the line, so it can appear in the effective action of the lattice line, and it corrects every level by a term proportional to $\Nm^{-(\Dh-1)/2}$. In the fits below we use $\Nm^{-0.26}$ as a representative slow correction, consistent with this exponent. The next neutral level, near $2.5$, is odd and cannot appear.

To see which observables this operator affects, we add the same pinning term on the next orbital at both poles and compute the first-order shift of each level. Over $\Nm=6$--$9$ in vector $O(2)$ the shift of the wall decreases as $\Nm^{-0.31}$, close to the slow power, while the shifts of the tilt level and of the displacement decrease as $\Nm^{-0.68}$ and $\Nm^{-0.69}$, and the wall and the tilt keep this ordering in $O(3)$ and $O(4)$. The wall is therefore the observable most exposed to the leading irrelevant operator, whereas the protected levels respond mainly through faster corrections. At $\Nm=9$ about three quarters of the wall's relative shift is carried by the curvature and one quarter by the shape factor. The amplitude of the slow term in the unperturbed sequences is not fixed at these sizes. A joint fit of the tilt level and the displacement, with the computed shifts as the slow component, determines it only if the bulk correction $\Nm^{-(\Delta_{\mathcal E'}-3)/2}\simeq\Nm^{-0.40}$ from the leading irrelevant singlet, $\Delta_{\mathcal E'}\simeq3.79$, is left out. Over $\Nm\le14$ that correction is nearly degenerate with $\Nm^{-0.26}$, and including it moves the implied continuum wall from about $0.9$ to above one.

The spin-one regulator constrains this amplitude, because its wall sequence is the best converged. Through $\Nm=14$ the local exponent $q_{\rm inc}$ of its increments, $\Delta_{+-}(\Nm)-\Delta_{+-}(\Nm-1)\propto\Nm^{-q_{\rm inc}}$, rises from $1.63$ to $1.92$, whereas a dominant term in $\Nm^{-0.26}$ would pull it toward $1.26$. Fits with the slow line or bulk term together with $\Nm^{-1}$ and $\Nm^{-2}$ corrections, or with three fast corrections, extrapolate to $0.899$, $0.914$ and $0.938$, all with residuals near $10^{-5}$. The fitted slow term is positive, so it lowers the limit. Forcing the limit to one requires a negative slow term and raises the residual to $1.3$--$1.7\times10^{-4}$, and to $8.8\times10^{-4}$ with fast corrections alone. Over the range $0.20$--$0.35$ of the slow exponent allowed by the dimension of $\hat\phi_1$, the spin-one fits give $0.886$--$0.910$, and a limit at one raises the residual by a factor of $11$--$16$. A common limit for the spin-one walls and the vector walls at $\Nm\ge8$, with separate corrections for the two regulators, gives $0.913$, $0.922$ and $0.935$ for the same three forms, and a limit at one doubles to octuples the residual. Including the vector sizes $\Nm=6$ and $7$, the furthest from their asymptotic convergence, changes these to $1.00$, $0.97$ and $0.92$. The vector sequence and that of Ref.~\cite{Yang:2026qtk} have increment exponents below $1.26$ at the sizes reached, so on their own they do not constrain the slow amplitude. Including both slow terms at once makes the fit ill-conditioned. Its free limit then lies above one, at $1.10$--$1.23$ over the same range of the line exponent and at $1.09$--$1.12$ in the common fit with the vector walls, and a limit at one fits as well as the preferred forms. The free fit reaches these values only through slow amplitudes of opposite sign, each an order of magnitude larger than in the single-slow fits, which largely cancel at the sizes reached. These data therefore favour a relevant wall without establishing it.

The factorized estimate of Table~\ref{tab:wallestimators} multiplies the fitted Ward stiffness by the shape-factor intercept. Because $\Dh_{\rm j}(\pi)=2\Ct^{\angle}\mathfrak r$ at every size, it differs from the measured wall through the continuation of $\Ct^{HQ}/\Ct^{\angle}$ and through the separate extrapolation of $\mathfrak r$. The ratio of the factorized to the measured midpoints is $1.05$, $1.05$, $0.94$ and $0.91$ for vector $O(2)$, spin-one $O(2)$, $O(3)$ and $O(4)$, against $1.02$, $1.03$, $0.96$ and $0.91$ for the ratios of the Ward and curvature stiffness continuations. For $N\ge3$ most of the difference comes from the curvature. The shape factor is defined with $\Ct^{\angle}$ at each size, so its extrapolation inherits the slow approach of the curvature to $\Ct$ (Appendix~\ref{app:tiltnorm}), and the factorized estimate lies low by about the same factor. In $O(2)$, where the two stiffnesses agree to $2$--$3\%$, the factorized estimate lies a further $2$--$3\%$ above the measured one, which measures the difference between extrapolating the shape factor and extrapolating the wall. In the spin-one regulator this difference appears as a tension, since the factorized estimate, $0.98$--$1.01$, remains above the spin-one wall extrapolations. With the Ward stiffness, the common limit corresponds to $\mathfrak r=\Delta_{+-}/2\Ct=1.36$--$1.43$ in both $O(2)$ regulators, so the larger split between their curvature-based shape factors in Table~\ref{tab:shape} reflects the extrapolation of the curvature, not the wall. The factorization thus adds these separate uncertainties to the extrapolation of the measured wall, and does not change the measured wall.

\begin{table}[t]\centering\small
\begin{tabular}{llrrrr}\toprule
model & $\Nm$ window & $1/\Nm$ & $(1/\Nm)^2$ & $x_t$ & $x_t^2$\\\midrule
$O(2)$ vector & 6--14 & 0.8504 & 0.8837 & 0.8608 & 0.9040\\
 & 10--14 & 0.8664 & 0.8960 & 0.8777 & 0.9127\\
 & 11--14 & 0.8692 & 0.8983 & 0.8805 & 0.9138\\
\midrule
$O(2)$ spin one & 6--14 & 0.9353 & 0.9416 & 0.9586 & 0.9622\\
 & 8--14 & 0.9371 & 0.9401 & 0.9594 & 0.9531\\
\midrule
$O(3)$ vector & 5--14 & 0.9323 & 0.9808 & 0.9388 & 0.9890\\
 & 9--14 & 0.9585 & 1.0093 & 0.9636 & 1.0227\\
 & 11--14 & 0.9685 & 1.0235 & 0.9748 & 1.0430\\
\midrule
$O(4)$ vector & 6--11 & 1.0149 & 1.0601 & 1.0067 & 1.0355\\
 & 7--11 & 1.0222 & 1.0551 & 1.0121 & 1.0246\\
\bottomrule\end{tabular}
\caption{Direct-wall extrapolations from the saved unrounded energies. The columns headed $1/\Nm$ and $x_t$ are linear fits; those headed $(1/\Nm)^2$ and $x_t^2$ include both linear and quadratic terms. The four-point quadratic fits have only one residual degree of freedom.}
\label{tab:wallfits}
\end{table}

The finite-size identity $\Delta_{+-}=2\Ct^\angle\mathfrak r$ also allows a prediction from the angle scans at $\Nm=5$--$8$. Refitting the saved ED values gives
\begin{equation}
 \Ct^\angle(\Nm)=0.352902+1.11289\Nm^{-1.97456},\qquad
 \mathfrak r(\Nm)=1.368592-0.899283[\Dh_{\rm tilt}(\Nm)-1].
 \label{eq:wallpredict}
\end{equation}
At each larger $\Nm$, the second expression takes the independently measured DMRG tilt level as input. The continuum product of these angle-based fits is $0.96596$. Table~\ref{tab:wallprediction} compares their finite-size prediction with the direct wall, and the upward residual grows from $0.0015$ at $\Nm=9$ to $0.0083$ at $\Nm=14$. This check tests the continuation of the measured wall in the angle normalization, not the split between stiffness and shape. At $\Nm=10$ and $12$ the DMRG angle scans measure both factors, $\Ct^\angle=0.36445$ and $0.3607$ against the forecasts $0.36470$ and $0.3611$, and $\mathfrak r=1.1814$ and $1.2160$ against $1.1778$ and $1.2078$, so the residual at these sizes lies in the shape factor.

\begin{table}[t]\centering\small
\begin{tabular}{rrrrr}\toprule
$\Nm$ & measured tilt & predicted wall & direct wall & direct $-$ predicted\\\midrule
9 & 1.234606 & 0.850687 & 0.852200 & +0.001513\\
10 & 1.212169 & 0.859087 & 0.861146 & +0.002059\\
11 & 1.193627 & 0.866413 & 0.869306 & +0.002893\\
12 & 1.178827 & 0.872340 & 0.877218 & +0.004878\\
13 & 1.165233 & 0.878233 & 0.883732 & +0.005500\\
14 & 1.154894 & 0.882573 & 0.890865 & +0.008291\\
\bottomrule\end{tabular}
\caption{Testing the $O(3)$ angle-based continuation of Eq.~\eqref{eq:wallpredict}. Only the curvature and shape coefficients are fitted at $\Nm\le8$; the tilt column is new input at each larger size.}
\label{tab:wallprediction}
\end{table}

\begin{table}[t]\centering\small
\begin{tabular}{lccc}\toprule
limit & $\Ct$ & $\mathfrak r$ & $\Delta_{+-}$\\\midrule
one loop, $N=2$ & $0.253+0.223\epsilon$ & $1+0.731\epsilon$ & $0.51+0.82\epsilon$\\
one loop, $N=3$ & $0.279+0.253\epsilon$ & $1+0.731\epsilon$ & $0.56+0.91\epsilon$\\
one loop, $N=4$ & $0.304+0.285\epsilon$ & $1+0.731\epsilon$ & $0.61+1.01\epsilon$\\
resummed shape + lattice stiffness, $O(2)$ & lattice, Table~\ref{tab:walls} & 1.34--1.37 & 0.87--0.92\\
$N\to\infty$ & $0.0888N$ & 1.551 & $0.276N$\\
$N=1$, Ising & --- & --- & $0.84(4)$; Pad\'e $0.85$\\
$N=2$, $d=2$ (Kosterlitz--Thouless) & $4/\pi^2$ & $\pi^2/4$ & 2\\
$N=2$, $[1/1]$ Pad\'e from $d=4$ to $d=2$ & 0.367 & 1.75 & 1.29\\
\bottomrule\end{tabular}
\caption{Analytic and limiting values of the stiffness, shape factor and wall dimension. The one-loop entries follow from the junction of Ref.~\cite{Cuomo:2024psk} (Appendix~\ref{app:pinning}). The resummed two-loop shape factor (Table~\ref{tab:shape}, Appendix~\ref{app:twoloop}) is combined with the measured $O(2)$ stiffness, since the two-loop stiffness is not known. The large-$N$ saddle, which also reproduces the one-point coefficient $a_\phi^2=0.5581N$ of Ref.~\cite{Cuomo:2021kfm}, is solved in Appendix~\ref{app:junctionlargeN}. The Ising entries are the fuzzy-sphere wall and the Pad\'e approximant anchored in two dimensions of Ref.~\cite{Cuomo:2024psk}. At the Kosterlitz--Thouless point the junction is the boundary-condition-changing operator with $\Dh_{\rm j}=2\eta^2/\pi^2$; the last row interpolates between it and the one-loop value and slope at $d=4$, separately for $\Ct$ and $\Delta_{+-}$, and its $\mathfrak r$ is their ratio $\Delta_{+-}/2\Ct$.}
\label{tab:anchors}
\end{table}

Near two dimensions the wall has a simple limit for $N\ge3$. There the pinned line becomes a Dirichlet line of the $O(N)$ sigma model, $S=(2g_\sigma)^{-1}\int(\partial\boldsymbol n)^2$ with $\boldsymbol n^2=1$, at its critical coupling $g_\sigma^*=2\pi(d-2)/(N-2)$.\footnote{The same sigma-model reasoning decides whether the boundary of the $O(N)$ model can order, where it leads to the extraordinary-log phase~\cite{Metlitski:2020cqy}.} A junction changes the boundary condition by the angle $\eta$ on both half-planes, and the configuration that rotates uniformly in the polar angle around it costs the action $\eta^2/(2\pi g_\sigma)$ per unit logarithm of scale on each half-plane. Hence
\begin{equation}
 \Dh_{\rm j}(\eta)\simeq\frac{\eta^2}{\pi g_\sigma^*}=\frac{(N-2)\,\eta^2}{2\pi^2(d-2)},\qquad \Delta_{+-}\simeq\frac{N-2}{2(d-2)},
 \label{eq:walld2}
\end{equation}
so the wall diverges as $d\to2$, and the shape factor approaches the rigid value $\pi^2/4$. For $N=2$ the coupling does not run, and the corresponding two-dimensional value is the Kosterlitz--Thouless anchor of Table~\ref{tab:anchors}.

The data above do not decide the threshold $N_c$, and three calculations could do so analytically. In the $\epsilon$ expansion the wall at second order requires the fixed-point pinning coupling to the same order, and hence the three-loop running of the pinning coupling. The defect beta function is known to next-to-leading order in the bulk coupling~\cite{Pannell:2023pwz}, which fixes the pinning coupling only to first order~\cite{Cuomo:2021kfm}, and the route through the Ward identity that relates the stiffness to the one-point and bulk-to-tilt coefficients~\cite{Girault:2025kzt} needs those coefficients at second order as well. The leading part of the missing term at large $N$ follows from the exact large-$N$ stiffness. At large $N$ the wall grows linearly in $N$, with slope $2\Ct\mathfrak r_\infty/N=0.28$ at $d=3$ from the saddle values of Appendix~\ref{app:junctionlargeN}, which would place the threshold between three and four. The first $1/N$ correction, a fluctuation determinant of the auxiliary field around the junction saddle, decides whether it stays there, and the lattice values show that this correction is large at small $N$. A bootstrap bound on the operator joining opposite pinnings would constrain the wall without an expansion. Once a second-order term is known, Eq.~\eqref{eq:walld2} anchors interpolations in $d$, whereas with first-order input alone they overshoot the lattice walls by more than half. On the lattice, an $O(3)$ regulator whose wall sequence converges as fast as that of the spin-one $O(2)$ model would test the correction law where the margin is smallest.

\subsection{The junction at large \texorpdfstring{$N$}{N}}\label{app:junctionlargeN}
At large $N$ the pinned line and its junctions are classical saddles of Eq.~\eqref{eq:largeMaction}, with a background $\bar\phi_a$ along the pinning directions. In the $S^1\times \mathbb H^2$ frame the straight line pinned along $\mathbf n_1$ has a constant field $\bar\phi_a=\sqrt N\phi_*n_{1a}$ and $\varsigma=\tfrac14$, which is $1/(4r^2)$ in flat space, since a constant solves the conformally coupled equation only for this mass. The transverse line modes then have $\Dh_s=\tfrac12+\sqrt{s^2+\tfrac14}$, so the tilt sits at $\Dh=1$. The constraint fixes $\phi_*^2$ to minus the coincident change of the propagator of one component,
\begin{equation}
 \phi_*^2=\frac1{4\pi^2}\sum_{s\in\mathbb Z}\Big[\operatorname{dg}\big(\tfrac12+\sqrt{s^2+\tfrac14}\big)-\operatorname{dg}\big(\tfrac12+|s|\big)\Big]=0.0444132.
 \label{eq:v0}
\end{equation}
The resulting one-point coefficient $a_\phi^2=4\pi N\phi_*^2=0.55811N$ agrees with the value $0.55813N$ of Ref.~\cite{Cuomo:2021kfm}.

For the junction we write the upper half plane of $\mathbb H^2$ as $\rho e^{\mathrm i\omega}$, with $0<\omega<\pi$, so that the line is the boundary and the junction sits at $\rho=0$. Dilatations about the junction translate $\log\rho$, and the saddle depends only on $\omega$. Writing $\bar\phi_a=\sqrt N[A(\omega)n_{1a}+B(\omega)n_{2a}]$, with $\mathbf n_1\cdot\mathbf n_2=\cos\eta$, the field equations are
\begin{equation}
 -A''+\frac{\varsigma-\frac14}{\sin^2\omega}A=0,\qquad -B''+\frac{\varsigma-\frac14}{\sin^2\omega}B=0,
\end{equation}
with $A(\pi)=B(0)=\phi_*$ and $A(0)=B(\pi)=0$. The constraint requires $\bar\phi^2/N+G_\varsigma(x,x)=\phi_*^2+G_{1/4}(x,x)$ at every point, with $\bar\phi^2/N=A^2+B^2+2AB\cos\eta$, $G_\varsigma$ the propagator of one transverse component in the background $\varsigma(\omega)$, and $G_{1/4}$ its value on the straight line. The junction dimension is $\Dh_{\rm j}(\eta)=N\mathcal F_{\rm j}(\eta)$, where $\mathcal F_{\rm j}$ is the saddle action per unit $\log\rho$ and per component.

Two statements are exact. First, varying the boundary direction gives
\begin{equation}
 \frac{\mathrm d\mathcal F_{\rm j}}{\mathrm d\eta}=-2\pi\phi_*\sin\eta\,B'(\pi).
\end{equation}
Second, at $\varsigma=\tfrac14$ the profiles $A$ and $B$ are linear in $\omega$, and $\mathcal F_{\rm j}=2\phi_*^2(1-\cos\eta)$. The saddle is stationary in $\varsigma$, so its relaxation changes $\mathcal F_{\rm j}$ only at order $\eta^4$, and $\Ct=2N\phi_*^2=0.0888N$. This agrees with the exact relation $\Ct=a_\phi^2/(\pi^2\beta_{\phi\hat t}^2)$ of Ref.~\cite{Girault:2025kzt}. At large $N$ the tilt term of the unit-normalized transverse propagator is $2\mathcal G_1(\chi)$, so $\beta_{\phi\hat t}^2=2/\pi$, and with $a_\phi^2=4\pi N\phi_*^2$ the relation gives $\Ct=2N\phi_*^2$. The shape factor of Eq.~\eqref{eq:rdef} is then
\begin{equation}
 \mathfrak r_\infty=-\frac{\pi}{2\phi_*}\int_0^\pi \mathrm d\eta\,\sin\eta\,B'(\pi),
\end{equation}
where $B'(\pi)$ is evaluated at each $\eta$. The unrelaxed profile has $B'(\pi)=-\phi_*/\pi$, which gives single exchange.

We solve the saddle with finite differences in $\omega$, circle modes $|s|\le s_{\max}$, a quadrature for the Fourier integral along $\log\rho$ with its local large-momentum tail added analytically, and Newton iteration with continuation in $\eta$. For a uniform shift of $\varsigma$ the propagator code gives the coincident change $-0.003644$, against the exact digamma value $-0.003645$. With $120$, $160$, $120$ and $240$ grid points, $s_{\max}=30$, $40$, $30$ and $50$, and $12$, $12$, $16$ and $12$ junction angles, we obtain $\mathfrak r_\infty=1.551255$, $1.551285$, $1.551256$ and $1.551348$, respectively, so $\mathfrak r_\infty=1.5513(1)$. The softest fluctuation, at $s=0$ and zero frequency along $\log\rho$, keeps a positive gap of $0.84$, $0.58$, $0.49$, $0.45$ and $0.44$ at $\eta/\pi=0.5$, $0.8$, $0.9$, $0.95$ and $1$. This holds even where $\varsigma<0$, so the saddle is locally stable.

The value lies between single exchange and the one-loop $1.73$ at $\epsilon=1$, and below the rigid $\pi^2/4$. Expanding $-\pi B'(\pi)/\phi_*=1+0.277(1-\cos\eta)+\cdots$, the quadratic term alone would give $\mathfrak r\simeq1.28$. The rest comes from a strong relaxation near $\eta=\pi$, where the antiparallel junction is an amplitude wall. The field vanishes on the equator $\omega=\pi/2$ and $\varsigma$ dips to $-0.27$ there, so the line is locally unpinned. At leading order the classical field obeys a linear equation for given $\varsigma$ and cannot rotate through the other components. At finite $N$ it can, and we expect, as a heuristic, that this lowers $\Delta_{+-}$. Altogether $\Delta_{+-}=2\Ct\mathfrak r_\infty=0.2756N$ at leading order. For comparison, the large-$N$ limit of the one-loop stiffness is $N(1+\tfrac32\epsilon)/4\pi^2$, or $0.0633N$ at $\epsilon=1$.

\subsection{The shape factor at two loops}\label{app:twoloop}

\begin{table}[t]\centering\small
\begin{tabular}{lcccc}\toprule
 & $O(2)$ & $O(3)$ & $O(4)$ & $N=\infty$\\\midrule
one loop, $\epsilon=1$ [Eq.~\eqref{eq:roneloop}] & 1.731 & 1.731 & 1.731 & 1.731\\
$\mathfrak r^{(2)}$ [Eq.~\eqref{eq:rtwoloop}] & $-0.8586$ & $-0.8441$ & $-0.8318$ & $-0.6795$\\
$[1/1]$ Pad\'e, $\epsilon=1$ & 1.336 & 1.339 & 1.342 & 1.379\\
large-$N$-corrected, $\epsilon=1$ & 1.372 & 1.387 & 1.399 & 1.551\\
lattice, tilt-excess fits & 1.38--1.41\,/\,1.49--1.50 & 1.380 & 1.361 & ---\\
\bottomrule\end{tabular}
\caption{The shape factor $\mathfrak r$. The two-loop coefficient and both resummations are derived in Appendix~\ref{app:twoloop}; at $N=\infty$ the large-$N$-corrected entry is the exact saddle value of Appendix~\ref{app:junctionlargeN}. The lattice entries are fits in the excess of the tilt level over one, using the small-angle curvature: linear over the fit windows of Appendix~\ref{app:pinning} for vector $O(2)$, linear and quadratic for the spin-one $O(2)$ regulator, and linear for the vector $O(3)$ and $O(4)$ models (Table~\ref{tab:junction} and Fig.~\ref{fig:junctionconvergence}).}
\label{tab:shape}
\end{table}

Through second order the junction dimension is a cubic polynomial in $1-\cos\eta$,
\begin{equation}
 \Dh_{\rm j}(\eta)=\Ct(1-\cos\eta)+\upsilon_2(1-\cos\eta)^2+\upsilon_3(1-\cos\eta)^3+O(\epsilon^3),
\end{equation}
which is the continuum counterpart of the angle fits above. The quadratic coefficient $\upsilon_2$ starts at one loop and the cubic $\upsilon_3$ at two, so $\mathfrak r=1+2\upsilon_2/\Ct+4\upsilon_3/\Ct$ follows from the first two orders of each coefficient.

A half-line source produces the potential $C_d\rho^{3-d}f(\omega)$, where $\rho$ is the distance from the junction, $\omega$ is the polar angle from the other half-line, $C_d=\Gamma(d/2-1)/(4\pi^{d/2})$ and
\begin{equation}
 f(\omega)=\sin^{3-d}\omega\int_0^\omega\!\mathrm d\omega'\,\sin^{d-4}\omega'.
\end{equation}
The graphs reduce to angular integrals on $S^{d-1}$ times a radial logarithm. Each graph is combined with its counterterm graph before the fixed-point couplings are substituted.

To specify the subtraction and the finite-$N$ coefficient, write $\lambda_{\rm q}=6\lambda_4$ for the coupling multiplying $(\phi_a\phi_a)^2/4!$ and $h_\ell$ for the continuum pinning coupling, distinct from the lattice fields. In the minimal-subtraction convention of Eqs.~(3.6) and (3.18) of Ref.~\cite{Cuomo:2021kfm},
\begin{equation}
 \begin{gathered}
 \lambda_{{\rm q}*}=\ell_1\epsilon+\ell_2\epsilon^2+O(\epsilon^3),\qquad
 h_{\ell*}^2=H_0+H_1\epsilon+O(\epsilon^2),\\
 \ell_1=\frac{3(4\pi)^2}{N+8},\quad
 \ell_2=\frac{9(4\pi)^2(3N+14)}{(N+8)^3},\quad
 H_0=N+8,\quad H_1=\frac{4N^2+45N+170}{2(N+8)}.
 \end{gathered}
 \label{eq:junctionfp}
\end{equation}
Let $\Omega_m=2\pi^{(m+1)/2}/\Gamma((m+1)/2)$ be the area of the unit $S^m$, set $f_a(\omega)=f(\omega)$, $f_b(\omega)=f(\pi-\omega)$, and abbreviate $f_P=f_af_b$ and $f_S=f_a+f_b$. The one-vertex angular integral is
\begin{equation}
 A_b(\epsilon)=-\frac{C_d^4\Omega_{d-2}}6
 \int_0^\pi\!\mathrm d\omega\,\sin^{d-2}\omega\,f_P(\omega)^2,
 \qquad A_b(0)=-\frac{\zeta(3)}{64\pi^6},\qquad d=4-\epsilon.
 \label{eq:junctionAb}
\end{equation}
Primes on $A_b$ below mean derivatives with respect to $\epsilon$ at zero.

The two-vertex tree is specified by the angular Green operator
\begin{equation}
 \mathcal L_d=-\sin^{2-d}\omega\,\partial_\omega
 (\sin^{d-2}\omega\,\partial_\omega)-\rho_c(\rho_c+d-2),
 \qquad \rho_c=11-3d.
\end{equation}
Its inverse uses solutions regular at the two endpoints and is continued in $d$. Define
\begin{equation}
 \begin{split}
 Q_d(F,G)&=\int_0^\pi\!\mathrm d\omega\,\sin^{d-2}\omega\,
 F(\omega)(\mathcal L_d^{-1}G)(\omega),\\
 A_c(\epsilon)&=\frac{C_d^6\Omega_{d-2}}{72}
 \left[4Q_d(f_Pf_S,f_Pf_S)+8Q_d(f_af_b^2,f_S^2f_a)\right]
 =\frac{c_{-1}}\epsilon+c_0+O(\epsilon).
 \end{split}
 \label{eq:junctionAc}
\end{equation}
For clarity, the endpoint-regular homogeneous solutions are
${}_2F_1(-\rho_c,\rho_c+d-2;(d-1)/2;\sin^2(\omega/2))$ and its reflection about $\pi/2$. Their weighted Wronskian fixes the Green function. At $d=4$, multiplication by $\sin\omega$ reduces this inverse to the Dirichlet Green function
$g_{\rm D}(\omega,\omega')=\min(\omega,\omega')[\pi-\max(\omega,\omega')]/\pi$.

The determinant graph has an equally explicit angular form. With $p=2(d-3)$, define the homogeneous convolution
\begin{equation}
 \begin{split}
 R_d(F,G)={}&\int_{S^{d-1}}\!\mathrm d\mathbf n\int_{S^{d-1}}\!\mathrm d\mathbf n'\,
 F(\mathbf n)G(\mathbf n')
 \int_0^\infty\!\frac{u^{d-1-p}\,\mathrm du}
 {(1+u^2-2u\mathbf n\cdot\mathbf n')^{d-2}},\\
 A_d(\epsilon;N)={}&\frac{C_d^6}{36}
 \left[(N+6)R_d(f_P,f_P)+2R_d(f_a^2,f_b^2)\right]
 =\frac{d_{-1}(N)}\epsilon+d_0(N)+O(\epsilon).
 \end{split}
 \label{eq:junctionAd}
\end{equation}
The angular functions are zonal about the line axis, and all integrals are defined first where they converge and then by dimensional continuation. This also fixes the finite parts $c_0$ and $d_0(N)$. The eigenvalue of the convolution in harmonic degree $\ell$ is
\begin{equation}
 \frac{\pi^{d/2}\Gamma(\epsilon/2)}{\Gamma(2-\epsilon)}
 \frac{\Gamma((\ell+6-d)/2)\Gamma((\ell+3d-10)/2)}
 {\Gamma((\ell+2d-6)/2)\Gamma((\ell+10-2d)/2)},
 \label{eq:junctionmultiplier}
\end{equation}
which provides a second representation for evaluating Eq.~\eqref{eq:junctionAd}. The pole residues are
$c_{-1}=\zeta(3)/(3072\pi^8)$ and $d_{-1}(N)=(N+8)c_{-1}$.

Write $\Ct=C_0+C_1\epsilon+\cdots$, $\upsilon_2=D_1\epsilon+D_2\epsilon^2+\cdots$ and $\upsilon_3=E_2\epsilon^2+\cdots$. The stiffness coefficients are
\begin{equation}
 C_0=\frac{N+8}{4\pi^2},\qquad
 C_1=\frac{3N^2+29N+106}{8\pi^2(N+8)}.
\end{equation}
Taking the simple-pole residue with
$-(\lambda_{\rm q}\partial_{\lambda_{\rm q}}+\tfrac12h_\ell\partial_{h_\ell})$
and including the counterterm graphs gives
\begin{align}
 D_1={}&-A_b(0)\ell_1H_0^2,\nonumber\\
 D_2={}&-A_b(0)(\ell_2H_0^2+2\ell_1H_0H_1)
 -\ell_1^2H_0^3\left[c_0+\frac{5A_b'(0)}{9(4\pi)^2}\right]\nonumber\\
 &-\ell_1^2H_0^2\left[d_0(N)+\frac{4(N+8)A_b'(0)}{9(4\pi)^2}\right].
 \label{eq:junctionD2}
\end{align}
The terms proportional to $A_b'(0)$ retain the dimensional dependence of the one-vertex graph in the counterterms. Dropping them changes the finite answer. For the cubic term define
\begin{equation}
 \mathcal P_3=\int_0^\pi\!\mathrm d\omega\int_0^\pi\!\mathrm d\omega'\,
 \frac{\omega(\pi-\omega)^2}{\sin^2\omega}\,
 g_{\rm D}(\omega,\omega')\,
 \frac{\omega'^2(\pi-\omega')}{\sin^2\omega'},\qquad
 E_2=\frac{N+8}{4\pi^7}\mathcal P_3.
 \label{eq:junctionE2}
\end{equation}
Thus the general-$N$ result, with every finite part specified above, is
\begin{equation}
 \mathfrak r^{(2)}(N)=\frac{2D_2+4E_2}{C_0}
 -\frac{2D_1C_1}{C_0^2}.
 \label{eq:junctionr2assembly}
\end{equation}
The unknown $C_2$ cancels from this ratio, while it remains in the wall dimension itself.

Three graphs enter at second order. The first has one quartic vertex and four pinning insertions. Its four-dimensional value gives Eq.~\eqref{eq:roneloop}, and the $O(\epsilon)$ part of its $d$-dimensional angular integral enters at second order. The second has two quartic vertices and six insertions, and comes from the second-order classical profile. It gives the cubic coefficient as the finite integral in Eq.~\eqref{eq:junctionE2}, $4\upsilon_3/\Ct=1.3489\,\epsilon^2$ for every $N$, together with a subdivergent part that renormalizes the pinning coupling. The third is the one-loop determinant around the pinned profile, with two quartic vertices and four insertions, which we reduce with Funk--Hecke multipliers on $S^{d-1}$. At $N=\infty$ the $\epsilon^2$ coefficient of $\mathfrak r$ is a sum of large terms that nearly cancel,
\begin{itemize}
\item the one-vertex graph with the $O(\epsilon)$ part of the squared pinning coupling gives $+2.923$;
\item the two-vertex tree gives $-10.124$, and the $d$-dependence of the one-vertex integral in its counterterm graph gives $+6.098$;
\item the determinant gives $-4.708$, and its counterterm graph $+4.879$;
\item the product of the one-loop terms gives $-1.096$, and the cubic term $+1.349$.
\end{itemize}
The total is $-0.680$.

Four checks support the result. It reproduces the one-loop tilt normalization of Ref.~\cite{Cuomo:2024psk}. The double poles of the two-loop graphs are fixed by one-loop data, as locality of the counterterms requires. The $N$-independent part of the determinant is evaluated in position space, and the same quadrature reproduces the angular-momentum sum on smooth test functions to 13 digits. Most stringently, the large-$N$ saddle of Appendix~\ref{app:junctionlargeN}, continued to $d=4-\epsilon$ in the frame $S^{d-2}\times \mathbb H^2$, is independent of every graph above. Fits at $\epsilon\le0.2$ give $(\mathfrak r-1)/\epsilon\to0.7305$, against $6\zeta(3)/\pi^2=0.730763$, and, with that slope imposed, an $\epsilon^2$ coefficient of $-0.681$, against $-0.6795$. The coefficients of the angular profile agree to $1$--$2\%$.

The $[1/1]$ Pad\'e approximant in Eq.~\eqref{eq:rtwoloop} is the ratio of two linear functions of $\epsilon$ that reproduces the series through $\epsilon^2$. The large-$N$-corrected estimate is $\mathfrak r_\infty(4-\epsilon)+[\mathfrak r^{(2)}(N)-\mathfrak r^{(2)}(\infty)]\epsilon^2$, with $\mathfrak r_\infty(3)=1.551$. It reproduces Eq.~\eqref{eq:rtwoloop} through order $\epsilon^2$ for every $N$ and uses the full leading large-$N$ curve at $N=\infty$. Its uncomputed terms begin with the $N$-dependent part at order $\epsilon^3$.

The wall itself needs the two-loop tilt normalization $\Ct^{(2)}$, the $\epsilon^2$ coefficient of $\Ct$. For $N=2$ the known part is $\Delta_{+-}=0.50661+0.81602\,\epsilon+(-0.10919+2\Ct^{(2)})\epsilon^2$. $\Ct^{(2)}$ depends on the pinning coupling at order $\epsilon^2$, and hence on the three-loop beta function of the pinning coupling. We have not found that beta function in the literature.

\subsection{Pinning a charged mode of the monodromy}\label{app:monopin}
The coupling of Eq.~\eqref{eq:monopinaction} preserves $L_z-sQ$, and Ref.~\cite{Kravchuk:2025evf} identifies its infrared endpoint as a monodromy pinning defect (Sec.~\ref{sec:monopin}). We test this identification at weak coupling, at strong coupling and at large $N$. Near marginality the coupling has the small scaling exponent $y_{\rm pin}=1-\Dh_s$, and its modulus runs as $\beta_{\rm RG}=-y_{\rm pin}|h_{\rm pin}|+\beta_{\rm RG}^{(3)}|h_{\rm pin}|^3$. Along the flow the two real components of the doublet then have $\Dh_{\rm Re}=1+\partial\beta_{\rm RG}/\partial|h_{\rm pin}|$ and $\Dh_{\rm Im}=1+\beta_{\rm RG}/|h_{\rm pin}|$, so their shifts at second order in the coupling are in the ratio $3:1$. This ratio is a unit-free test of the single-coupling structure of the flow.

\emph{(a) Weak coupling.} We pin the lowest charged mode, $s=\alpha$, at both poles with the time-reversal-even combination $\mathsf R_1+\mathsf R_2$ of the two charge-raising bilinears, $\mathsf R_1=\mathsf f_0^\dagger\mathsf f_{12-}$ and $\mathsf R_2=\mathsf f_{12+}^\dagger\mathsf f_0$, each acting on the pole-most orbitals with matching $L_z-sQ$. Here $\mathsf f_{12\pm}$ are the charge-$\pm1$ combinations of the twisted pair. A defect primary inserted with equal strength at both poles, which lie at unit distance on either side of the origin of the line, creates only even $SL(2)$ descendants of the doublet. The weight of the first descendant in $\mathsf R|0_\Dc\rangle$, where $\mathsf R$ is the charge-raising part of the pin, therefore measures how unequally the two half-lines are coupled. At $\Nm=8$ and $\alpha=0.45$ it is $4\times10^{-4}$ for this pin, against $0.28$ when a single bilinear sits at each pole and $0.10$ for a pin displaced from the pole by $\vartheta_0=0.35$. Those alternatives perturb the two half-lines differently, like a junction, and we do not use them.

Write the ratio of the second-order shifts of the two levels as $(1+\mathfrak q)/(1-\mathfrak q)$, so that the prediction is $\mathfrak q=\tfrac12$. In second-order perturbation theory $\mathfrak q$ is the off-diagonal element that mixes $\hat\psi_s$ with $\hat\psi_s^\dagger$, divided by the diagonal element minus the shift of the vacuum. With the pin at both poles, the vacuum alone, as an intermediate state, gives $\mathfrak q=\tfrac13$, a ratio of two. The remaining states cancel most of that contribution, completely so for a generalized free field, so the universal ratio resides in the interacting remainder, which approaches its continuum value slowly on the lattice. Table~\ref{tab:monopin} lists $\mathfrak q$ at $\alpha=0.45$, where $y_{\rm pin}\approx0.1$, from exact diagonalization and from DMRG that conserves only the fermion number and $L_z-sQ$. The DMRG values at $\Nm=11$ and $12$ lie within $2\times10^{-4}$ of the $1/\Nm$ trend of the exact diagonalization. Extrapolating in $1/\Nm$, with or without a quadratic term, gives $\mathfrak q_\infty=0.487$--$0.492$, a ratio of $2.90$--$2.94$. Varying the angle and the mode, $s=\alpha$ for $\alpha=0.35$--$0.5$ and $s=\alpha-1$ for $\alpha=0.42$--$0.48$, moves $y_{\rm pin}$ between $-0.06$ and $0.15$ at the sizes used, and the extrapolated $\mathfrak q$ by about $0.01$. The weak-coupling flow therefore has the single-coupling structure of Ref.~\cite{Kravchuk:2025evf}, with no sign of an additional marginal singlet coupling.

\begin{table}[t]\centering\small
\begin{tabular}{llcc}\toprule
$\Nm$ & method & $\mathfrak q$ & ratio\\\midrule
6 & ED & 0.3037 & 1.872\\
7 & ED & 0.3289 & 1.980\\
8 & ED & 0.3482 & 2.069\\
9 & ED & 0.3635 & 2.142\\
10 & ED\,/\,DMRG & 0.3758 & 2.204\\
11 & DMRG & 0.3860 & 2.257\\
12 & DMRG & 0.3944 & 2.303\\\midrule
$\infty$ & extrapolated & 0.487--0.492 & 2.90--2.94\\
 & Ref.~\cite{Kravchuk:2025evf}, $y_{\rm pin}\to0$ & $1/2$ & 3\\
\bottomrule\end{tabular}
\caption{Second-order splitting of the charged doublet of the $O(2)$ monodromy at $\alpha=0.45$ under the time-reversal-even pole pin. The entries are evaluated at pinning strength $0.4$ in bulk units; quartic corrections raise $\mathfrak q$ by $5\times10^{-4}$, and the extrapolated row includes them. The ratio is $(1+\mathfrak q)/(1-\mathfrak q)$ of the gap shifts of the two levels. Exact diagonalization and DMRG agree at $\Nm=10$ to $3\times10^{-4}$ in the ratio.}
\label{tab:monopin}
\end{table}

\emph{(b) Strong coupling.} Increasing the pinning strength drives the line to the endpoint of the flow. We use $h_{\rm pin}=1000$ in bulk units, at which the low levels have saturated, and attach a phase $e^{\mathrm i\eta}$ to the pin at the south pole. A $U(1)$ rotation or a transverse rotation shifts the phases at the two poles equally, so their difference $\eta$ is the internal angle of the junctions of Sec.~\ref{sec:junction}, and $\eta=\pi$ joins the lines pinned with $\pm h_{\rm pin}$. The pin conserves the fermion number and $L_z-sQ$. Exact diagonalization covers $\Nm=6$--$10$ at $\alpha=0.2$, $0.3$ and $0.5$, and DMRG, which agrees with it at $\Nm=8$ to six digits in the levels and junction energies, extends $\alpha=0.3$ and $0.5$ to $\Nm=12$. As an untwisted reference at equal size we use the same pin at $\alpha=0$, with the orbital assignment of the limit $\alpha\to0^+$, in which the second bilinear sits one orbital away from the pole. It is a second lattice realization of the ordinary pinned $O(2)$ line of Eq.~\eqref{eq:pin}, with its own finite-size corrections, and we compute it by exact diagonalization for $\Nm\le9$ and by DMRG for $\Nm=10$--$12$.

On any conformal line on which $Q$ is broken, the broken charge creates a tilt at $\Dh=1$, and the displacement sits at $\Dh=2$ in the sector $L_z-sQ=\pm1$. Table~\ref{tab:monopinlattice} collects the measurements. The lowest excitation at $\eta=0$ in the sector $L_z-sQ=0$ falls steadily with $\Nm$. Its extrapolations lie within $0.04$ of one for $\alpha\le0.3$, comparable to the spread of the untwisted reference, and $0.08$--$0.09$ above it at $\alpha=0.5$.\footnote{At half twist the sizes $\Nm=11$ and $12$ leave this extrapolation where $\Nm\le10$ had put it. Over $\Nm\ge8$ the increments of the level fall as $\Nm^{-2}$, as at the smaller twists, so no slower correction is visible at these sizes. The excess grows with the twist, from about $0.03$ at $\alpha=0.3$, and is largest at half twist, where the charge-one modes $s=\alpha$ and $s=\alpha-1$ are degenerate. Its origin is not settled at these sizes, and at half twist the identification rests on continuity in $\alpha$ rather than on the tilt level.} The Ward estimator of Eq.~\eqref{eq:tiltHQ} rises and the angle curvature falls with $\Nm$, as for the ordinary lines. Changing the relative weight of the two bilinears by factors $0.7$ and $1.4$ moves the levels by at most $6\times10^{-4}$ and the junction energies by at most $3\times10^{-4}$ over $\Nm=6$--$8$, so the data describe the endpoint rather than the microscopic pin. The second level of the sector $L_z-sQ=\pm1$, the candidate displacement, decreases with $\Nm$ at the same rate as in the untwisted reference, where it reaches $2.06$ at $\Nm=12$, but at $\Nm=11$ and $12$ it stays $0.11$ above that reference at $\alpha=0.3$ and $0.17$--$0.18$ above it at $\alpha=0.5$. Level repulsion offers a possible finite-size explanation for this offset. At $\alpha=0$ the pin of Eq.~\eqref{eq:pin} preserves the reflection $\phi_2\to-\phi_2$, under which the displacement is even and the lowest level of its sector is odd, so the two cannot mix. The reference pin breaks this reflection through its second bilinear, and at nonzero twist no lattice symmetry separates the two levels. The lowest level then repels the candidate upward, while the candidate itself carries almost no Ward weight, as a displacement should. If the endpoint is conformal and the selected level approaches its displacement, its gap must tend to two. The sizes reached show no decay of the offset, however, so the mixing explanation and the displacement identification remain untested by this sequence. The weak-coupling splitting, the tilt approaching its protected value for $\alpha\le0.3$, the finite stiffness and the weak dependence on the microscopic pin support identifying the endpoint with the monodromy pinning defect of Ref.~\cite{Kravchuk:2025evf}, which becomes the ordinary pinned line as $\alpha\to0$.

The junction data then follow as in Sec.~\ref{sec:junction}. We take $\Ct^{\angle}$ from the junction energy at $\eta=\pi/16$ divided by $1-\cos\eta$, which differs from the curvature by less than $1\%$ at this angle, $\mathfrak r$ from Eq.~\eqref{eq:rdef}, and $\Delta_{+-}=\Dh_{\rm j}(\pi)$ directly. The wall dimension is below one at every twist and decreases with $\alpha$. At $\alpha=0.3$ it changes by less than $0.01$ between $\Nm=6$ and $12$ and stays well below one throughout this range. At $\alpha=0$ the extrapolations of this pin lie several percent below those of Tables~\ref{tab:shape} and~\ref{tab:walls} for the pin of Eq.~\eqref{eq:pin}, with $\mathfrak r=1.32$ against $1.38$--$1.41$ and a measured wall of $0.80$--$0.81$ against $0.85$--$0.91$. This difference measures how the extrapolated values depend on the lattice pin. Ratios at equal $\Nm$ to the untwisted reference cancel much of it, and they are the comparison used in Table~\ref{tab:monojunction}. Rescaling the measured $O(2)$ vector walls of Table~\ref{tab:walls} by these ratios gives $0.80$--$0.88$, $0.74$--$0.81$ and $0.55$--$0.63$ at $\alpha=0.2$, $0.3$ and $0.5$, again below one. The Ward estimator gives a smaller reduction of the stiffness than the angle curvature (Table~\ref{tab:softening}).

\begin{table}[t]\centering\small
\begin{tabular}{lcccccc}\toprule
 & & continuum & \multicolumn{2}{c}{largest size} & \multicolumn{2}{c}{continuum}\\
$\alpha$ & $\Nm$ & $\Dh_{\rm tilt}$ & $\Ct^{HQ}$ & $\Ct^{\angle}$ & $\mathfrak r$ & $\Delta_{+-}$\\\midrule
0 (reference) & 6--12 & 0.99--1.01 & 0.289 & 0.322 & 1.32 & 0.80--0.81\\
0.2 & 6--10 & 0.99--1.02 & 0.282 & 0.320 & 1.27--1.31 & 0.75--0.78\\
0.3 & 6--12 & 1.03--1.04 & 0.286 & 0.305 & 1.26--1.28 & 0.70--0.71\\
0.5 & 6--12 & 1.08--1.09 & 0.265 & 0.270 & 1.18--1.21 & 0.53--0.56\\
\bottomrule\end{tabular}
\caption{The time-reversal-even pinned $O(2)$ monodromy line at strong coupling, $h_{\rm pin}=1000$. The two stiffness columns are the values at the largest size; the Ward estimator rises and the angle curvature falls with $\Nm$. The other entries are continuum extrapolations: the tilt level linearly in $1/\Nm$ over the last three sizes and in $1/\Nm$ and $1/\Nm^2$ over $\Nm\ge8$, and $\mathfrak r$ and $\Delta_{+-}$ with the same two forms and linearly in the tilt excess $x_t$ over the last three and four sizes. Ranges span these fit choices and are not confidence intervals. The $\alpha=0$ row is the same pin without the twist, used as the reference at equal size.}
\label{tab:monopinlattice}
\end{table}

\begin{table}[t]\centering\small
\begin{tabular}{lccccccc}\toprule
 & & \multicolumn{2}{c}{$\mathfrak r/\mathfrak r|_{\alpha=0}$} & \multicolumn{2}{c}{$\Ct^{\angle}/\Ct^{\angle}|_{\alpha=0}$} & \multicolumn{2}{c}{$\Delta_{+-}/\Delta_{+-}|_{\alpha=0}$}\\
$\alpha$ & $\Delta_{+-}$ & $O(2)$ & $N=\infty$ & $O(2)$ & $N=\infty$ & $O(2)$ & $N=\infty$\\\midrule
0.2 & 0.75--0.78 & 0.96--0.98 & 0.981 & 0.96--1.00 & 0.945 & 0.94--0.96 & 0.927\\
0.3 & 0.70--0.71 & 0.94--0.96 & 0.954 & 0.91--0.94 & 0.871 & 0.87--0.89 & 0.831\\
0.5 & 0.53--0.56 & 0.88--0.90 & 0.835 & 0.74--0.79 & 0.570 & 0.65--0.69 & 0.476\\
\bottomrule\end{tabular}
\caption{Junction data of the pinned $O(2)$ monodromy line at twist $\alpha$. The wall dimension $\Delta_{+-}$ is the extrapolated junction energy at $\eta=\pi$ (Table~\ref{tab:monopinlattice}). The $O(2)$ ratios compare the line with the same lattice pin at $\alpha=0$ at equal $\Nm$, extrapolated linearly in $1/\Nm$ over the last three sizes, in $1/\Nm$ and $1/\Nm^2$ over all sizes, and linearly in the tilt excess; the stiffness ratio uses the angle curvature. Ranges span these fit choices and are not confidence intervals. The $N=\infty$ entries are the saddle values of Table~\ref{tab:monopinlargeN}. Sizes are $\Nm=6$--$12$ at $\alpha=0.3$ and $0.5$, and $\Nm=6$--$10$ at $\alpha=0.2$.}
\label{tab:monojunctiondetails}
\end{table}

\emph{Regulator dependence of the twist ratios.} We repeat the equal-size ratios of Table~\ref{tab:monojunctiondetails} in the spin-one regulator, with the same time-reversal-even pin, by exact diagonalization at $\Nm=6$--$10$ and DMRG at $\Nm=12$ (Table~\ref{tab:softening}). The ratios of the shape factor, the angle curvature and the wall overlap the vector ranges at every twist, but the ratio of the Ward stiffness does not. It stays close to one in the spin-one regulator and falls less than the angle curvature in both, so the division of the wall reduction between stiffness and shape depends on the estimator at these sizes. In the spin-one regulator the Ward estimator also rises above the angle curvature at $\Nm=12$ for $\alpha\ge0.3$, which Eq.~\eqref{eq:towerS} allows once $\mathcal Q$ exceeds $2-\Dh_{\rm tilt}$, so the bracketing of the stiffness by the two estimators seen in the vector data is not general.

\begin{table}[t]\centering\small
\begin{tabular}{lccc}\toprule
ratio to $\alpha=0$ & $\alpha=0.2$ & $\alpha=0.3$ & $\alpha=0.5$\\\midrule
$\mathfrak r$, vector & 0.962--0.979 & 0.943--0.959 & 0.876--0.902\\
$\mathfrak r$, spin one & 0.971--0.979 & 0.944--0.957 & 0.865--0.897\\
$\Ct^{\angle}$, vector & 0.962--1.001 & 0.912--0.944 & 0.739--0.793\\
$\Ct^{\angle}$, spin one & 0.960--0.998 & 0.915--0.970 & 0.766--0.862\\
$\Delta_{+-}$, vector & 0.941--0.964 & 0.871--0.890 & 0.647--0.694\\
$\Delta_{+-}$, spin one & 0.936--0.970 & 0.866--0.918 & 0.651--0.758\\
$\Ct^{HQ}$, vector & 0.970--0.990 & 0.957--0.972 & 0.875--0.901\\
$\Ct^{HQ}$, spin one & 1.002--1.017 & 0.992--1.014 & 0.953--0.983\\\bottomrule
\end{tabular}
\caption{Twist ratios of the pinned $O(2)$ monodromy line at equal $\Nm$ in the two regulators, extrapolated as in Table~\ref{tab:monojunction}, linearly in $1/\Nm$ over the last three sizes, in $1/\Nm$ and $1/\Nm^2$ over all sizes, and linearly in the tilt excess. The vector ratios use the data of Table~\ref{tab:monojunctiondetails}. The spin-one ratios use $\Nm=6$--$10$ by exact diagonalization and $\Nm=12$ by DMRG. Ranges span these fit choices and are not confidence intervals.}
\label{tab:softening}
\end{table}

\emph{(c) Large $N$.} For a large-$N$ counterpart we group the $N$ components into $N/2$ complex fields, each of which acquires $e^{2\pi\mathrm i\alpha}$ around the defect, and pin the mode $s=\alpha$ of one of them. In the $S^1\times \mathbb H^2$ frame of Appendix~\ref{app:junctionlargeN} the straight line carries the classical field $\sqrt N\phi_*e^{\mathrm i\alpha\theta}$, constant on $\mathbb H^2$. It solves the conformally coupled equation only for $\varsigma_*=\tfrac14-\alpha^2$, so the twisted modes have $\Dh_s=\tfrac12+\sqrt{s^2+\tfrac14-\alpha^2}$ with $s\in\mathbb Z+\alpha$, and the mode $s=\alpha$ sits at the tilt value $\Dh=1$ for every twist. The constraint gives
\begin{equation}
 \phi_*^2(\alpha)=\frac1{4\pi^2}\Big\{\sum_{s\in\mathbb Z+\alpha}\Big[\operatorname{dg}\big(\tfrac12+\sqrt{s^2+\tfrac14-\alpha^2}\big)-\log|s|\Big]+\log(2\sin\pi\alpha)\Big\},
 \label{eq:phistaralpha}
\end{equation}
where $\log(2\sin\pi\alpha)$ is the regularized difference between $\sum_{s\in\mathbb Z+\alpha}\log|s|$ and the untwisted sum $\sum_{s\in\mathbb Z}\operatorname{dg}(\tfrac12+|s|)$ of Eq.~\eqref{eq:v0}. As $\alpha\to0$, Eq.~\eqref{eq:phistaralpha} reproduces Eq.~\eqref{eq:v0}, and at $\alpha=\tfrac12$, where $\varsigma_*=0$, it equals $1/4\pi^2$ to the $25$ digits evaluated. Stationarity in $\varsigma$ again gives $\Ct=2N\phi_*^2$.

The junction saddle has the structure of Appendix~\ref{app:junctionlargeN}, with the classical field $e^{\mathrm i\alpha\theta}[A(\omega)+e^{\mathrm i\eta}B(\omega)]$, the factor $\varsigma-\tfrac14$ in the profile equations replaced by $\varsigma-\varsigma_*$, and the circle modes of the propagator running over $s\in\mathbb Z+\alpha$, so that their potential at $\varsigma=\varsigma_*$ is $(s^2-\alpha^2)/\sin^2\omega$. With $120$ grid points, $s_{\max}=30$ and twelve junction angles we obtain Table~\ref{tab:monopinlargeN}. At $\alpha=0.3$, $160$ points and $s_{\max}=40$ change $\mathfrak r_\infty$ by $4\times10^{-5}$, and $\alpha=0$ reproduces $\mathfrak r_\infty=1.551255$. At $\eta=\pi$ the field again vanishes on the equator, where $\varsigma$ dips to between $-0.29$ and $-0.26$ at all twists, so the wall remains an amplitude wall. The twist lowers $\Ct$, $\mathfrak r_\infty$ and $\Delta_{+-}$ monotonically. Since $\Delta_{+-}/N$ decreases, the twist also raises the number of components at which walls become irrelevant, although the leading slope does not locate that number.

Comparing with Table~\ref{tab:monojunctiondetails}, the reduction of the shape factor in $O(2)$ matches the large-$N$ saddle at $\alpha=0.2$ and $0.3$ and is smaller at $\alpha=0.5$. The stiffness and the wall, which grow with $N$, fall more slowly with the twist in $O(2)$ than at large $N$.

The partial twist of Sec.~\ref{sec:entropy}, with one twisted complex field and $N-2$ untwisted spectators, gives a different large-$N$ line. There the spectators carry the constraint, and their modes $s\in\mathbb Z$ see the background $\varsigma_*=\tfrac14-\alpha^2$ set by the pinned field, so that
\begin{equation}
 \phi_*^2(\alpha)=\frac1{4\pi^2}\sum_{s\in\mathbb Z}\Big[\operatorname{dg}\big(\tfrac12+\sqrt{s^2+\tfrac14-\alpha^2}\big)-\operatorname{dg}\big(\tfrac12+|s|\big)\Big].
\end{equation}
The twist now lowers $\Ct$, $\mathfrak r_\infty$ and $\Delta_{+-}$ faster. At $\alpha=0.3$ their ratios to $\alpha=0$ are $0.826$, $0.912$ and $0.754$, against $0.871$, $0.954$ and $0.831$ in Table~\ref{tab:monopinlargeN}. At half twist the spectator mode at $s=0$ reaches the Breitenlohner--Freedman value $\Dh=\tfrac12$ and $\phi_*^2$ vanishes, so at leading order this line has no condensate.

The next order decides whether the line survives. The broken charge keeps the tilt, which is the mode $s=\alpha$ of the pinned field, at $\Dh=1$ to all orders. Near half twist the pinned background is weak, so the order-$1/N$ shift of this mode equals its shift on the unpinned partial twist up to terms of order $N^{-3/2}$. The tilt then fixes the spectator background to $\varsigma_*=1-\Dh_{1/2}$, where $\Dh_{1/2}$ is the dimension of the pinning mode on the unpinned line. Within this near-half-twist saddle expansion, a nonzero condensate therefore requires the pin to be relevant, and Eq.~\eqref{eq:gap} with the condensate added reads $N\phi_*^2=\mathcal A_\alpha/2\pi^2+(N-2)\sqrt{\varsigma_*}/8$ on the regular branch. On the unpinned line the spectators screen the tadpole of the twisted pair to order $N^{-2}$, so at order $1/N$ only the exchange of $\varsigma$ shifts the twisted modes. Projecting this self-energy onto the modes of $S^1\times\mathbb H^2$ gives, for the two lowest twisted modes of the unpinned partial twist,
\begin{equation}
 \Dh_s=\tfrac12+|s|+(4|s|-3)\big(\Delta_\phi-\tfrac12\big)+O(N^{-2}),\qquad |s|\in\{\alpha,1-\alpha\},
 \label{eq:pinlaw}
\end{equation}
where $\Delta_\phi-\tfrac12=4/(3\pi^2N)$ is the anomalous dimension of $\phi$ at this order.\footnote{We evaluate the self-energy on $S^1\times\mathbb H^2$ in two schemes with independent codes, a chordal cutoff with a finite-part fit and analytic continuation in the exponent of the auxiliary-field propagator. On untwisted lines both reproduce the shift $\Delta_\phi-\tfrac12$ for every mode with $s\neq0$, which fixes the normalization and the sign. The coefficients in Eq.~\eqref{eq:pinlaw} are numerical. In the continuation scheme they hold to $3\times10^{-11}$ for $\alpha=0.1$--$0.5$, and the cutoff scheme reproduces them at $\alpha=0.3$ and $\tfrac12$ within its fit uncertainty of about $10^{-3}$ in units of $\Delta_\phi-\tfrac12$. The higher modes do not follow this form. Relative to the untwisted shift, the defect part is $-4(\Delta_\phi-\tfrac12)\alpha(1-\alpha)/|s|$, which has the same dependence on $\alpha$ and $s$ as the exchange term of Eq.~\eqref{eq:oneloop}, whereas the tadpole term proportional to $n_t$ is screened.} At half twist $\Dh_{1/2}=\tfrac32-\Delta_\phi$, so the pin stays relevant, and
\begin{equation}
 \varsigma_*=\Delta_\phi-\tfrac12,\qquad \phi_*^2=\frac1{4\pi\sqrt{3N}}+O(N^{-1}).
 \label{eq:halftwistpin}
\end{equation}
The spectators thus stay above the Breitenlohner--Freedman bound, and the squared amplitude $\phi_*^2$ scales as $N^{-1/2}$ instead of vanishing. The leading-order loss of the condensate is therefore an artifact of the truncation, and the expansion in $1/N$ is nonuniform only for $\tfrac12-\alpha$ of order $1/N$. Since this condensate is parametrically small and is supported by the spectators' response, which $O(2)$ does not have, the model with every component twisted, which contains $O(2)$ at $N=2$, remains the large-$N$ counterpart of the lattice data.

\begin{table}[t]\centering\small
\begin{tabular}{lccccccc}\toprule
$\alpha$ & $\phi_*^2$ & $\Ct/N$ & $\mathfrak r_\infty$ & $\Delta_{+-}/N$ & $\Ct/\Ct|_{\alpha=0}$ & $\mathfrak r_\infty/\mathfrak r_\infty|_{\alpha=0}$ & $\Delta_{+-}/\Delta_{+-}|_{\alpha=0}$\\\midrule
0 & 0.044413 & 0.08883 & 1.5513 & 0.2756 & 1 & 1 & 1\\
0.1 & 0.043815 & 0.08763 & 1.5441 & 0.2706 & 0.987 & 0.995 & 0.982\\
0.2 & 0.041969 & 0.08394 & 1.5216 & 0.2554 & 0.945 & 0.981 & 0.927\\
0.3 & 0.038688 & 0.07738 & 1.4800 & 0.2290 & 0.871 & 0.954 & 0.831\\
0.4 & 0.033535 & 0.06707 & 1.4108 & 0.1892 & 0.755 & 0.909 & 0.687\\
0.5 & 0.025330 & 0.05066 & 1.2950 & 0.1312 & 0.570 & 0.835 & 0.476\\
\bottomrule\end{tabular}
\caption{The pinned monodromy line at large $N$, with all components twisted in complex pairs and the mode $s=\alpha$ of one pair pinned. $\phi_*^2$ is Eq.~\eqref{eq:phistaralpha}, $\Ct=2N\phi_*^2$, and $\mathfrak r_\infty$ and $\Delta_{+-}=2\Ct\mathfrak r_\infty$ come from the junction saddle at leading order. At $\alpha=\tfrac12$, $\Ct/N=1/2\pi^2$.}
\label{tab:monopinlargeN}
\end{table}
\subsection{The entropy of the pinned line at large \texorpdfstring{$N$}{N}}\label{app:pinentropy}

\begin{table}[t]\centering\small
\begin{tabular}{lccc}\toprule
 & walls relevant & \multicolumn{2}{c}{disordering allowed for $N$ below}\\
 & for $N$ below & leading order & with order $N^0$\\\midrule
$\alpha=0$, large $N$ & 3.6 & 4.5 & 4.3\\
$\alpha=0.2$, large $N$ & 3.9 & 5.0 & 4.9\\
$\alpha=0.3$, large $N$ & 4.4 & 5.7 & 5.8\\
$\alpha=0.4$, large $N$ & 5.3 & 7.6 & 8.0\\
$\alpha=0.5$, large $N$ & 7.6 & 14.6 & 15.8\\\midrule
$\alpha=0$, one loop at $\epsilon=1$ & none & \multicolumn{2}{c}{3.1}\\\bottomrule
\end{tabular}
\caption{Threshold estimates from walls and entropy for a candidate ordered line built from pinned monodromy lines with every component twisted. The first column solves $\Delta_{+-}=1$ with the saddle values of Table~\ref{tab:monopinlargeN}, at leading order. The other two solve $\log\mathfrak g_{\rm u}-\log\mathfrak g_{\rm pin}=\log2$, at leading order with Eq.~\eqref{eq:entropygap} and with the entropy term of order $N^0$ of Table~\ref{tab:pinentropy} added (Appendix~\ref{app:pinentropy}), and below them an ordered line may disorder into the unpinned line. The wall correction of order $N^0$ is not known. At one loop the wall of Table~\ref{tab:anchors} exceeds one at $\epsilon=1$ for every $N$, so that row has no wall threshold, and its entropy entry uses the entropy of the ordinary pinned line, $\log\mathfrak g_{\rm pin}=-(N+8)\epsilon/16$~\cite{Cuomo:2021kfm}, which holds for every $N$. Corrections of relative order $1/N$ or $\epsilon$ are otherwise large at these values of $N$.}
\label{tab:order}
\end{table}

On $S^1\times\mathbb H^2$ with the regularized volume $\mathrm{Vol}(S^1\times\mathbb H^2)=-4\pi^2$, the free energy of the fields at a constant background $\varsigma$ obeys $\partial_\varsigma F=\tfrac12\mathrm{Vol}\sum_a\delta g_a(\varsigma)$, where $\delta g_a$ is the coincident shift of component $a$ relative to the bulk. For one twisted complex component, $\delta g=-\mathcal A_\alpha(\varsigma)/4\pi^2$ per real component, with
\begin{equation}
 \mathcal A_\alpha(\varsigma)=\sum_{s\in\mathbb Z+\alpha}\Big[\operatorname{dg}\big(\tfrac12+\sqrt{s^2+\varsigma}\big)-\log|s|\Big]+\log(2\sin\pi\alpha),
 \label{eq:Asigma}
\end{equation}
and the untwisted limit $\mathcal A_0(\varsigma)=\sum_{n\in\mathbb Z}[\operatorname{dg}(\tfrac12+\sqrt{n^2+\varsigma})-\operatorname{dg}(\tfrac12+|n|)]$. Eq.~\eqref{eq:Asigma} reduces to Eq.~\eqref{eq:Aalpha} at $\varsigma=0$, its zero is the gap equation~\eqref{eq:largeNgap}, and at $\varsigma_{\rm pin}=\tfrac14-\alpha^2$ it gives $4\pi^2\phi_*^2$ of Eq.~\eqref{eq:phistaralpha}. Since $\mathcal A_\alpha$ increases with $\varsigma$, it is positive between its zero and $\varsigma_{\rm pin}$. Hence Eq.~\eqref{eq:dloggdsigma}, and per complex component $\log\mathfrak g(\varsigma)=\log\mathfrak g_{\rm free}(\alpha)-\int_0^\varsigma\mathcal A_\alpha$. On the pinned saddle the classical action is $\tfrac12\mathrm{Vol}\,N\phi_*^2(\alpha^2+\varsigma_{\rm pin}-\tfrac14)=0$, and the fluctuations of the pinned pair enter only at $O(N^0)$, which gives Eq.~\eqref{eq:entropygap}.

The evaluation reproduces the full-twist values $\varsigma_{\rm u}=-0.168374$ and $\log\mathfrak g_{\rm u}/(N/2)=0.1617$ of Appendix~\ref{app:largeM} at half twist, the $\phi_*^2$ column of Table~\ref{tab:monopinlargeN} at every twist, and at $\alpha=0$ the large-$N$ entropy $\log\mathfrak g_{\rm pin}=-0.153673\,N$ of Ref.~\cite{Cuomo:2021kfm} to five digits. Table~\ref{tab:pinentropy} lists the entropy released by pinning. A flow from the unpinned line into the ordered line would be triggered by an operator that is even under the exchange of the two pinnings and breaks the continuous symmetry. On the trivial line the lowest one is the traceless symmetric tensor, with $\Delta_t>1$~\cite{Chester:2019ifh,Chester:2020iyt}. On the $O(2)$ monodromy line it carries charge two, and in the spin-one regulator at $\Nm=9$ the lowest charge-two level lies between $1.44$ at $\alpha=0.02$ and $1.94$ at half twist in displacement units.

\begin{table}[t]\centering\small
\begin{tabular}{lcccccc}\toprule
$\alpha$ & 0 & 0.1 & 0.2 & 0.3 & 0.4 & 0.5\\\midrule
\multicolumn{7}{l}{Unpinned saddle background}\\\midrule
$\varsigma_{\rm u}$ & 0 & $-0.00996$ & $-0.03929$ & $-0.08543$ & $-0.13905$ & $-0.16837$\\
\midrule\multicolumn{7}{l}{Leading entropy difference per component, $(\log\mathfrak g_{\rm u}-\log\mathfrak g_{\rm pin})/N$}\\\midrule
full twist & 0.153675 & 0.150325 & 0.139905 & 0.12111 & 0.09132 & 0.04749\\
partial twist, spectators & 0.153675 & 0.14498 & 0.11985 & 0.081345 & 0.035805 & 0\\\midrule
\multicolumn{7}{l}{Order-$N^0$ correction to $\log\mathfrak g_{\rm u}-\log\mathfrak g_{\rm pin}$}\\\midrule
full twist & $+0.0294$ & $+0.0255$ & $+0.0131$ & $-0.0092$ & $-0.0399$ & $-0.0549$\\\bottomrule
\end{tabular}
\caption{The unpinned saddle background and the entropy released by pinning, $(\log\mathfrak g_{\rm u}-\log\mathfrak g_{\rm pin})/N$, at leading order in $1/N$, from Eqs.~\eqref{eq:entropygap} and~\eqref{eq:Asigma}, with every component twisted (full twist) and for the spectator part of a partial twist. For the partial twist the spectators carry the part proportional to $N$, which vanishes at half twist together with the leading condensate. There the background $\varsigma_*=\Delta_\phi-\tfrac12$ of Eq.~\eqref{eq:halftwistpin} leaves a classical part of order $N^{-1/2}$. The last row is the term of order $N^0$ of $\log\mathfrak g_{\rm u}-\log\mathfrak g_{\rm pin}$ with every component twisted, not divided by $N$, from the fluctuation determinants of both lines.}
\label{tab:pinentropy}
\end{table}

For the full twist, the next-order entropy can be written in the same bubble normalization as Eq.~\eqref{eq:spectral}. For a background $\varsigma$, put
\begin{equation}
 \Delta_s(\varsigma)=\tfrac12+\sqrt{s^2+\varsigma},\qquad
 U_\kappa^{(\alpha,\varsigma)}(\nu)=
 \sum_{s\in\mathbb Z+\alpha}\mathcal B_{\Delta_s(\varsigma),\Delta_{s-\kappa}(\varsigma)}(\nu).
\end{equation}
The unpinned determinant has no condensate mixing, while the pin winds with the mode $s=\alpha$. In terms of the untwisted kernel $\widetilde\Pi_{0,\kappa}$ of Eq.~\eqref{eq:spectral}, the two ratios are
\begin{equation}
 \begin{split}
 \mathscr R_{{\rm u},\kappa}&=
 \frac{U_\kappa^{(\alpha,\varsigma_{\rm u})}}{\widetilde\Pi_{0,\kappa}},\\
 \mathscr R_{{\rm pin},\kappa}&=
 \frac{U_\kappa^{(\alpha,\varsigma_{\rm pin})}+2\pi\phi_*^2
 \left[\dfrac1{\nu^2+(\kappa+\alpha)^2+\varsigma_{\rm pin}}
 +\dfrac1{\nu^2+(\kappa-\alpha)^2+\varsigma_{\rm pin}}\right]}
 {\widetilde\Pi_{0,\kappa}}.
 \end{split}
 \label{eq:pindetratios}
\end{equation}
For $X={\rm u},{\rm pin}$ the term of order $N^0$ is
\begin{equation}
 (\log\mathfrak g_X)^{(1)}=
 \frac12\lim_{\Lambda\to\infty}\sum_{\kappa\in\mathbb Z}
 \int_0^\infty\!\mathrm d\nu\,\nu\tanh(\pi\nu)
 F_{\rm cut}\!\left(\frac{\kappa^2+\nu^2}{\Lambda^2}\right)
 \log\mathscr R_{X,\kappa}(\nu).
 \label{eq:pindettrace}
\end{equation}
This fixes the defect-free subtraction, the mixing normalization and the common isotropic cutoff prescription. At zero twist the unpinned ratio is one, and the mixing term of the pinned ratio becomes $4\pi\phi_*^2/(\nu^2+\kappa^2+\tfrac14)$. The prescription has the same ultraviolet qualification as Appendix~\ref{app:analytic}; the two-cutoff comparison below tests its numerical evaluation.

For the partial twist the spectators carry the part proportional to $N$, so near half twist the entropy of the pinned partial twist, which survives by Eq.~\eqref{eq:halftwistpin}, approaches that of the unpinned line, at least in the classical part of the saddle. Near four dimensions the one-loop entropy of the ordinary pinned line is known for every $N$, $\log\mathfrak g_{\rm pin}=-(N+8)\epsilon/16$~\cite{Cuomo:2021kfm}. Its term of order $N^0$ has the sign of the leading term, and at $\epsilon=1$ the entropic threshold falls to $16\log2-8\simeq3.1$. In three dimensions the term of order $N^0$ is the fluctuation determinant of $\varsigma$ around the pinned saddle, $-\tfrac12\log\det(\tfrac12G^2+\phi_*^2G)/\det(\tfrac12G_0^2)$, where $G$ is the propagator at background $\varsigma_{\rm pin}$, $G_0$ the propagator without the defect, and the term $\phi_*^2G$ the mixing of $\varsigma$ with the condensate. In each circle mode it is a sum of AdS$_2$ bubbles, as in Appendix~\ref{app:largeM}. The gap equation cancels its leading large-momentum term, and the next term is a traceless stress-tensor structure that falls as the third power of the momentum, so a regulator that depends on momentum only through the $S^1\times\mathbb H^2$ Laplacian gives a finite result without counterterms, as for Eq.~\eqref{eq:cresult}. At zero twist we find $\log\mathfrak g_{\rm pin}=-0.153673\,N-0.0294(1)+O(1/N)$, with Gaussian and power-law regulators agreeing to the quoted digit, and a one-percent change of the mixing coefficient destroys the plateau in the cutoff. The correction has the sign of the one-loop term $-\epsilon/2$ but is much smaller, and it moves the entropic threshold of the ordinary pinned line from $4.5$ to $4.3$. At nonzero twist both lines contribute at this order. The unpinned line has the same determinant without the mixing term, at its background $\varsigma_{\rm u}$, and on the pinned line the condensate winds, so the mixing couples $\varsigma$ in the circle mode $\kappa$ to the twisted propagator in the modes $\kappa\pm\alpha$ and averages the two. Both cutoff dependences reach a plateau, the two regulators agree to $3\times10^{-4}$ or better, and at zero twist the same evaluation reproduces $-0.0294$. The resulting term of order $N^0$ of the entropy released by pinning, in the last row of Table~\ref{tab:pinentropy}, changes sign between $\alpha=0.2$ and $0.3$. It lowers the entropic threshold at small twist and raises it at large twist, and the threshold stays above the leading wall threshold at every twist evaluated (Table~\ref{tab:order}). At $N=2$ the entropy released by pinning is then $0.307+0.029=0.337$, about half of $\log2=0.693$. The fuzzy-sphere overlap values of Ref.~\cite{Yang:2026qtk}, $\mathfrak g_{\rm pin}=0.585$, $0.490$ and $0.439$ for $N=2$, $3$ and $4$, give $\log\mathfrak g_{\rm pin}=-0.536$, $-0.713$ and $-0.823$. These lie below the large-$N$ values with the term of order $N^0$ by about $0.2$ at each $N$, so the higher orders in $1/N$ are large at these values of $N$, or the overlap values carry systematic errors that are not quoted. The corresponding entropy release, $-\log\mathfrak g_{\rm pin}$, crosses $\log2$ between $N=2$ and $N=3$.

\subsection{Pins along the two branches}\label{app:pinbranches}
Two pins of the $O(3)$ half-twist can be followed along both branches of the anisotropy flow. The charged pin is that of Appendix~\ref{app:monopin}. The neutral pin couples to the untwisted component at the poles,
\begin{equation}
 -h_3\left[(\mathsf f_{{\rm N}0}^\dagger\mathsf f_{{\rm N}3}+{\rm h.c.})+\eta_3(\mathsf f_{{\rm S}0}^\dagger\mathsf f_{{\rm S}3}+{\rm h.c.})\right],\qquad \eta_3=\pm1,
 \label{eq:pin3}
\end{equation}
in the notation of Eq.~\eqref{eq:pin}, and we take $h_3\to\infty$ by projecting both pole orbitals onto the aligned combination. This pin conserves the $U(1)$ charge of the twisted pair and $L_z$, and $\eta_3=-1$ creates the wall. The Hamiltonian is that of the pinned flow of Sec.~\ref{sec:monopin}, on the tuned trajectory with the local potential subtracted, supplemented by the frozen easy-axis limit $g=3$ of Sec.~\ref{sec:core} with the bulk unit taken from $\sigma$ and its first descendant at the same couplings. At the $O(3)$ point the construction reproduces the spectator-pinning energies of Appendix~\ref{app:pinning} to machine precision. Two references fix what the endpoint should be. The same pin without the flux measures what remains of the twist, and the pinned line of the Ising Hamiltonian of Eq.~\eqref{eq:isingH}, with both pole orbitals fixed to spin up~\cite{Hu:2023ghk,Zhou:2023fqu} and the bulk unit defined in the same way, is an independent realization of the endpoint. Table~\ref{tab:pinbranches} collects the results. The easy-plane ratios in Fig.~\ref{fig:pins}(c,d) use the pinned $O(2)$ line at the same size. Applying the same code to the $O(2)$ model reproduces its saved data to all printed digits. Keeping the local potential at the flux, or reweighting the second pin bilinear by $0.7$ or $1.4$ in both models, changes the endpoint ratios by at most $0.1\%$.

\begin{table}[t]\centering\small
\begin{tabular}{ccccccccc}\toprule
 & \multicolumn{2}{c}{twist effect on wall} & \multicolumn{2}{c}{wall, frozen limit} & \multicolumn{2}{c}{lowest level} & \multicolumn{2}{c}{easy plane, $g=-0.09$}\\
$\Nm$ & $g=0$ & $g=0.09$ & $O(3)$ & Ising & $O(3)$ & Ising & wall & charge-one excess\\\midrule
6 & 0.0016 & 0.0003 & 0.750 & 0.774 & 1.534 & 1.572 & 0.139 & 0.204\\
7 & 0.0022 & 0.0004 & 0.751 & 0.777 & 1.536 & 1.577 & 0.095 & 0.178\\
8 & 0.0028 & 0.0005 & 0.753 & 0.781 & 1.540 & 1.582 & 0.067 & 0.158\\
9 & 0.0033 & 0.0005 & 0.755 & 0.783 & 1.543 & 1.587 & 0.048 & 0.143\\\bottomrule
\end{tabular}
\caption{The $\phi_3$-pinned half-twist along the two branches, by exact diagonalization with infinite pole fields, in bulk units. The twist effect is the wall of the $\phi_3$-pinned half-twist minus that of the same pin without the flux. In the frozen easy-axis limit the two agree to $2\times10^{-5}$ in the wall and in the levels. The frozen-limit columns compare the $\phi_3$-pinned half-twist at $g=3$ with the pinned line of Eq.~\eqref{eq:isingH}, and the lowest level is the first excitation of the pinned line at zero transverse spin. The last two columns follow the easy-plane branch, where the charge-one excess is the lowest charge-one level of the pinned line minus that of the unpinned line at the same couplings.}
\label{tab:pinbranches}
\end{table}

On the easy-axis branch the $\phi_3$ pin survives and the twist disappears. The twist effect on the wall falls along the flow and vanishes in the frozen limit, where the pinned line no longer depends on the flux. Its wall and lowest level lie about $3\%$ below those of the Ising Hamiltonian at every size, and both drift upward together. The offset is about the amount by which the Ising Hamiltonian, whose couplings are not tuned to the light-field condition of Eq.~\eqref{eq:flowtune}, places $\sigma$ above its dimension, $0.532$ against $0.519$ in the same unit. The ratio of the wall to the lowest level, $0.489$ against $0.493$ at $\Nm=8$, does not depend on the unit and agrees to $1\%$. The Ising wall still rises at $\Nm=12$, where it reaches $0.790$, toward the overlap value $0.84(5)$ of Ref.~\cite{Zhou:2023fqu}. The lowest level of the Ising pinned line, $1.598$ at $\Nm=12$, likewise approaches the dimension of the leading defect operator, $1.6081$ from the endpoint bootstrap~\cite{Lanzetta:2025xfw} and $1.63(6)$ on the fuzzy sphere~\cite{Hu:2023ghk}.

On the easy-plane branch the $\phi_3$ pin decouples. Its wall falls with $\Nm$ at fixed coupling. At a finite field $h_3=\gamma_B$ the charge-one level at $g=-0.09$ lies within $3.5\%$ of the unpinned line, whereas the same field raises it by a quarter at the $O(3)$ point. At infinite field the locked pole orbitals leave the charge-one excess of Table~\ref{tab:pinbranches}, which falls roughly as $\Nm^{-0.9}$.

The charged pin behaves the other way round. Along the easy-axis branch at $h_{\rm pin}=1000$ its wall falls from $0.48$ at $g=0.01$ to $0.043$ in the frozen limit at $\Nm=8$, and in the frozen limit it falls as $0.21$, $0.10$ and $0.043$ at $\Nm=6$, $7$ and $8$. Both stiffness estimators vanish, and the Ward matrix element is exactly zero, so the $U(1)$ is no longer broken. At this infinite lattice field the pole orbitals are locked into states of the twisted pair, which removes them from the Ising dynamics. The remnant is even under $\phi_3\to-\phi_3$ and couples to $\varepsilon$, which is irrelevant on the line, but it lifts the lowest even level to $3.2$--$3.6$ at these sizes, and its effect decays only as $\Nm^{-(\Delta_\varepsilon-1)/2}$. A weaker pin isolates the decoupling. At $h_{\rm pin}=3$ the same pin opens a wall of $0.54$--$0.58$ at the $O(3)$ point for $\Nm=6$ and $7$, and the wall falls along the easy-axis flow to about $10^{-4}$ in the frozen limit, where the even spectrum agrees with the defect-free one to $0.02$.

\section{Four-sphere construction and higher-dimensional tests}\label{app:s4}
The five-dimensional construction of Sec.~\ref{sec:five} needs a target and a regulator. The target is the exact free correlator of the regular half-twist, and the regulator is a finite twisted projector on the fuzzy four-sphere with local interactions.

\subsection{Exact regular Green function}\label{app:green}
Regular free modes give $\Dh_s=3/2+|s|$ by Eq.~\eqref{eq:freespec}, hence $\Dh_{1/2}=2$ and $\Dh_{3/2}=3$. Parallel derivatives raise the dimension by one, and the displacement has $\Dh_D=4$. The full free correlator tests the spectrum and its mode normalizations together. With
\begin{equation}
 \xi=\frac{|y_1-y_2|^2+(r_1-r_2)^2}{4r_1r_2},\qquad
 Z=\frac{\cos(\theta_{12}/2)}{\sqrt{1+\xi}},
 \label{eq:crossratios}
\end{equation}
where $\theta_{12}=\theta_1-\theta_2$ is the relative azimuth, the regular half-twist ratio in general dimension is
\begin{equation}
 \frac{G_\Dc}{G_0}=F_d(Z)=
 \frac{\int_0^Z(1-t^2)^{(d-4)/2}\,\mathrm dt}
 {\int_0^1(1-t^2)^{(d-4)/2}\,\mathrm dt}.
 \label{eq:exact}
\end{equation}
The signed integral implements the two sheets and gives
\begin{equation}
 F_3=\frac2\pi\arcsin Z,\qquad
 F_5=\frac2\pi\big(\arcsin Z+Z\sqrt{1-Z^2}\big).
 \label{eq:green35}
\end{equation}
The derivation below uses only the free equation and the regular boundary condition~\cite{Giombi:2021uae,Dowker:2022mex}, and Fig.~\ref{fig:fiveconstruction}(a) shows how the ratio changes with dimension while the monodromy stays the same.

For a canonically normalized real free scalar,
\begin{equation}
 G_0(x_1,x_2)=\frac{\Gamma(d/2-1)}{4\pi^{d/2}|x_1-x_2|^{d-2}}.
\end{equation}
Write $G_\Dc=(r_1r_2)^{-\Delta_{\rm free}}\widetilde G(\xi,\theta)$, with $\Delta_{\rm free}=(d-2)/2$ the free-field dimension, $\xi$ the cross-ratio of Eq.~\eqref{eq:crossratios} and $\theta=\theta_{12}$. Away from coincidence the free equation becomes
\begin{equation}
 \left[\xi(1+\xi)\partial_\xi^2+(d-1)(\xi+\tfrac12)\partial_\xi
 +\partial_\theta^2+\Delta_{\rm free}^2\right]\widetilde G=0.
\end{equation}
Substituting $\widetilde G\propto[1+\xi-\cos^2(\theta/2)]^{-\Delta_{\rm free}}F(Z)$ leaves
\begin{equation}
 F''(Z)+\frac{(d-4)Z}{1-Z^2}F'(Z)=0.
\end{equation}
The regular half-twist fixes $F(0)=0$, while the coincident singularity fixes $F(1)=1$, and integrating gives Eq.~\eqref{eq:exact}, including its signed continuation to the other sheet.

\subsection{The finite twisted projector and local interactions}
\begin{figure}[t]\centering
\includegraphics[width=\textwidth]{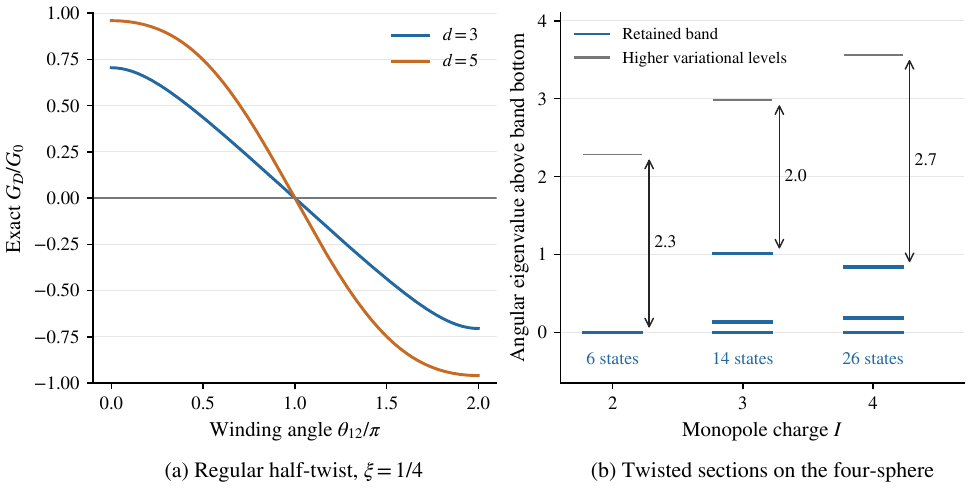}
\caption{Regular half-twist on the four-sphere. (a) Exact free correlator ratios at fixed cross-ratio in three and five dimensions, whose sign a full winding reverses. (b) Angular-form eigenvalues in the twisted trial spaces, measured from each band bottom. Blue levels are retained as orbitals, and arrows mark the gap to the next variational level. These finite bands define the projectors used in the many-body Hamiltonian.}
\label{fig:fiveconstruction}
\end{figure}

Write $x=(y,w)$ with $y\in\mathbb R^3$ and $w=x_3+\mathrm ix_4$ transverse to the defect. The untwisted Yang-monopole lowest Landau level is
\begin{equation}
 \Hilb_I=\operatorname{Sym}^I\mathbb C^4,
 \qquad N_I=\binom{I+3}{3},
 \label{eq:s4lll}
\end{equation}
where $I/2$ is the monopole isospin~\cite{Qian:2026fbx}. We define the half-twist on the complement of the great sphere $S^2_\Dc=\{x\in S^4:x_3=x_4=0\}$. The odd flavour is a section of the monopole bundle tensored with the flat line bundle of holonomy $-1$ around $S^2_\Dc$, while the even flavour is untwisted, and the regular domain retains $r^{|s|}e^{\mathrm is\theta}$ near $S^2_\Dc$ and excludes $r^{-|s|}$. Using Hopf coordinates $z\in S^7$ with $x_i=z^\dagger\mathbf\Gamma_iz$, where $\mathbf\Gamma_i$ are the five Euclidean Clifford matrices, we take finite trial sections
\begin{equation}
 \begin{aligned}
 \Phi(z)&=\sum_{|\mathbf k|=I}z^{\mathbf k}\,\Phi_{\mathbf k}(x),\\
 \Phi_{\mathbf k}(x)&\in\operatorname{span}\{w^{1/2}x^{\mathbf m},\bar w^{1/2}x^{\mathbf m}\},
 \end{aligned}
 \label{eq:trialsections}
\end{equation}
where $x^{\mathbf m}$ runs over monomials of bounded degree in the base coordinates and the torus charges are those required by the bundle. The square-root factors fix the monodromy and the regular behaviour, and since they depend only on the base coordinates they preserve the fibre transformation law of the untwisted sections. For a trial basis $\{\Phi_a\}$, the Gram matrix and covariant angular form
\begin{equation}
 \mathcal S_{ab}=\langle \Phi_a,\Phi_b\rangle,\qquad
 \mathcal J_{ab}=\sum_{i<i'}\langle J_{ii'}\Phi_a,J_{ii'}\Phi_b\rangle,
 \qquad \mathcal Jv=\varpi\mathcal Sv,
 \label{eq:galerkin}
\end{equation}
define a Hermitian variational problem for the coefficient vectors $v$ and eigenvalues $\varpi$ in the prescribed twisted domain, where $J_{ii'}$ are the covariant angular derivatives for the monopole bundle. A separated band of this problem defines the finite orbital projector, with $6$, $14$ and $26$ orbitals at $I=2,3,4$ [Fig.~\ref{fig:fiveconstruction}(b)], and local interactions specified before projection give a finite Hermitian many-body Hamiltonian with the prescribed holonomy.

The square-root factors also have finite first-derivative norm, so they are compatible with the angular kinetic form. For the lowest mode the gradient has size $r^{-1/2}$, whose squared norm is integrable with the transverse measure $r\,\mathrm dr\,\mathrm d\theta$. The Gram quotient and angular form therefore define a Hermitian variational problem before any many-body approximation. The lowest selected bands at $I=2,3,4$ have widths $0,1.014,0.838$ and gaps of about $2.3,2.0,2.7$ to the next variational level, and they are flattened after selection. These bands contain complete multiplets of the parallel rotations, and the projected generators satisfy their rotation algebra. The quoted orbital counts refer to these finite variational bands; their continuation to larger $I$ requires a band-convergence study. A twist built from polynomials on $\CP^3$ instead fails for a geometric reason. A homogeneous polynomial of positive degree restricts on each twistor fibre to a section of a line bundle of positive degree on $\CP^1$, which either vanishes identically or has a zero, so its zero set projects onto all of $S^4$ rather than only the desired $S^2$, and its square root cannot localize the twist on the great sphere.

For isospin components $M=-I/2,\ldots,I/2$, let
\begin{equation}
 \mathcal P_{I_pM_p}(x)=\sum_{M_1,M_2}
 \langle\tfrac I2M_1;\tfrac I2M_2|I_pM_p\rangle
 \mathsf f_{\uparrow,M_1}(x)\mathsf f_{\downarrow,M_2}(x),
\end{equation}
where $I_p$ and $M_p$ label the pair isospin and its component. The interactions are
\begin{align}
 \mathcal V_{I_p}&=\int \mathrm d\Omega\sum_{M_p}\mathcal P_{I_pM_p}^\dagger\mathcal P_{I_pM_p},\nonumber\\
 \widetilde{\mathcal V}_{I_p}&=\int \mathrm d\Omega\sum_{i<i',M_p}
 (J_{ii'}\mathcal P_{I_pM_p})^\dagger J_{ii'}\mathcal P_{I_pM_p},\nonumber\\
 \mathcal V_\nabla&=\int \mathrm d\Omega\,{:}[\mathsf n^0(-\nabla^2)\mathsf n^0-\mathsf n^z(-\nabla^2)\mathsf n^z]{:},\nonumber\\
 H_{S^4}&=\sum_{I_p}(U_{I_p}\mathcal V_{I_p}+\widetilde U_{I_p}\widetilde{\mathcal V}_{I_p})
 +U_\nabla\mathcal V_\nabla-h\sum_m(\mathsf n_{m+}-\mathsf n_{m-}).
 \label{eq:s4H}
\end{align}
Here $\mathsf f_{\uparrow,\downarrow}=(\mathsf f_+\pm\mathsf f_-)/\sqrt2$, $m$ labels the orbitals and $\mathsf n_{m\pm}$ their occupations in the two flavours, the $U$'s are couplings, and $\mathsf n^0,\mathsf n^z$ sum the corresponding densities over the isospin index. The measure is the unit-$S^4$ measure divided by $4\pi^2$, and the derivative terms are the checked first-derivative quadratic forms. Higher-derivative local products would require a new domain analysis, because square-root sections need not have square-integrable second derivatives. The principal $I=3$ model, denoted A, has $(U_3,U_2,U_1,U_\nabla)=(1,0.38,2,0.163)$, and at $I=4$ model A has $(U_4,U_3,U_2,U_\nabla)=(1,0.38,2,0.0864)$. Model B at $I=3$ replaces $U_1$ by $\widetilde U_1=0.4$ and uses $U_\nabla=0.3375$, and all unlisted couplings vanish. The filling is $N_e=N_I$, with the even flavour on $N_I$ orbitals and the odd flavour on the selected twisted band. At the tested $I=2$ cutoff the map from six local couplings to bulk pseudopotentials has rank five, so equal untwisted pseudopotentials do not determine the defect Hamiltonian uniquely, and a comparison across cutoffs must track the local interaction.

\subsection{Spatial resolution and the fibre modes}\label{app:s4locality}
The untwisted lowest Landau level supplies a check of spatial resolution independent of the many-body dynamics. Let $\mathbb P_I$ project onto $\Hilb_I$ and let $Y_n$ be any scalar harmonic of degree $n$ on the unit $S^4$, normalized by $\int\mathrm d\mu\,|Y_n|^2=1$, with $\mathrm d\mu=\mathrm d\Omega/\operatorname{Vol}(S^4)$. Its projected multiplication operator is $\mathsf Y_n^{(I)}=\mathbb P_IY_n\mathbb P_I$. The normalized density weight is
\begin{equation}
 f_n(I)\equiv\frac1{N_I}\Tr_{\Hilb_I}\!\left[\mathsf Y_n^{(I)\dagger}\mathsf Y_n^{(I)}\right]
 =\frac{I!\,(I+3)!}{(I-n)!\,(I+n+3)!},\qquad 0\le n\le I.
 \label{eq:s4densityweight}
\end{equation}
In particular $f_0=1$, and at fixed $n$,
\begin{equation}
 f_n(I)=1-\frac{n(n+3)}{I}+O(I^{-2}).
 \label{eq:s4densitylimit}
\end{equation}
The coefficient is the scalar Laplacian eigenvalue on $S^4$. Thus every fixed harmonic recovers its continuum weight as $I$ grows.\footnote{For a zonal harmonic take $Y_n$ proportional to the Gegenbauer polynomial $C_n^{(3/2)}(2t-1)$, with $t=|z_1|^2+|z_2|^2$ on $S^7$. A normalized monomial with $k$ powers in the first two components induces the beta distribution with parameters $(k+2,I-k+2)$, and there are $(k+1)(I-k+1)$ such monomials. Averaging the polynomial in each distribution and summing the squared diagonal matrix elements gives Eq.~\eqref{eq:s4densityweight}. The harmonic norm uses the marginal measure $6t(1-t)\,\mathrm dt$.} This check concerns the untwisted projection and its spatial resolution; it does not determine the many-body spectrum in the twisted band.

The full matrix algebra on $\Hilb_I$ also contains harmonics of the $\CP^1$ fibre in the twistor description $\CP^3\to S^4$~\cite{Medina:2002pc}. Their energies are dynamical. To recover the intended scalar theory, these additional modes must decouple from its low-energy sector as the cutoff is removed. The local pair-isospin couplings of Eq.~\eqref{eq:s4H} distinguish interactions that act differently on these modes while preserving the rotations of the untwisted four-sphere. Establishing their decoupling requires tracking the representations and matrix elements along a sequence of critical Hamiltonians, in addition to checking the scalar levels.

\subsection{Spectra, correlators and the Ward identity}\label{app:s4tests}
The free real scalar in five dimensions has defect primaries at $2$ and $3$ for transverse spins $1/2$ and $3/2$, and parallel descendants of the lowest mode at $3$ and $4$ for parallel spins $s_\parallel=1,2$. Table~\ref{tab:fivelevels} lists selected levels with the displacement candidate fixed at four.\footnote{This calibration alone can make the lowest mode look conformal. Whenever $E_D-E_0\simeq2(E_{\hat\psi_{1/2}}-E_0)$, it gives $4(E_{\hat\psi_{1/2}}-E_0)/(E_D-E_0)\simeq2$, even without a conformal descendant tower. The bulk calibration, parallel-spin assignments and matrix elements therefore supply independent tests.} The bulk field is selected near $(E_\phi-E_0)/(E_{\partial\phi}-E_\phi)=3/2$, which gives $h\simeq4.5,8.2,5.7$ for $I=3$ A, $I=3$ B and $I=4$ A, and the $I=4$ spin search was performed at the nearby field $h=6$. The parallel Casimir $L_\parallel^2=s_\parallel(s_\parallel+1)$ distinguishes a spin-two state from a spin-three state with the same azimuthal charge.

\begin{table}[t]\centering\small
\begin{tabular}{lccccc}\toprule
model & field $h$ & $\hat\psi_{1/2}$ & $\hat\psi_{3/2}$ & $s_\parallel=1$ candidate & $s_\parallel=2$ candidate\\\midrule
free target & --- & 2 & 3 & 3 & 4\\
$I=3$, A & 4.5 & 2.047 & 3.050 & 3.316 & 4.085\\
$I=3$, B & 8.2 & 2.06 & 2.88 & 3.14 & 3.80\\
$I=4$, A & 6.0 & 2.018 & 3.005 & 3.374 & 4.999\\\bottomrule
\end{tabular}
\caption{Selected five-dimensional half-twist levels with the candidate displacement fixed at four. The spin-two entries for model A use a parallel-Casimir search. At $I=4$ a lower state near $4.619$ has $L_\parallel^2\simeq11.983$ and is spin three, whereas the displayed candidate has $L_\parallel^2\simeq6.045$.}
\label{tab:fivelevels}
\end{table}

The primaries are close to their targets, but the states are not yet those of the free defect. The squared bulk-to-defect coefficients $\beta_s^2=\Gamma(3/2+|s|)/[\Gamma(3/2)\Gamma(1+|s|)]$ have ratio $\beta_{3/2}^2/\beta_{1/2}^2=4/3$, while the lattice overlaps at $I=3$ give about $0.78$, and the squared overlaps grow with the transverse distance $r$ from $S^2_\Dc$ as $r^{1.72}$ and $r^{3.53}$ instead of $r$ and $r^3$. The normalized equal-time defect correlator deviates from Eq.~\eqref{eq:green35} by a median $0.10$ at $I=2$, and by $0.15$--$0.22$ at $I=3$ away from the region where the lattice bulk correlator passes through zero. The tilt carries $3.8\%$ of its exact weight on the displacement family at $I=2$ by exact diagonalization, and $19\%$ and $11\%$ at $I=3$ and $4$ by DMRG.

\subsection{A parallel conformal-generator test}\label{app:s4generator}
A conformal generator tests which states belong to a descendant tower. On the unit cylinder, let $\mathfrak e_I(x)$ be a microscopic energy density satisfying $H_{S^4}=\int\mathrm d\Omega\,\mathfrak e_I(x)$ and form its first moment along a direction parallel to the defect,
\begin{equation}
 \mathsf G_5=\int\mathrm d\Omega\,x_5\mathfrak e_I(x).
 \label{eq:s4generator}
\end{equation}
Here $x_5$ is one of the three embedding coordinates along the great $S^2$. When the density approaches the conformal stress tensor, this moment approaches a common normalization times $P_5+K_5$, where $P_5$ and $K_5$ generate translations and special conformal transformations parallel to the defect~\cite{Fardelli:2024qla}. For a unit-normalized parallel-scalar primary $|\hat\psi\rangle$, $K_5|\hat\psi\rangle=0$ and $\|P_5|\hat\psi\rangle\|^2=2\Dh_\psi$, so the exact generator produces only its first parallel descendant, at $\Dh_\psi+1$.

For a finite-cutoff primary candidate we measure the fraction of the off-diagonal weight carried by each other eigenstate $|k\rangle$,
\begin{equation}
 p_k^{(5)}=\frac{|\langle k|\mathsf G_5|\hat\psi\rangle|^2}
 {\langle\hat\psi|\mathsf G_5^2|\hat\psi\rangle-|\langle\hat\psi|\mathsf G_5|\hat\psi\rangle|^2},
 \qquad |k\rangle\ne|\hat\psi\rangle.
 \label{eq:s4generatorshare}
\end{equation}
This fraction removes the overall normalization of the moment. Table~\ref{tab:s4generator} compares the state carrying the largest resolved weight with the first-descendant level expected from the measured primary. The weight is shared among several states, and the largest contribution is displaced from the expected level. At the smaller cutoff, exact diagonalization and DMRG agree on this distribution. At $I=3$ the energy levels can also be compared with the larger-bond-dimension calculation of Table~\ref{tab:fivelevels}, but convergence of the matrix elements has not yet been tested.

\begin{table}[t]\centering\small
\begin{tabular}{llrrrr}\toprule
cutoff & energy-density moment & $\bar E_\psi$ & $\bar E_\psi+1$ & $\bar E_{k_{\max}}$ & $p_{k_{\max}}^{(5)}$\\\midrule
$I=2$ & original & 1.983 & 2.983 & 2.715 & 0.486\\
 & vacuum-improved & 1.983 & 2.983 & 2.673 & 0.579\\
$I=3$, A & original & 2.045 & 3.045 & 3.306 & 0.215\\\bottomrule
\end{tabular}
\caption{Parallel-generator diagnostic. Here $\bar E_k=4(E_k-E_0)/(E_D-E_0)$ uses the displacement candidate, and $k_{\max}$ is the state with the largest resolved fraction in Eq.~\eqref{eq:s4generatorshare}. The $I=2$ control has $(U_2,U_1,\widetilde U_1,h)=(1.595622,1.668390,-0.134185,8.3)$, with all other couplings zero, and uses the complete exact-diagonalization spectrum. Its improved moment is fixed using only the defect vacuum, leaving the $\hat\psi$ sector as a separate test. The $I=3$ run uses model A at $h=4.5$ and DMRG bond dimension $400$, resolving nine excited states that together carry $0.448$ of the total off-diagonal weight. Its quoted maximum is therefore a maximum within this resolved set.}
\label{tab:s4generator}
\end{table}

This diagnostic depends on the local energy density as well as on the Hamiltonian. A scalar improvement $\delta\mathfrak e_I=u_{\rm imp}\nabla^2\mathcal O$ integrates to zero when its surface terms vanish, but on the unit $S^4$ it changes the moment by
\begin{equation}
 \delta\mathsf G_5=-4u_{\rm imp}\int\mathrm d\Omega\,x_5\mathcal O,
 \label{eq:s4improvement}
\end{equation}
since $\nabla^2x_5=-4x_5$. Thus a density improvement can change the descendant weights without moving any energy level.\footnote{The original moment uses the first-derivative quadratic forms of Eq.~\eqref{eq:s4H} as its density prescription. For example, replacing the density-gradient form by the symmetrized density-Laplacian form changes $\mathsf G_5$ by $2U_\nabla\int\mathrm d\Omega\,x_5\,{:}[(\mathsf n^0)^2-(\mathsf n^z)^2]{:}$, while leaving $H_{S^4}$ unchanged. The integration by parts requires vanishing boundary flux at the excluded great sphere; the regular sections used here satisfy this condition for these density bilinears.} In the $I=2$ control we minimize the off-vacuum norm of the moment by adding the $x_5$ moments of $-(\mathsf n_+-\mathsf n_-)$ and ${:}[(\mathsf n^0)^2-(\mathsf n^z)^2]{:}$, where $\mathsf n_\pm$ are the local flavour densities. These are improvements of the form in Eq.~\eqref{eq:s4improvement}. This reduces the vacuum leakage, but the independent primary test still distributes its weight among several levels, as Table~\ref{tab:s4generator} shows. Improving the density is therefore useful, while the descendant spectrum itself still requires control. The unresolved high-energy weight in the $I=3$ test also prevents interpreting its fractions alone as a continuum Ward-identity violation.

\section{Numerical coverage}\label{app:numerics}
Table~\ref{tab:coverage} summarizes the system sizes and controls underlying the comparisons in the paper. Agreement in a shared sector is a solver check for that sector, and the finite sets of fit variations quoted in the text quantify fit sensitivity. Control of omitted operators and of the continuum extrapolation requires additional data.

\begin{table}[!htbp]\centering\footnotesize\setlength{\tabcolsep}{4.5pt}
\begin{tabular}{lll}\toprule
calculation & principal range & relevant control\\\midrule
Ising defect & ED $\Nm\le14$; selected DMRG $\Nm\le20$ & shared gaps agree to four digits\\
spin-one $O(2)$ & ED $\Nm\le10$; DMRG $\Nm\le14$ & shared ratios within $6\times10^{-5}$\\
bulk and defect flows & mainly ED $\Nm=4$--$9$ & endpoint and trajectory comparisons\\
free displacement & ED $\Nm\le10$; DMRG $\Nm=12,14$ & DMRG equals ED at $\Nm=6$\\
pinned junctions & ED $\Nm\le10$; DMRG $\Nm\le14$ & DMRG equals ED at $\Nm=8$\\
monodromy pinning & ED $\Nm\le10$; DMRG $\Nm\le12$ & DMRG equals ED at $\Nm=8,10$\\
pinned flow & ED $\Nm=6$--$8$ & reproduces the pinned $O(2)$ data\\
large-$N$ trace & $|\kappa|\le160$, levels $\le8000$ & two cutoffs, two truncations\\
large-$N$ junction & $120$--$240$ points, $s_{\max}=30$--$50$ & grid, mode and angle variations\\
large-$N$ mode shifts & $\alpha=0.1$--$0.5$, two lowest modes & two schemes, independent codes\\
pin branches & ED $\Nm=6$--$9$, infinite pole fields & reproduces Appendix~\ref{app:pinning} at $g=0$\\
$O(N^0)$ entropy & $|\kappa|\le160$ ($\alpha=0$), $\le120$ ($\alpha>0$) & two regulators, two codes at $\alpha=0$\\
four-sphere defect & $I=2,3,4$ & band, spin and bond-dimension checks\\\bottomrule
\end{tabular}
\caption{Coverage of the calculations used here.}
\label{tab:coverage}
\end{table}

\section{Notation}\label{app:notation}
Table~\ref{tab:notation} lists the main symbols, grouped by topic, with the place where each is defined. Symbols used only within a single paragraph are defined there and are not repeated here.

{\small
\setlength{\tabcolsep}{5pt}
\begin{longtable}{>{$}l<{$}p{0.60\textwidth}l}
\caption{Main symbols.}\label{tab:notation}\\
\toprule
\text{symbol} & meaning & defined in\\\midrule
\endfirsthead
\toprule
\text{symbol} & meaning & defined in\\\midrule
\endhead
\bottomrule
\endfoot
\multicolumn{3}{l}{\emph{Symmetry, flow and regularization}}\\
N & number of real components of the $O(N)$ vector & Sec.~\ref{sec:intro}\\
d,\ \epsilon & spacetime dimension, and $\epsilon=4-d$ & Eq.~\eqref{eq:bulkaction}\\
(y,r,\theta) & coordinates along the defect ($y\in\mathbb R^{d-2}$) and transverse to it & Eq.~\eqref{eq:monodromy}\\
\alpha & twist: a field of charge $q$ acquires $e^{2\pi\mathrm iq\alpha}$ around the defect; the half twist is $\alpha=\frac12$ & Eq.~\eqref{eq:monodromy}\\
s & transverse spin, $s\in\mathbb Z+q\alpha$ & Eq.~\eqref{eq:monodromy}\\
g_\pi & the $O(3)$ half twist $\operatorname{diag}(-1,-1,+1)$ & Eq.~\eqref{eq:axialtwist}\\
n_t,\ n_{\rm tw} & number of twisted vector planes, and of twisted real components & App.~\ref{app:estimators}, Sec.~\ref{sec:entropy}\\
g,\ \widehat W & anisotropy coupling and operator; $g<0$ is the easy-plane and $g>0$ the easy-axis branch & Eq.~\eqref{eq:amp_ops}\\
S,\ \lambda_4 & bulk action and its quartic coupling & Eq.~\eqref{eq:bulkaction}\\
t_W,\ \Delta_t,\ y_t & traceless tensor that the anisotropy couples to, its dimension $\Delta_t=1.20954$~\cite{Chester:2020iyt}, and $y_t=3-\Delta_t=1.79046$ & Eq.~\eqref{eq:anisoaction}\\
\mathcal Y & scaling variable $|g|\Nm^{y_t/2}$ & Sec.~\ref{sec:flows}\\
\sigma,\ \varepsilon & Ising spin and energy operators, the latter distinct from the expansion parameter $\epsilon$ & Sec.~\ref{sec:memory}\\
\mu_t,\ \mu_{\mathcal E} & continuum couplings of the tensor and the singlet $\mathcal E$ & Eq.~\eqref{eq:anisoaction}\\
\Nm,\ j & number of orbitals of the untwisted lowest Landau level on $S^2$, $\Nm=2j+1$, with monopole spin $j$ & Sec.~\ref{sec:method}, Eq.~\eqref{eq:orb}\\
N_e,\ R & fermion number, $N_e=\Nm$, and sphere radius, $R^2=\Nm$ & App.~\ref{app:three}\\
(\vartheta,\theta) & polar and azimuthal angles on $S^2$ & Eq.~\eqref{eq:orb}\\
\Phi_m^{(\alpha)},\ \mathfrak n_m & regular orbital at fractional flux, and its normalization & Eq.~\eqref{eq:orb}\\
\mathsf f_a,\ \mathsf n_{ma} & fermion flavours ($\mathsf f_0$ singlet, $\mathsf f_1,\ldots,\mathsf f_N$ vector) and orbital occupations & Eq.~\eqref{eq:amp_ops}\\
h & thermal tuning field of the critical Hamiltonian $H_{\rm WF}(h)$ & Eq.~\eqref{eq:amp_ops}\\
\lambda,\ V_{\rm core} & core-potential prescription, with $\lambda=1$ the original and $\lambda=0$ the subtracted Hamiltonian & Eq.~\eqref{eq:coreknob}\\
\gamma_B,\ \bar E_n & bulk energy unit fixed by the light field $\phi_{\rm L}$, and calibrated energies & Eq.~\eqref{eq:flowtune}\\[3pt]
\multicolumn{3}{l}{\emph{Defect data}}\\
\Dc & the defect & Sec.~\ref{sec:defects}\\
|0_\Dc\rangle & ground state on the cylinder in the presence of the defect & Eq.~\eqref{eq:tiltHQ}\\
\hat\psi_s,\ \Dh_s & defect primary of transverse spin $s$ (of charge one for a complex twist) and its dimension & Eq.~\eqref{eq:freespec}\\
\alpha_c & marginal crossing, the twist at which the lowest charged mode reaches dimension one & App.~\ref{app:three}\\
\Delta_\phi & dimension of the bulk order parameter & Sec.~\ref{sec:entropy}\\
\beta_{\mathcal O\hat{\mathcal O}},\ \beta_s & bulk-to-defect coefficient, with $\beta_s=\beta_{\phi\hat\psi_s}$ & Sec.~\ref{sec:flows}\\
b_1 & bulk-response slopes & App.~\ref{app:cpt}\\
\mathfrak v,\ \mathfrak v_{\rm sp},\ \tilde s & strength of the tadpole potential $\mathfrak v/r^2$, its spectator value, and effective spin $\sqrt{s^2+\mathfrak v}$ & Sec.~\ref{sec:entropy}, Eq.~\eqref{eq:betanu}\\
N_{\rm sp},\ \mathfrak w & number of spectators $N-2n_t$, and their self-response $(N_{\rm sp}+2)\epsilon/(N+8)$ & Eq.~\eqref{eq:spgap}\\
\epsilon_{\rm c} & zero of the discriminant of Eq.~\eqref{eq:spgap}; locates the merger of the spectator fixed points only at large $N$ & App.~\ref{app:entropyeps}\\
\operatorname{dg},\ \gamma_{\rm E} & digamma function $(\log\Gamma)'$ and Euler's constant & Eq.~\eqref{eq:betaeps}\\
\hat\phi_{\rm sp},\ \Dh_{\rm sp} & lowest spectator line mode and its dimension & Eq.~\eqref{eq:sqrta}\\
D,\ D_i,\ C_D & displacement operator, its transverse components, and its normalization & App.~\ref{app:estimators}\\
\mathfrak g & defect expectation value; $\log\mathfrak g$ is the defect entropy & Sec.~\ref{sec:results}\\
\mathfrak g_{\rm u},\ \mathfrak g_{\rm pin} & expectation values of the unpinned and the pinned line & Sec.~\ref{sec:order}\\
\mathfrak h & conformal weight, from $\langle T_{ab}\rangle=-\mathfrak h\,\delta_{ab}/(2\pi r^d)$ & Sec.~\ref{sec:entropy}\\
\mathfrak g_{\rm free},\ \mathfrak h_{\rm free} & the same quantities for a free twisted complex scalar & Eqs.~\eqref{eq:loggfree}, \eqref{eq:hfree}\\
\mathfrak S_\Dc & defect entanglement entropy, $\log\mathfrak g-2\pi\mathfrak h$ & Eq.~\eqref{eq:Sdefect}\\
\mathcal A_\alpha & coincident deficit of a twisted component, $\delta g_{\rm tw}=-\mathcal A_\alpha/4\pi^2$, with $\mathcal A_\alpha=\frac\pi2(1-2\alpha)\tan\pi\alpha$ & Eq.~\eqref{eq:Aalpha}\\
\mathcal A_\alpha(\varsigma) & the same deficit at background $\varsigma$ on $S^1\times\mathbb H^2$ & Sec.~\ref{sec:order}, Eq.~\eqref{eq:Asigma}\\[3pt]
\multicolumn{3}{l}{\emph{Large $N$ and the hyperbolic frame}}\\
S^1\times \mathbb H^{d-1} & Weyl frame in which the defect sits at the boundary of $\mathbb H^{d-1}$ & Sec.~\ref{sec:entropy}\\
\chi,\ \nu & geodesic distance and spectral parameter on hyperbolic space & Sec.~\ref{sec:entropy}, Eq.~\eqref{eq:G0H2}\\
\mathcal F,\ \Theta & free energy on $S^1\times \mathbb H^2$, and length of the circle & Sec.~\ref{sec:entropy}\\
\widetilde F & defect free energy continued in $d$, $\sin(\frac\pi2(d-2))\log\langle\Dc\rangle$ & App.~\ref{app:entropyeps}\\
\varsigma,\ \varsigma_* & auxiliary field and its saddle, $\langle\varsigma\rangle=\varsigma_*/r^2$ & Eq.~\eqref{eq:largeMaction}\\
\varsigma_{\rm u},\ \varsigma_{\rm pin} & backgrounds of the unpinned and the pinned line with every component twisted & Sec.~\ref{sec:order}\\
\delta g_{\rm tw},\ \delta g_{\rm sp} & coincident changes of one twisted and one spectator propagator & Eq.~\eqref{eq:gap}\\
\delta g^{(4)}_{\rm tw} & coincident change of one twisted propagator near $d=4$ & Eq.~\eqref{eq:dPi4d}\\
G_0,\ G_\alpha,\ G_\Dc & untwisted, twisted and defect propagators in three dimensions & Sec.~\ref{sec:entropy}\\
\mathcal G_\Delta & AdS$_2$ propagator of dimension $\Delta$ & Eq.~\eqref{eq:AdS2modes}\\
\Pi_0,\ \delta\Pi & $\varsigma$ self-energy kernel without the defect, and its change & Eqs.~\eqref{eq:ctrace}, \eqref{eq:deltaPi}\\
\Sigma_\varsigma,\ \Sigma_a & full $\varsigma$ kernel, and sunset self-energy of component $a$ & Apps.~\ref{app:largeM}, \ref{app:estimators}\\
\mathfrak c & coefficient of the $1/N$ correction to $\log\mathfrak g$ & Eqs.~\eqref{eq:ctrace}, \eqref{eq:cresult}\\
\kappa & circle mode of a kernel on $S^1\times \mathbb H^2$ & Sec.~\ref{sec:entropy}\\
\mathcal B_{\Delta_1\Delta_2},\ e_\ell & bubble of two AdS$_2$ propagators, and its coefficients & Eq.~\eqref{eq:bubbles}\\
\Pi_{0,\kappa},\ W_\kappa & circle mode of $\Pi_0$, and of its inverse & App.~\ref{app:sector}\\
\mathcal I_\kappa & $\nu$ integral of the sector $\kappa$ & Eq.~\eqref{eq:Iposition}\\
\Xi,\ \varphi_i & antiderivative of the bubble integrand, and the radial solutions it is built from & Eq.~\eqref{eq:antider}\\
\mathfrak b_\kappa,\ \mathfrak s_\kappa & closed-form bubble on the lattice lines, and its lattice sum & Eq.~\eqref{eq:bkappa}\\
\mathcal T,\ \mathcal C_{ij} & stress-tensor part of $\delta\Pi$, and its large-momentum coefficients & App.~\ref{app:largeM}\\[3pt]
\multicolumn{3}{l}{\emph{Pinned lines and junctions}}\\
\Dc_\pm & lines pinned in opposite directions & Sec.~\ref{sec:junction}\\
\eta & internal angle between the pinning directions at the two poles & Sec.~\ref{sec:monopin}\\
\Dh_{\rm j}(\eta),\ \Delta_{+-} & junction dimension, and the wall dimension $\Dh_{\rm j}(\pi)$ & Eq.~\eqref{eq:junction}\\
\hat t_a,\ \Ct & tilt operator and its normalization & Eq.~\eqref{eq:junction}\\
\mathfrak r,\ \mathfrak r_\infty & shape factor $\Dh_{\rm j}(\pi)/2\Ct$, and its large-$N$ value & Eq.~\eqref{eq:rdef}, App.~\ref{app:junctionlargeN}\\
\mathfrak r^{(2)} & $\epsilon^2$ coefficient of the shape factor & Eq.~\eqref{eq:rtwoloop}\\
\upsilon_2,\ \upsilon_3 & coefficients of $(1-\cos\eta)^2$ and $(1-\cos\eta)^3$ in $\Dh_{\rm j}$ & App.~\ref{app:twoloop}\\
\Ct^{HQ},\ \Ct^{\angle} & Ward and angle-curvature estimators of $\Ct$ & Eq.~\eqref{eq:tiltHQ}, Sec.~\ref{sec:junction}\\
Q_{12} & generator of the broken internal rotation & Eq.~\eqref{eq:tiltHQ}\\
\tau,\ \mathbf x_\perp & coordinate along a line defect, and the transverse position & Eqs.~\eqref{eq:monopinaction}, \eqref{eq:pinaction}\\
h_{\rm pin} & pinning coupling, in the continuum and on the lattice & Eqs.~\eqref{eq:monopinaction}, \eqref{eq:pin}\\
\Dh_{\rm tilt},\ x_t & lattice tilt level in bulk units, and the tilt excess $\Dh_{\rm tilt}-1$ used in finite-size fits & App.~\ref{app:wallroutes}\\
V'_{\rm N},\ V'_{\rm S} & rotations of the north and south pins at zero angle & Eq.~\eqref{eq:towerS}\\
w_n,\ P_n & single-pole weight of a level in the curvature, and its reflection parity & Eq.~\eqref{eq:towerS}\\
\mathcal Q & descendant part of the curvature sum, equal to one in the continuum & Eq.~\eqref{eq:towerS}\\
\Delta_{\mathcal E'} & dimension of the leading irrelevant bulk singlet & App.~\ref{app:wallroutes}\\
g_\sigma,\ \boldsymbol n & sigma-model coupling and unit order-parameter field near two dimensions & Eq.~\eqref{eq:walld2}\\
N_c & number of components above which walls are irrelevant & Sec.~\ref{sec:junction}\\
a_\phi & one-point coefficient of the pinned line & Eq.~\eqref{eq:v0}\\
\phi_* & large-$N$ amplitude of the pinned field & Sec.~\ref{sec:twistwalls}, Eq.~\eqref{eq:phistaralpha}\\
L_z & transverse rotation generator; $L_z-sQ$, with $Q=Q_{12}$ the $U(1)$ charge, is conserved on the pinned monodromy line & Sec.~\ref{sec:monopin}\\
(\rho,\omega),\ A,\ B & polar coordinates on $\mathbb H^2$ about the junction, and the large-$N$ profiles & App.~\ref{app:junctionlargeN}\\
y_{\rm pin},\ \mathfrak q & scaling exponent $y_{\rm pin}=1-\Dh_s$ of the coupling to a pinned charged mode, and the splitting variable, with $(1+\mathfrak q)/(1-\mathfrak q)$ the ratio of the second-order shifts of its doublet & App.~\ref{app:monopin}\\[3pt]
\multicolumn{3}{l}{\emph{Five dimensions}}\\
I,\ N_I & Yang-monopole charge, with isospin $I/2$, and number of lowest-Landau-level orbitals on $S^4$ & Eq.~\eqref{eq:s4lll}\\
S^2_\Dc & the great sphere that carries the defect & App.~\ref{app:s4}\\
w & complex coordinate transverse to the defect & Sec.~\ref{sec:five}\\
\mathcal S,\ \mathcal J,\ \varpi & Gram matrix, covariant angular form, and its eigenvalue & Eq.~\eqref{eq:galerkin}\\
\end{longtable}
}

\bibliographystyle{JHEP}
\bibliography{references}
\end{document}